\documentclass{article}
\usepackage[a4paper, total={6.5in,9in}]{geometry}
\usepackage[utf8]{inputenc}
\usepackage{amsmath}
\usepackage{amssymb}
\usepackage{graphicx}
\usepackage{authblk}
\usepackage{url}
\usepackage[authoryear]{natbib}
\usepackage{algorithm2e}
\usepackage{appendix}
\usepackage{adjustbox}
\usepackage{rotating}
\usepackage{appendix}
\usepackage{comment}
\usepackage{longtable}
\usepackage{xcolor}
\usepackage[english]{babel}
\usepackage[autostyle]{csquotes}

\vspace{-40pt}
\title{Topological Data Analysis Ball Mapper for Finance}
\vspace{-30pt}
\author[1]{Pawe{\l} D{\l}otko\thanks{Full Address: Dioscuri Centre in Topological Data Analysis, Mathematics Institute, Polish Academy of Sciences, Warsaw, 01-2000, Poland. Email:pdlotko@impan.pl}}
\affil[1]{Dioscuri Centre in Topological Data Analysis, Warsaw, Poland}
\vspace{-40pt}
\author[2]{Wanling Qiu \thanks{Full Address: Hawkes Centre for Empirical Finance, Accounting and Finance Department, School of Management, Swansea University, Bay Campus, Swansea, SA1 8EN, United Kingdom. Email:wanling.qiu@swansea.ac.uk}}
\affil[2]{Hawkes Centre for Empirical Finance, Swansea University, UK}
\vspace{-30pt}

\author[3]{Simon Rudkin \thanks{\textbf{Corresponding Author}. Full Address: Institute for Data Science and Artificial Intelligence, University of Manchester, Oxford Road, Manchester, M13 9PL, United Kingdom. Tel: +44 (0)1792 606325 Email:s.t.rudkin@swansea.ac.uk}}
\affil[3]{Institute for Data Science and Artificial Intelligence, University of Manchester, UK}
\vspace{-30pt}

\begin{document}

\maketitle

%\begin{center}
%    \begin{Large}
%    Topological Data Analysis Ball Mapper for Finance
%    \end{Large}
%\end{center}

\begin{abstract}
    Finance is heavily influenced by data driven decision making. Meanwhile our ability to comprehend the full informational content of data sets remains impeded by the tools we apply in analysis, especially where the data is high-dimensional. Presenting the Topological Data Analysis Ball Mapper algorithm this paper illuminates a new means of seeing the detail in data from data shape. With comparisons to existing approaches and illustrative examples, the value of the new tool is shown. Directions for employing Ball Mapper in practice are given and the benefits reviewed.  
\end{abstract}

\section{Introduction}

Topological Data Analysis (TDA) is a data science methodology which allows users to fully appreciate the shape of data. Ball mapper (BM) is a TDA tool which goes further to create a map of the data's shape and facilitates the visualisation of multiple dimensional data. By understanding the appearance of the joint distribution through BM, the analyst may extract otherwise hidden insight into the behaviour of outcomes of interest across that joint distribution. This paper unlocks the potential for finance by demonstrating the method, the robustness of inference upon BM and through an exposition of examples that inspire the research agenda.

Financial data is regularly understood through its' appearance in graphs, such as the candlestick charts described in \cite{malkiel1999random} and evaluated as important in \cite{marshall2006candlestick}. \cite{lo2000foundations} argue that there is power in the work on charting by \cite{malkiel1999random} that can be harnessed for asset pricing. More recently exploitation of graph shape for forecasting in \cite{jiang2020re} demonstrates how seeing the appearance of data, in that case via machine, has value. Whether implicitly or explicitly, it is the visual information from the data shape that is being used. This paper discusses a new means of understanding the shape of data in multiple dimensions, with applications in credit risk following \cite{qiu2020refining} and stock market direction forecasting inspired by \cite{nyberg2011forecasting}. We demonstrate how additional information may be drawn from continuous financial variables using  Topological Data Analysis (TDA) and the BallMapper(BM) algorithm of \cite{dlotko2019ball}. Through a systematic exploration of the parameters an important consistency is shown which may be used by researchers in Finance to leverage the power of the shape of data.

TDA considers data through it's shape, exploiting measures of the space to inform on content. Exploring data through pattern visualisation is not new, the scatter plot can be thought of as a cloud of points on two axes. As typically taught, elementary statistics begins with notions of correlation and association that are visible from these two-dimensional clouds. It stands to reason that the same information would be contained within a three-dimensional cloud, a four-dimensional cloud etc. All which challenges the realisation of this potential is that visualisation over multiple dimensions sits beyond our usual capabilities. We show how BM overcomes these limitations.

Value in visualisation is well understood on the broader scale, fundamental statistics tells us that simply imposing relationships on data is insufficient to ensuring that we have a good model. An excellent demonstration of the disconnect between models, summary statistics and the visualisation of data is provided by \cite{anscombe1973graphs}\footnote{Anscombe's quartet \citep{anscombe1973graphs} are for data sets which have the same regression line when regressing upon a single explanatory variable. Viewing the scatterplot with the independent variable on the horizontal axis and dependent variable on the vertical axis shows that in fact the relationships are often non-linear and that therefore the imposition of a linear relationship is erroneous.}. When we see data and the model which has been fitted in the same image we appreciate far better the quality of the model fit. A second motivation for visualisation stems from the datasaurus example of \cite{matejka2017same}, which proves that data sets with identical summary statistics can have very different shapes. Again, by creating a scatter plot those shape differences are shown clearly, allowing the analyst to derive far greater insight than is provided by the summary statistics alone. The subsequent debate about the importance of graphical analysis highlights the importance of ensuring ease of accurate interpretation and cautions against overuse of the functionality of the scatter plot \citep{cleveland1984graphical,cleveland1984many}. 

%\cite{cleveland1984graphical} identifies the circle is a more valuable portrayer of information since the eye is better at spotting circles than other shapes. BM employs solid circles as the natural representation of higher dimensional balls and therefore benefits from the ability of the reader to interpret solid circles best.

Scatter plots are by nature two-dimensional with the additional overlay of information through colour, size and shape offering up to five dimensions. Notwithstanding the arguments of \cite{cleveland1984many} and others about the dangers of presenting all of these within one plot, we may also recognise that data sets move well beyond five dimensions. In this paper we use artificial examples with six dimensions, five inputs and one output for computational efficiency, this takes us one dimension beyond what a single scatter plot could show. To facilitate more dimensions researchers have employed panels of scatter plots with each row and column representing a pairwise comparison\footnote{\cite{cleveland1984many} notes that the work of \cite{tukey1981graphical} was amongst the first to use this matrix approach, with \cite{carr1987scatterplot} one of the early extenders into multiple dimensions.}. Such can be combined with the outcome to add a third dimension to each plot. However, this does not show the full nature of the data within a single plot. More dimensions mean the page space for each is also necessarily reduced. BM offers a solution to enable the full comparison to remain.

Challenges of visualisation appear throughout the discussion, including in \cite{cleveland1984graphical}, \cite{cleveland1984many} and more recently in \cite{correll2018looks}. One challenge not presented by these papers is that of axes. In scatter plots the axes are well defined since there are two and they follow the familiar Cartesian coordinate design. Extensions to get multiple axes within the two dimensional constraint change the focus to having each observation represented by a set of points on various axes radiating from the centre. First documented in \cite{mayr1877gesetzmassigkeit}, these radar plots are formalised in \cite{kolence1973software}. Each co-ordinate can be imagined as a point on a shape and many will join these points with lines to represent each observation by the shape the co-ordinates define. Very quickly the overlaying of observations will obscure which observation is which and the data will become confused. One solution is to show the polygon for all observations separately,  but in big data sets this will quickly become impractical. The existence of these plots may tempt the reader to believe there are axes in the BM graph; there are not. BM is presenting an abstract representation of multiple dimensions and therefore does not have the natural axes that scatter plots have. In BM if we want to see each axis we must colour the balls according to the axis value, this is discussed further later.

Exploiting the benefits outlined, early examples of the application of BM within Finance are \cite{qiu2020refining} and \cite{dlotko2019financial}. In these works the focus is on understanding how outcomes, firm failure and stock returns respectively, may be visualised on the whole space of common explanatory factors. There are many such situations in which we may wish to see the way that outcomes vary across the space. Immediately non-linearities, the importance of interactions, and potential anomalies within the outcome variable are apparent. Often there are real world implications to be drawn from the visualisation, such as the conclusion identified in \cite{qiu2020refining} that failing firms occupy just a small subset of the full danger zone first developed in \cite{altman1968financial}.

BM is just one of the algorithms through which TDA is used to construct visualisations of data. Most closely related is the mapper algorithm of \cite{singh2007topological}. The first of the mapper algorithms, the approach has a functional instability which makes small perturbations of the data produce notably different graphs \citep{dlotko2019ball, belchi2020numerical}. \cite{carriere2018structure} offers further analysis of the ability of the original mapper algorithm to represent data, providing suggested solutions to some of the instability issues. Despite this adoption of mapper within the finance literature may not be found\footnote{The only published exception identified at the time of writing is the recent work by \cite{kim2020investigation}.}. As discussed in \cite{baastopological}, commercial firms are using mapper as a way to identify potential fraud. As our aim is to produce robust visualisations the inherent appeal of the BM algorithm is its employment of a single parameter, the ball radius, and a simple to interpret approach to the construction of the cover. This paper represents both a demonstration of the value of TDA and reaffirmation of BM as the means through which that value is best delivered.

Contributions of the paper are threefold. Firstly, this paper provides the methodological introduction of TDA, and specifically BM, as a means of representing data within Finance.  Secondly, comparisons with established techniques included herein are as yet undeveloped outwith the natural science literature. Finally, this work functions as a guide for the implementation of BM to derive more detailed understanding of data across Finance and its' sub-disciplines. 

%To further promote the contribution all of the R code required to reproduce the results in this paper is provided within the supplementary material. [****] ADD CODE IN A LATER REVISION

The remainder of the paper is organised as follows. Section \ref{sec:rep} briefly reviews the ways in which newly assembled data is first approached in Finance. Providing the overview of the BM approach Section \ref{sec:bm} gives the methodological advancement to the set of data analysis tools. To illustrate the method Section \ref{sec:bmeg} shows example BM graphs for our two artificial clouds. Section \ref{sec:meas} considers metrics for the understanding of these BM graphs and captures the effects of the number of points in the data set, average correlation, numbers of variables to be plotted, and the role of the ball radius. In Section \ref{sec:colour} we consider alternative ways of colouring the BM plot, and in Section \ref{sec:label} labelling of plots is introduced. Both serve to convey further information. Presenting two applications, Section \ref{sec:app} shows BM in action, before Section \ref{sec:discuss} concludes.

\section{Representing Data}
\label{sec:rep}

Understanding of a data set is regularly targeted in the production of graphs and summary statistics. Alternatively data may first be grouped into clusters and the analysis performed thereupon. Presentation of the intuition behind these established phases in this section serves to motivate the adoption of BM advocated in the remainder of the paper. 

\subsection{Artificial Data Sets}

Herein the data being visualised consists of five variables, with each drawn randomly from a normal distribution of given mean and variance. To assess the properties of BM graphs there will be two clouds considered. In the first the draws for each axis are made from standard normal distributions of mean 0 and variance 1. This cloud is essentially noise since there are no relationships between any of the axis variables. We may thus refer to it as the ``noise cloud''. Our second cloud consists of five smaller noise clouds each with a different mean. To almost separate the points the means are set at 0,2,4,6 and 8, whilst the variance remains 1. Because of it's composition from five smaller clouds we refer to the second cloud as being the ``five-part cloud''. In the analyses that follow we explore the impact of imposing correlation structures within these noise clouds, changing the number of points and changing the number of variables. Unless otherwise indicated the number of points will be 500 and the number of dimensions will be 5. Consequently each sub cloud of the five-part cloud has 100 observations contained within it. Use of normal distributions in this paper follows from the commonality of a normal distribution assumption in empirical finance. Qualitatively similar results emerge if we use a uniform distributions to develop the point clouds. 

The BM graph also requires an outcome variable. For this purpose we will simply assume that the outcome is the sum of all of the co-ordinates plus a noise term. Formally for point $i$ the outcome $Y_i$ is given by:
\begin{align}
    Y_i = X_{1i} +X_{2i}+X_{3i}+X_{4i}+X_{5i}+\nu_i \label{eq:out}
\end{align}
where $X_{ji}$ is the value of $X_j$, $j\in[1,5]$ associated with point $i$. Where applied $\nu_i$ is a random term drawn from a normal distribution of mean 0 and variance 0.1. The presence of this noise avoids the overly simplistic summing of all axis values. We may also impose other rules on the colouration and this is discussed in Section \ref{sec:colour}.

\subsection{Summary Statistics}

When tasked with introducing a data set, convention dictates that authors report summary statistics for each variable. Typically this means the mean and standard deviation as measures of place and dispersion. These may be supplemented by measures of the minimum and maximum, or quantiles of the variables, as a means to indicate the distribution of the variables. In Finance it is also common to report the skewness and kurtosis owing to the importance of distributions within the discipline. Our exemplar data for both the noise cloud and five part cloud is normally distributed, but the skewness and kurtosis are reported for completeness. 

Table \ref{tab:sumstats} reports summary statistics for the data set defined in the opening of this section. Panel (a) shows the noise cloud to have the expected properties in that the mean of all of the $X$ variables is 0 and the standard deviation is 1. Minima and Maxima for the respective variables are also similar, reflecting the fact that all of the values are drawn from the same distribution. There is little within panel (a) which suggests that any variable is not normally distributed, the skewness is close to 0 and the kurtosis is close to 3.

\begin{table}
 \begin{center}
     \caption{Basic Data Summary Statistics}
     \label{tab:sumstats}
     \begin{tabular}{ll c c c c c c cc c}
         \hline
         &Variable & Mean & sd & Min & q25 & q50 & q75 & Max & Skew & Kurtosis \\
         \hline
         \multicolumn{11}{l}{Panel (a): Noise Cloud}\\
         &$X_1$ &0.002&1.035&-3.396&-0.672&-0.021&0.661&3.196&0.046&3.243\\
&$X_2$&-0.055&0.959&-2.907&-0.673&-0.063&0.589&2.706&-0.084&3.180\\
&$X_3$&0.032&0.938&-2.93&-0.592&0.067&0.67&2.691&-0.176&2.960\\
&$X_4$&-0.003&1.024&-3.122&-0.67&-0.052&0.68&3.168&0.022&2.867\\
&$X_5$&0.051&1.042&-3.095&-0.664&0.102&0.75&3.022&-0.028&2.868\\
&$Y$&0.028&2.33&-7.081&-1.626&-0.025&1.664&6.829&-0.008&2.882\\

         \\
         \multicolumn{11}{l}{Panel (b): Five Part Cloud}\\
         &$X_1$ &3.978&2.98&-3.206&1.465&4.099&6.31&11.34&-0.028&2.040\\
         &$X_2$&4.03&2.959&-2.455&1.529&4.047&6.338&11.42&0.064&2.022\\
&$X_3$&3.91&3.017&-2.526&1.433&3.912&6.292&10.17&-0.034&1.944\\
&$X_4$&3.941&2.989&-2.2&1.361&3.995&6.474&10.96&0.019&1.923\\
&$X_5$&3.988&3.03&-2.971&1.482&3.944&6.571&10.45&-0.046&1.988\\
&$Y$&19.84&14.3&-4.585&8.259&19.97&31.2&46.55&0.007&1.772\\

%         \\
%          \multicolumn{11}{l}{Panel (c): Skewness Cloud}\\
%         &$X_1$ & 0.049&1.031&-2.737&-0.712&0.037&0.773&3.848&0.163&3.058\\
%&$X_2$&-0.035&1.045&-3.639&-0.608&0.197&0.768&1.501&-0.969&3.365\\
%&$X_3$&0.016&0.993&-1.621&-0.756&-0.150&0.57&3.424&0.899&3.584\\
%&$X_4$&-0.077&1.139&-9.137&-0.558&0.073&0.509&5.411&-1.494&13.89\\
%&$X_5$&-0.018&1.036&-4.210&-0.624&-0.075&0.48&7.033&1.017&9.315\\
%&$Y_1$&-0.067&2.333&-8.539&-1.523&-0.06&1.348&9.025&0.146&4.205\\

\hline
     \end{tabular}
 \end{center}
\raggedright
\footnotesize{Notes: Clouds have 500 data points and are constructed using random draws from normal distributions. In all three cases there a five axes $X_1$,$X_2$,$X_3$,$X_4$ and $X_5$. Outcome variables are calculated such that $Y_1=X_1+X_2+X_3+X_4+X_5+\nu$ with $\nu \sim (0,0.1)$. In panel (a), the Noise Cloud, all five axes, $X_1$ to $X_5$, are drawn at random from a normal distribution of mean 0 and variance 1. In panel (b), the Five Part Cloud, each of $X_1$ to $X_5$ are specified to create five equally sized groups, with means of 0, 2, 4, 6, and 8. It is assumed that there is zero correlation between any of the variables within either cloud 1, or the five sub-clouds of cloud 2.}
\end{table}

Panel (b) of Table \ref{tab:sumstats} represents our second point cloud, the five part cloud. Here the focus is on five sub clouds, each of which would have similar properties to panel (a). However, because each sub cloud has a different mean when we bring everything back together the overall distribution no longer has normal properties. In this case we cannot infer much from the means and standard deviations since there are inevitably very low and very high values within the sample. The large variation in $Y_1$ is likewise not particularly informative. An unintended consequence of the data construction is that kurtosis is lower than the noise cloud. Panel (b) is suggestive of the limitations of simply understanding data from the summary statistics.

\begin{table}
    \begin{center}
        \caption{Correlation for Artificial Clouds}
        \label{tab:cor}
        \begin{tabular}{l c c c c c l l c c c c c}
             \hline
             \multicolumn{6}{l}{Panel (a): Noise Cloud} && \multicolumn{6}{l}{Panel (b): Five Part Cloud}\\
             & $X_1$ & $X_2$ & $X_3$ & $X_4$ & $X_5$ && & $X_1$ & $X_2$ & $X_3$ & $X_4$ & $X_5$\\
             \hline
             $X_1$ & 1 &&&&& & $X_1$ & 1 &&&&\\
             $X_2$ & 0.08 & 1 &&&&& $X_2$ & 0.89 & 1 &&&\\
             $X_3$ & 0.08 & -0.03 & 1 &&&& $X_3$ & 0.89 & 0.89 & 1 &&\\
             $X_4$ & -0.00 & 0.03 & -0.01 & 1 &&& $X_4$ & 0.89 & 0.89 & 0.89 & 1 &\\
             $X_5$ & -0.01 & 0.03 & 0.01 & 0.03 & 1 && $X_5$ & 0.89 & 0.90 & 0.90 & 0.89 & 1\\
             \hline
        \end{tabular}
    \end{center}
\raggedright
\footnotesize{Notes: Clouds have 500 data points and are constructed using random draws from normal distributions. In all three cases there a five axes $X_1$,$X_2$,$X_3$,$X_4$ and $X_5$. Outcome variables are calculated such that $Y_1=X_1+X_2+X_3+X_4+X_5+\nu$ with $\nu \sim (0,0.1)$. In panel (a), the Noise Cloud, all five axes, $X_1$ to $X_5$, are drawn at random from a normal distribution of mean 0 and variance 1. In panel (b), the Five Part Cloud, each of $X_1$ to $X_5$ are specified to create five equally sized groups, with means of 0, 2, 4, 6, and 8.}
\end{table}

A further useful insight comes from the correlation statistics. We know that the noise cloud should have zero correlation between the $X$ variables, but the comparatively small sample size means that some non-zero correlation is likely. Panel (a) of Table \ref{tab:cor} informs us that $X_1$ is correlated with $X_2$ and $X_3$ at 0.08, which is still very close to 0. The process of drawing random clouds from the normal distribution means that another draw could easily produce -0.08 as the correlation. Panel (a) gives broad support to the idea that cloud 1 is the noise cloud. In panel (b) the way that the variables comprise five sub clouds creates a strong correlation when aggregated together. Each sub cloud would have a similar correlation matrix to panel (a) otherwise. These correlation matrices provide useful information on top of the summary statistics table, but there is still a benefit to understanding more about the respective data clouds.

\subsection{Data Visualisation}

\begin{figure}
    \begin{center}
        \caption{Scatter Plots of Selected Series}
        \label{fig:scatter}
        \begin{tabular}{c c}
             \includegraphics[width=7cm]{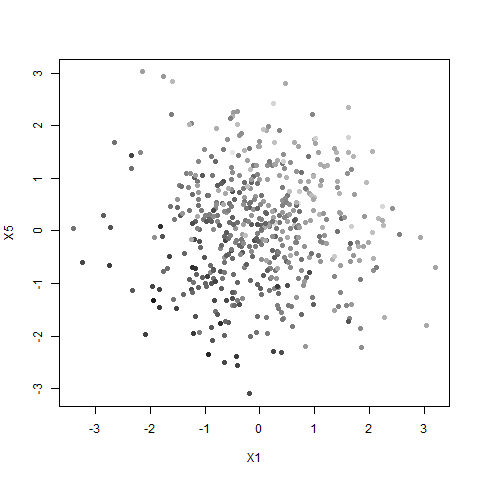} &
             \includegraphics[width=7cm]{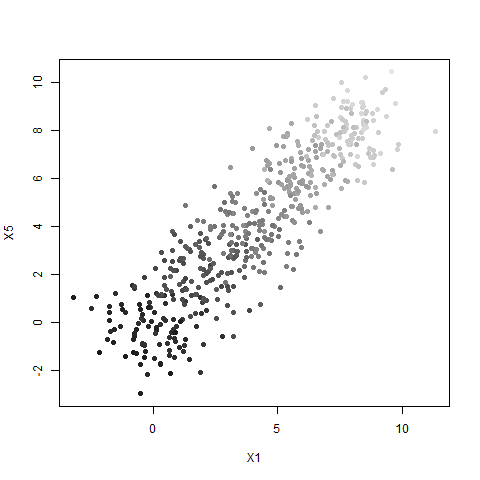} \\
             Noise Cloud ($X_1,X_5$) & Five Part Cloud ($X_1,X_5$) \\
        \end{tabular}
    \end{center}
\raggedright
\footnotesize{Notes: Scatter plots showing the 500 points within each data cloud. Horizontal axis is the variable $X_1$, with the vertical axis in each case being $X_5$. Colouration is on a grey scale from low to high values of the outcome variable $Y$. Clouds have 500 data points and are constructed using random draws from normal distributions. In all three cases there a five axes $X_1$,$X_2$,$X_3$,$X_4$ and $X_5$. Outcome variables are calculated such that $Y_1=X_1+X_2+X_3+X_4+X_5+\nu$ with $\nu \sim (0,0.1)$. In panel (a), the noise cloud, all five axes, $X_1$ to $X_5$, are drawn at random from a normal distribution of mean 0 and variance 1. In panel (b), the five part cloud, each of $X_1$ to $X_5$ are specified to create five equally sized groups, with means of 0, 2, 4, 6, and 8. It is assumed that there is zero correlation between any of the variables within either the noise cloud, or the five sub-clouds of the five part cloud.}
\end{figure}

Directly related to point clouds are our two dimensional scatter plots from which we draw inference on relationships, correlations and underlying distributions. Beyond merely showing the locations of each data point these plots can be extended with colours and shapes to convey more details about the observation(s) they represent. Figure \ref{fig:scatter} plots examples from the three data sets using $X_1$ and $X_5$

\begin{figure}
    \begin{center}
        \caption{Scatter Plots Showing Clustering Groups (Five Part Cloud)}
        \label{fig:cluster}
        \begin{tabular}{c c c}
             \includegraphics[width=5.5cm]{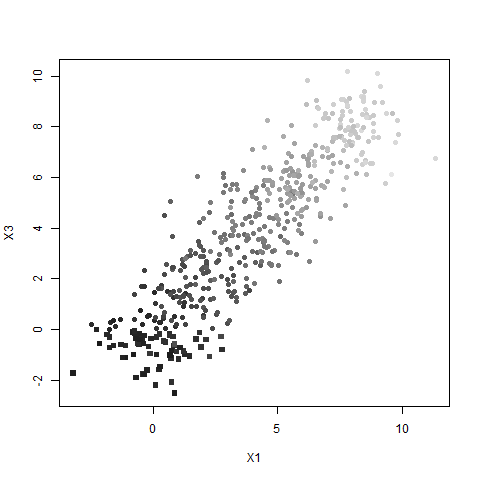} &
             \includegraphics[width=5.5cm]{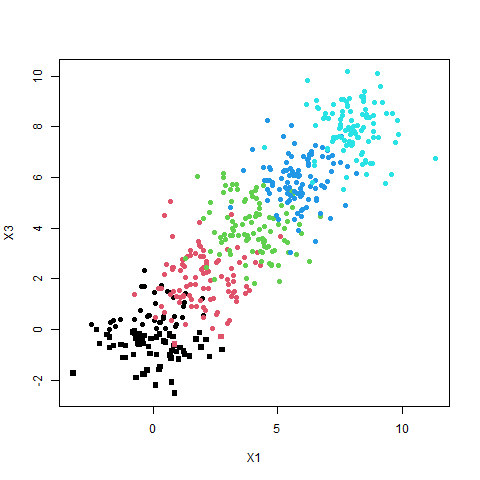} &
             \includegraphics[width=5.5cm]{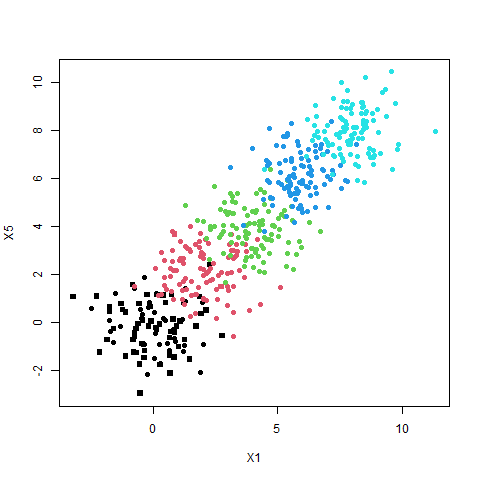} \\
             (a) Original Scatter ($X_1,X_3$) & (b) Grouped ($X_1,X_3$) & (c) Grouped ($X_1,X_5$) \\
        \end{tabular}
    \end{center}
\raggedright
\footnotesize{Notes: Plots all show data from the five part cloud. Here the data is constructed using 5 smaller clouds of 100 points each with a different mean. Panel (a) is coloured according to the outcome variable $Y_1$ with lighter grey implying higher values. Panels (b) and (c) are coloured according to which of the five subsets of the data each point is taken from. Panels (a) and (b) show $X_1$ and $X_3$ for comparison, whilst panel (c) shows that the same effect appears when plotting $X_1$ and $X_5$. Squares denote those points for which $X_3<0$}
\end{figure}

As a second consideration let us apply colouration to the data points to understand more of the process that generates the cloud. Let us also apply a shape distinction by having squares represent any point where $X_3$ is less than 0. Figure \ref{fig:cluster} is designed to show the data generation process for the five part cloud. By colouring the points according to which of the five groups the individual point is drawn from, we are able to see clearly the different means. We may understand also that there is some overlap between each group. Panel (a) is coloured by the outcome variable $Y_1$, lighter colours again showing the higher values. Panels (b) and (c) show $X_1,X_3$ and $X_1,X_5$ respectively, with the colouration being the sub cloud to which the point belongs. It is further apparent from panels (b) and (c) that although the overall cloud has stronger correlation the points within each of the five groups bear strong resemblance to the full point cloud. 

It is further understood that the means of each part of the cloud are only a distance of 2 apart for each variable. Since the standard deviation of the distribution from which points are drawn is 1 it is very possible to have overlap between the colours in panels (b) and (c) of Figure \ref{fig:cluster}. On any axis the distance between two means of the sub-clouds is 2, and hence with a standard deviation of 1 there is an overlap whenever a point in the lower sub-cloud is 1 standard deviation above the mean, and a point in the higher sub-cloud is 1 standard deviation below the mean. The design of the five part cloud is to illustrate how BM can recover information on true group membership amongst multi-dimensional point clouds.

\subsection{Clustering}

Data may be grouped with the aim of understanding the properties of the outcomes of specific groups. Later the similarity between BM and clustering is explored, here we simply look at potential clustering within the three artificial data sets. The k-means clustering algorithm of \cite{hartigan1979algorithm} is used, as implemented within the base package of R \citep{rbase}. K-means targets the division of the data set into a set number of clusters, $k$, such that the distance from every point to the average point within the cluster is minimised. Choice of $k$ is the primary input, but over time a number of algorithms have been devised that suggest optimal levels and optimal centre points \citep{khan2004cluster,steinley2006k,chiang2010intelligent}. Our aim in this paper is to discuss data visualisation, the lack of overlap within k-means clustering renders it unsuitable for the construction of graphs of the BM type. Inclusion of k-means here is therefore as a demonstration of the differentiation of the approach from BM.

\begin{figure}
    \begin{center}
        \caption{K-Means Clustering of Artificial Data Sets}
        \label{fig:clusterart}
        \begin{tabular}{c c}
             \includegraphics[width=7cm]{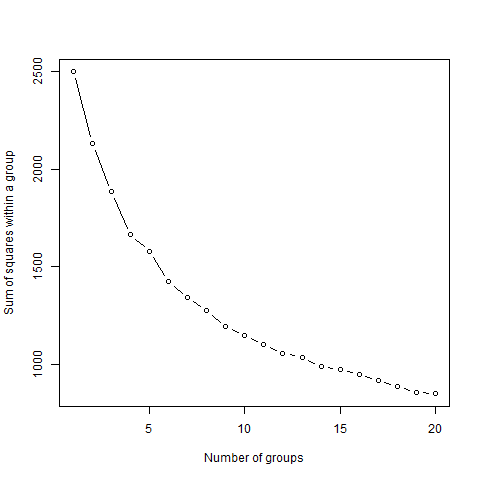} &
             \includegraphics[width=7cm]{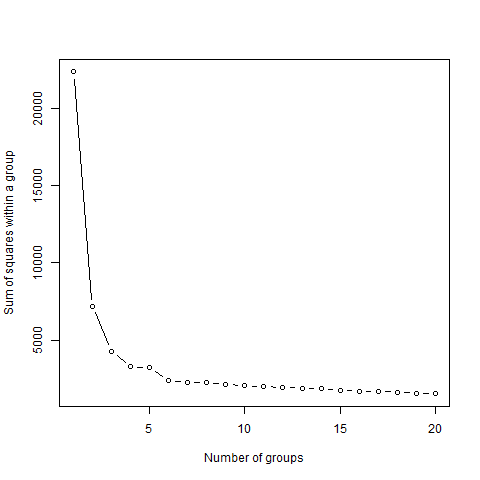}\\
             (a) Noise Cloud Elbow & (b) Five Part Cloud Elbow \\
             \includegraphics[width=7cm]{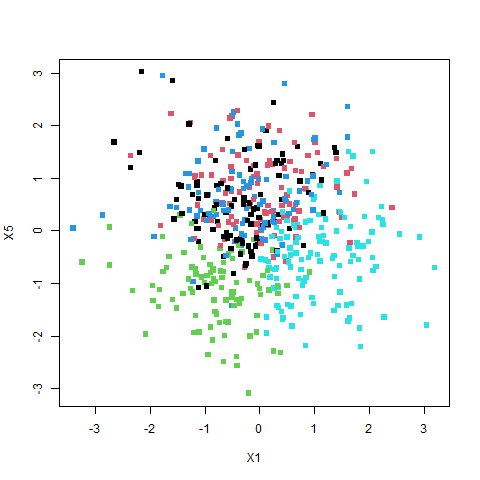}&
             \includegraphics[width=7cm]{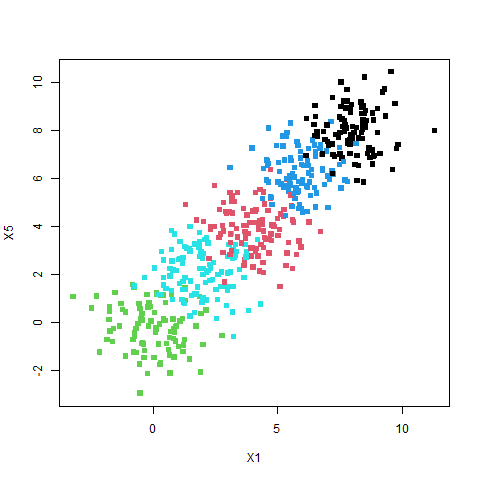}\\
             (c) Noise Cloud Clusters & (d) Five Part Cloud Clusters \\
        \end{tabular}
    \end{center}
\raggedright
\footnotesize{Notes: Panels (a) and (b) represent elbow plots of the within cluster sum of squares for given numbers of clusters. Panels (c) and (d) then plot $X_1$ and $X_5$ as a representative pair of variables from the cloud. Colouration in panels (c) and (d) is according to the cluster to which the data point is assigned. Colouration and labels are applied without loss of generality. Clouds have 500 data points and are constructed using random draws from normal distributions. In both cases there a five axes $X_1$,$X_2$,$X_3$,$X_4$ and $X_5$. Outcome variables are calculated such that $Y_1=X_1+X_2+X_3+X_4+X_5+\nu$ with $\nu \sim (0,0.1)$. In panel (a), the noise cloud, all five axes, $X_1$ to $X_5$, are drawn at random from a normal distribution of mean 0 and variance 1. In panel (b), the five part cloud, each of $X_1$ to $X_5$ are specified to create five equally sized groups, with means of 0, 2, 4, 6, and 8.}
\end{figure}

In the first data set, our noise cloud, there are no natural clusters; all variables are centered on 0 and have normal distributions. Segregation of the data follows simply because the algorithm is targeting a user specified number of clusters. By contrast, there is natural clustering within the second data set and this is picked up by the k-means clustering algorithm. The elbow plot\footnote{Elbow plots here are constructed using the function \textit{wssplot} described in \textit{https://www.r-bloggers.com/2013/08/k-means-clustering-from-r-in-action/}.} of panel (b) shows the within cluster variation falls quickly as the number of clusters increases towards 5. After 5 the variation falls much slower. This contrasts markedly with the elbow plot for the noise cloud. Within a small distance of the origin you may find points in any one of the five clusters. This is a feature of K-means clustering that we may contrast with the BM representations later. For this demonstration it is panel (d) which recovers much of the information of the underlying data generation process that k-means has the most value to represent the point cloud.

Here we may also wish to retain knowledge about the values of $Y_1$ that are associated with each data point. There are many ways to do so, but the simplest is simply to colour according to the average value within a cluster. Because each of these is just a single number for the cluster the effect is qualitatively similar to panels (c) and (d), but with the colour now representative of a specific value. To aid interpretation Table \ref{tab:clustery} reports the colour and associated average $Y$ values for each of the five identified clusters on each cloud. We may note the correspondence of the five part cloud to the underlying means in that the five clusters were specified to have average input values of 0, 2, 4, 6 and 8 respectively, making the average sum of the axes five times these values. Hence we would expect the average $Y$ in the limit to be 0, 10, 20, 30, and 40 respectively. Since the sub-clouds contain just 100 points the estimates seen in Table \ref{tab:clustery} are consistent. The values for the noise cloud are less informative since there is no natural segmentation and the algorithm is producing groups with high values for some axes and low for others; the sum therefore depends precisely on the combination.

\begin{table}
    \begin{center}
        \caption{Cluster Average $Y$ Values}
        \label{tab:clustery}
        \begin{tabular}{l l c c c c c}
        \hline
        &&\multicolumn{5}{l}{Cluster Number}\\
 && 1 & 2 & 3 & 4 & 5  \\
 \hline
 \multicolumn{2}{l}{Colouration used in Figure \ref{fig:clusterart}} & Green & Light Blue & Red & Blue & Black \\
 Noise Cloud & Average $Y$ & -2.73 & 1.33 & -0.51 & 1.24 & 0.43 \\
 & Points in Cluster & 93 & 121 & 97 & 92 & 97\\
 Five Part Cloud & Average $Y$ & -0.31 & 9.63 & 20.13 & 30.09 & 40.07 \\
 & Points in Cluster & 96 & 104 & 104 & 100 & 96\\
 \hline
        \end{tabular}
    \end{center}
\raggedright
\footnotesize{Notes: Colours represent the colouration used in Figure \ref{fig:clusterart}, Numbers are for labelling purposes only. Values are reported for the average value of $Y$ for all members within the cluster and the number of data points which are in each cluster. Clouds have 500 data points and are constructed using random draws from normal distributions. In both cases there a five axes $X_1$,$X_2$,$X_3$,$X_4$ and $X_5$. Outcome variables are calculated such that $Y_1=X_1+X_2+X_3+X_4+X_5+\nu$ with $\nu \sim (0,0.1)$. In the noise cloud, all five axes, $X_1$ to $X_5$, are drawn at random from a normal distribution of mean 0 and variance 1. In the five part cloud, each of $X_1$ to $X_5$ are specified to create five equally sized groups, with means of 0, 2, 4, 6, and 8.}
\end{table}

BM is often likened to a clustering algorithm, though there are some very important differences. Clustering algorithms target segmentation of the data and allow large differentials in cluster size. Further clustering targets inputs, such as the minimisation of the within sum of squares in k-means, and therefore may not identify all of the interesting pictures in the outcomes. BM by contrast captures the features on the input data set in a consistent way that is agnostic to the distances between points. The next section of this paper will set out the BM algorithm to highlight these key differentials and the benefits brought to analysis.

\section{Topological Data Analysis Ball Mapper}
\label{sec:bm}
%
%\SR{PD TO REWRITE - NOTE THAT THE EXISTING DRAFT IS A PARAPHRASE FROM ANOTHER PAPER SO OBVIOUSLY I GOT THE PARAPHRASING WRONG :) }
%
Topological Data Analysis Ball Mapper (BM), as developed in \cite{dlotko2019ball}, builds a \emph{graph--based model} of the considered space $X$ oftentimes equipped with a scalar valued function $f : X \rightarrow \mathbb{R}$. As an intermediate step, for a fixed metric (typically a standard Euclidean metric), it creates a \emph{cover} of $X$ using a collection of balls\footnote{In the two-dimensional setting a ball is a region bounded by a circle. In general dimension, a ball centered in $x \in X$ is composed of points $y \in X$ such that the distance between $x$ and $y$ is less or equal the radius of the ball.} of a fixed radius $\epsilon$. Every point in $X$ belongs to at least one element of a cover, more formally a cover of the space $X$ is obtained by constructing a set of balls, $B(X)$ having the property that $X \subset B(X)$. This property is enforced via a greedy algorithm based on $\epsilon$-net construction. We start from an empty cover, $B(X) = \emptyset$. Then interactively a point $p \in X$ that is not covered by any ball in $B(X)$ is selected at random. A ball, $B(p,\epsilon)$ is drawn and the points $x \in B(p,\epsilon)$ are now considered to be covered and the ball $B(p,\epsilon)$ is added to $B(X)$. The loop ends when there are no longer any uncovered points.

The collection of balls in $B(X)$ will correspond to abstract vertices $V$ of the constructed Ball Mapper graph. Note that repetition and random selection means that there are several different covers that may appear for the same data. Within the BM algorithm as implemented in \cite{dlotko2019R}, the computer simply allocates the first point in the dataset that is not covered according to the order that the cloud is provided at the input of the algorithm. Hence, it is always good to re-run the algorithm a number of times for a permuted data to verify if the information we get is stable. 

Of importance to the result is the choice of the radius of the ball, $\epsilon$. Small radii give detailed pictures but risk becoming over focused on local phenomenon and unable to show the bigger picture. By contrast a too large ball will miss the details and risks missing critical inference. The way that the BM graph is constructed means that the maximum separation on one axis is achieved if two data points from the same ball have identical values all but one variables. In this case, they can be separated by $2\epsilon$ on that one variable on which they are different. In practice, as with the unit circle, the fixed radius will determine how wide a combination of values from different axes may be included. Unlike clustering algorithms, where the number of cluster is discrete, obtaining an optimal radius presents more of a challenge. It is left to the researcher to determine $\epsilon$, though a recommendation to use more than one, and to present some exploration of the role of changing $\epsilon$ is made. Our approach to the examples in Section \ref{sec:app} may be taken as a guide.

Graphically representing the cover with BM requires further steps. Taking the cover, $B(X)$, the cloud $X$ and the radius $\epsilon$, we may identify edges $E$ of the graph. An edge between two vertices of the cover, corresponding to balls centered in $p_1$ and $p_2$, is drawn if and only if there is a point $x$ which sits in $B(p_1,\epsilon) \cap B(p_2,\epsilon)$\footnote{Note that the fact that a distance between $p_1$ and $p_2$ is bounded by $2\epsilon$ is a necessary, but not a sufficient condition for the existence of an edge.}. Through consideration of all possible pairs of vertices the set of edges, $E$ is formed. A BM graph, $G(V,E)$, may then be presented.

Because of the means of production of the BM graphs, full knowledge of which points act as the vertexes means that we may add further detail to the BM plot. Firstly the vertexes are represented by discs which change size according to the number of points of $X$ within that specific ball. Secondly, these varied size balls are then coloured according a given function $f : X \rightarrow \mathbb{R}$ on the points that comprise the ball. To achieve this for every ball $B \in B(X)$, an \emph{aggregation} of values of $f$ at $B$ is computed. Typically this is an average accross the value associated with each point in the ball. The value of the aggregation is then represented in an appropriate colouring scale. The $f$ may be the outcome, the value of any of the axis variables, or some further value which is linked through knowledge of the data points. A discussion of possibilities follows in Section \ref{sec:colour}. An ability to interrogate membership aligns BM with clustering techniques, though the analogy ends there owing to clustering methodologies having algorithms to optimise the membership. BM does not seek to optimise, merely to showcase a representation of the data that is present.

An important final point is to note that BM is extending the balls across every axis. Hence if a certain axis has a dominant span, it will be dominant to the distance between points. This effect can be balanced by normalization of all of the variables before employing the algorithm\footnote{Alternatively, a re-normalization may be used to indicate the levels of importance of various variables.}. In the event that all variables are on the same scale, as is the case for our example clouds, then normalisation will not be essential. For the example clouds of Section \ref{sec:app}, normalisation is used on both datasets.

To summarize, good practice in using the Ball Mapper algorithm requires:
\begin{enumerate}
    \item Verification if the span of each variable for a point cloud $X$ is comparable. If it is not, the user should determine if the difference in spans is desirable, as for instance it reflects the importance of variables, or if this is not the case. In the former case, a normalization of variables may be performed.
    \item A number of constructions for different values of $\epsilon$ should be performed to locate the desired range of $\epsilon$.
    \item In case of using Ball Mapper to visualize the value of the function $f$, it is desirable to check if the variation of the function is not large. That can be achieved by checking the ''relative error'' of the mean value, i.e. to compare the mean value of the function on the ball with the standard deviation of the function. if the first is considerably larger than the second, the Ball Mapper representation of function can be trusted. 
    \item Once a final Ball Mapper plot is obtained, its stability should be validated by performing a number of constructions for similar values of $\epsilon$ as well as permutations of the input points. If any inference persists across the permutations, it may be taken as valid. 
\end{enumerate}

BM requires two parameters for the user to select; the radius $\epsilon$ and the distance between points. In the latter case, most typically, the Euclidean distance is used. The $\epsilon$ choice ensures that the representation is agnostic to the density of the point cloud in any given sub-region, delivering a fair comparison across balls that traditional clustering methods do not. However the process of selecting points from the uncovered set at random means that inevitably there is scope for different representations to emerge from the same data and the same $\epsilon$. Wheresoever there are random draws from a population, the use of bootstrapping facilitates better representation of the data. As we discuss in Section \ref{sec:meas}, there are many ways to capture the messages from the BM graph. This paper demonstrates that having a single BM is still sufficient for the development of inference, provided appropriate thought is given to the selection of $\epsilon$. 

%\SR{I note the comment at the end of the section in handwritten form which says that the user should look at distances between graphs. As we have not yet implemented that code it may be better not to talk too formally about it here. We can always update the paper later if it is ready before journal submission.}

\section{Ball Mapper Representations}
\label{sec:bmeg}

\begin{figure}
   \begin{center}
       \caption{Example Ball Mapper Graphs: Noise Cloud ($\epsilon=2$)}
       \label{fig:egbm1}
       \begin{tabular}{c c c}
            \multicolumn{3}{c}{\includegraphics[width=8cm]{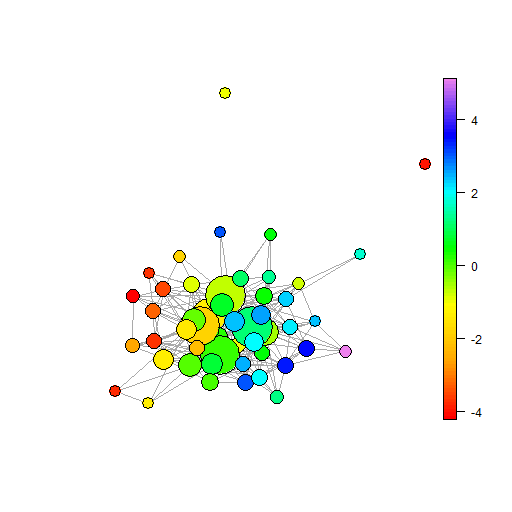}}\\
            \multicolumn{3}{c}{(a) Noise Cloud: Coloured by $Y$} \\
            \includegraphics[width=5cm]{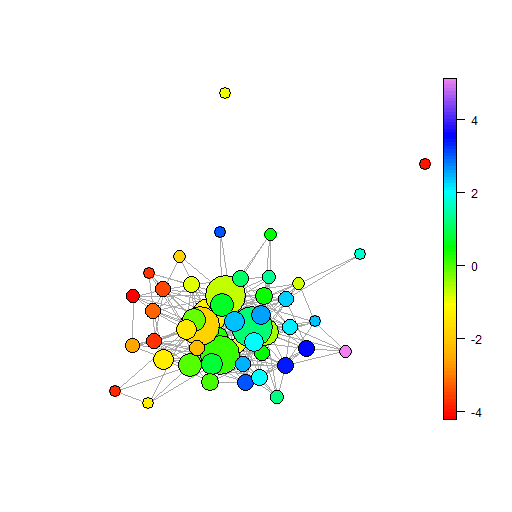} &
            \includegraphics[width=5cm]{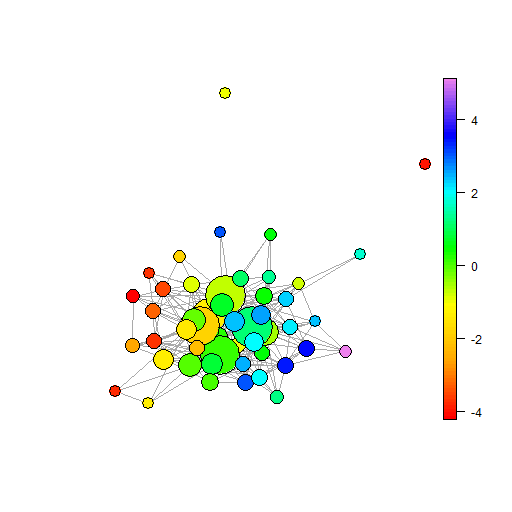} &
            \includegraphics[width=5cm]{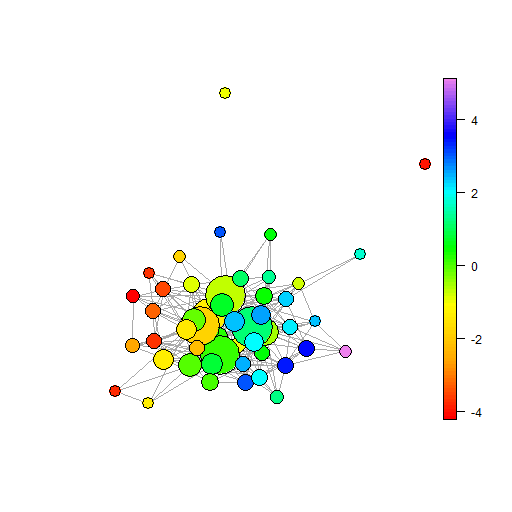} \\
            (b) $X_1$ & (c) $X_2$ & (d) $X_3$ \\
            \includegraphics[width=5cm]{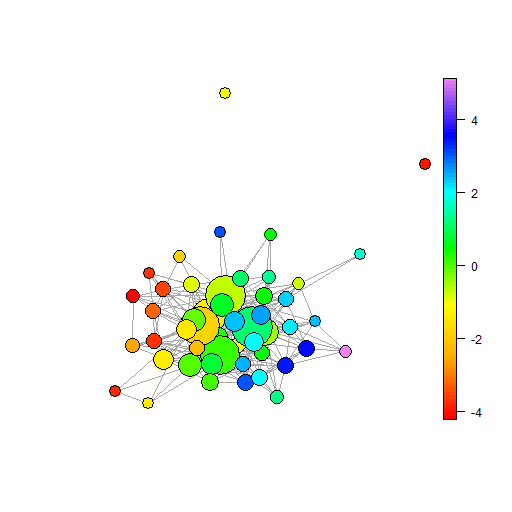} &
            \includegraphics[width=5cm]{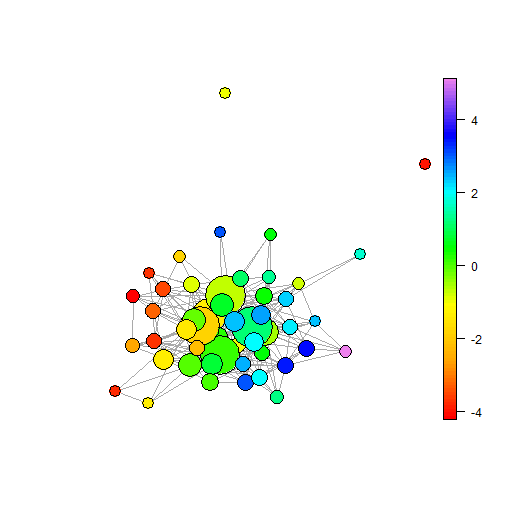} &
            \includegraphics[width=5cm]{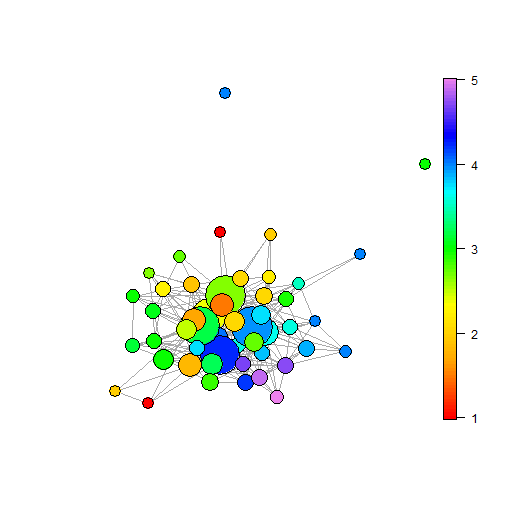}
            \\
            (e) $X_4$ & (f) $X_5$ & (g) Cluster Group\\
       \end{tabular}
   \end{center}
\raggedright
\footnotesize{Notes: Noise cloud comprises 500 data points with 5 variables. Each variable, $X_1$ to $X_5$, is drawn randomly from a standard normal distribution of mean 0 and variance 1. An outcome variable $Y$ is added which is set equal to the sum of the $X$ values plus a noise term $\nu$ with $\nu \sim N(0,0.1)$. Panel (a) is coloured according to the outcome variable, with panels (b) to (f) then coloured by the average value of the respective $X$ variables. Panel (g) is coloured according to the cluster number from Section \ref{sec:rep}. BM graphs created using the R package \textit{BallMapper} \citep{dlotko2019R}}
\end{figure}

As a primary illustration of the function of BM let us first plot graphs for the two clouds introduced in Section \ref{sec:rep}. Figure \ref{fig:egbm1} shows cloud 1 as a main shape of highly connected balls with two outliers. This may be expected since at the ends of a normal distribution the tails become very thin. Even with the ball radius set at 2 the combination of differences in other co-ordinates is likely to leave outliers. Below the main plot of panel (a) there are five plots, panels (b) to (f), that are coloured according to the value of each of the axis variables. A final panel, panel (g), gives the group from the k-means clustering. Such is the density of the cloud in panel (a) that it is hard to determine specific regions of interest, but we may note a broad pattern of increasing values of $Y$ moving towards the right of the shape. 

\begin{figure}
   \begin{center}
       \caption{Example Ball Mapper Graphs: Five Part Cloud}
       \label{fig:egbm2}
       \begin{tabular}{c c c}
            \multicolumn{3}{c}{\includegraphics[width=8cm]{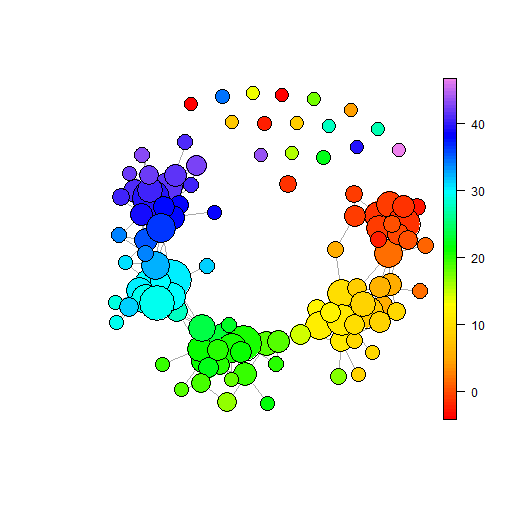}}\\
            \multicolumn{3}{c}{(a) Cloud 2: Coloured by $Y$} \\
            \includegraphics[width=5cm]{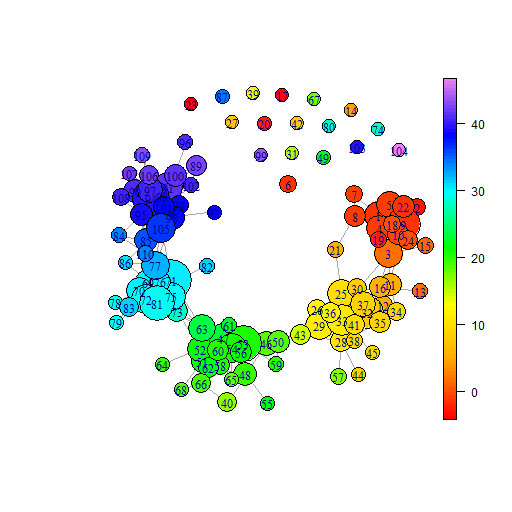} &
            \includegraphics[width=5cm]{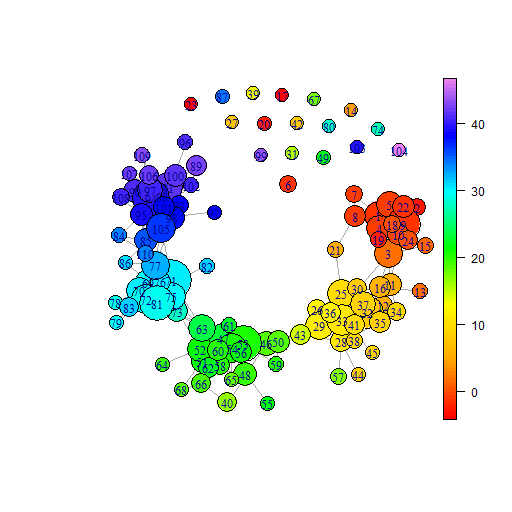} &
            \includegraphics[width=5cm]{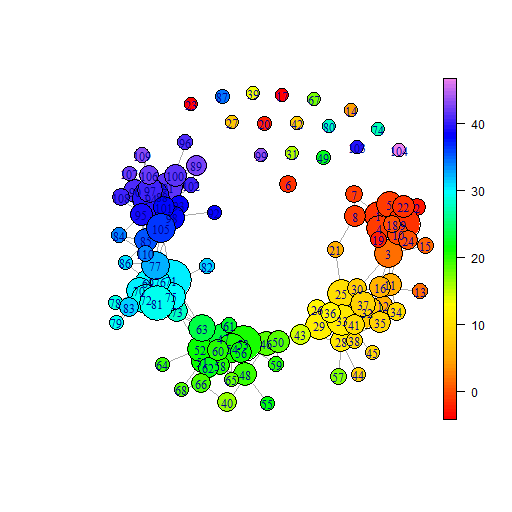} \\
            (b) $X_1$ & (c) $X_2$ & (d) $X_3$ \\
            \includegraphics[width=5cm]{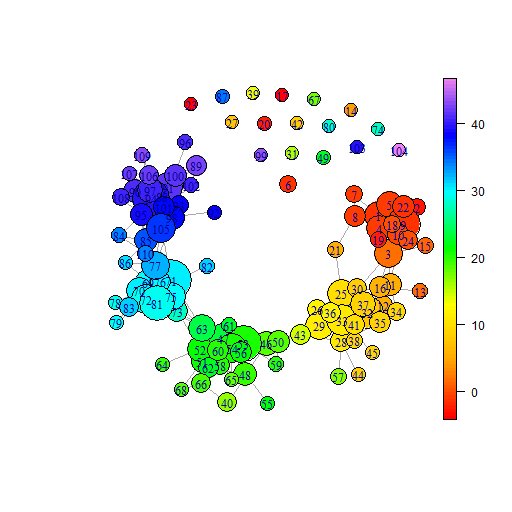} &
            \includegraphics[width=5cm]{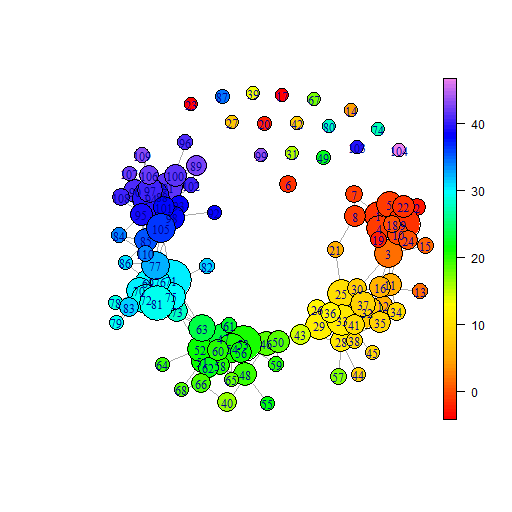} &
            \includegraphics[width=5cm]{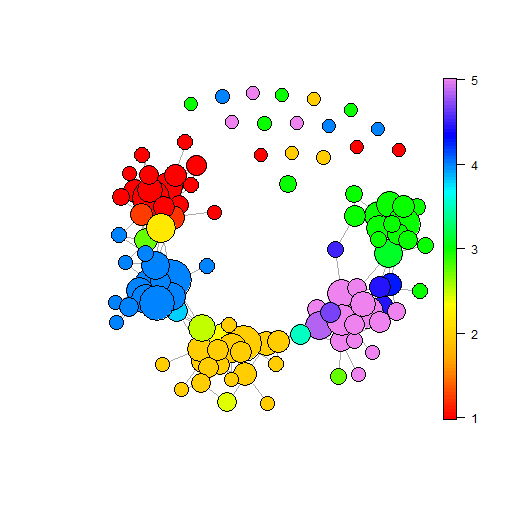}
            \\
            (e) $X_4$ & (f) $X_5$ & (g) Cluster Group \\
       \end{tabular}
   \end{center}
  \raggedright
  \footnotesize{Notes: Five part cloud comprises 500 data points with 5 variables. Each variable, $X_1$ to $X_5$, is drawn randomly from a standard normal distribution of a given mean and variance 1. For the first 100 points the given mean is 0, for the next 100 the mean is 2, for the next 100 the mean is 4 and for the fourth 100 points the mean is 6. For the final 100 points the mean is 8. An outcome variable $Y$ is added which is set equal to the sum of the $X$ values plus a noise term $\nu$ with $\nu \sim N(0,0.1)$. Panel (a) is coloured according to the outcome variable, with panels (b) to (f) then coloured by the average value of the respective $X$ variables. Panel (g) is coloured according to the cluster number from Section \ref{sec:rep}. BM graphs created using the R package \textit{BallMapper} \citep{dlotko2019R}.}
\end{figure}

Plotting the five part cloud, Figure \ref{fig:egbm2} shows distinctly the five groups that make up the whole cloud. Each set of observations appears as a dense set of balls with a few connections then leading into the next dense set. Recalling that the colouration is the sum of the five axes and therefore in the highest cloud each axis having mean value 8 will produce a total of around 40. The heavy blue area is then the fifth of the sub-clouds. At the other end of the connected shape, to the right of the BM graph, is a set of points coloured red and corresponding to mean 0. Between these lie dense regions with averages around 10, 20 and 30, corresponding to the axes being set to have means 2, 4 and 6 respectively. An important point here is that the axes plots of both figures are very similar. Because there are no relationships between the variable, and the distributions from which the values are drawn are identical, we would expect strong similarity and hence it is confirmatory to see BM showing consistency.

In Section \ref{sec:rep} links between BM and clusters were discussed. We may now use BM to visualise the clustering that results. Here colouration is according to the cluster number so the lowest level, the red, is for balls where every point is in cluster 1, and the the highest level, the purple, is for balls where every point is in cluster 5. What we see is that BM is not attempting to split the data, there are balls which sit in the overlap of the clusters and hence have colouration that is not a whole number. We may see this readily from the scatter plots in Figure \ref{fig:clusterart} and the overlaps of groups therein. We may thus view BM as a means of clustering data but it is not one which targets an optimal segmentation on input values, rather it is allowing us to see the space in a transparent and consistent way. In the very rare case where the BM cover and the clustering are concordant there is still benefit in seeing the data shape\footnote{An extreme example is that in which there is no overlap between any of the clusters. This may be achieved in artificial form by following a similar process to our five part cloud and setting the means to be sufficiently far apart. In such cases the five parts would appear as smaller versions of the noise cloud. When the ball radius is sufficiently large one ball can cover one whole sub-cloud but not create overlap with any others. This would produce one ball for each cluster as soon as the largest sub-cloud was covered. Such a case is the only one where the k-means clustering and the BM would give the same result.}. 

\section{Measuring BM Graphs}
\label{sec:meas}

To this point our illustrations have served simply to visualise the impacts of the data set features that drive the appearance of a BM graph. Numbers of points, numbers of axes, correlation and the radius of the balls, have all been shown to cause notable differences to the graphs. We now propose a set of measures that will allow those BM graphs to be better understood. Using the suggested metrics we give a more formal exploration of the role of the number of points an the ball radius. As we saw in the previous section correlation will also have an effect, but the process of constructing appropriate data sets means that the results are not directly comparable. As \cite{matejka2017same} show in two dimensions, a given average correlation can produce some very different point clouds. Here we impose the normal distribution for each axis to create a central mass of points on each dimension. Nonetheless correlation between axes impacts the shape of the cloud just as we understand from basic statistics. Correlations between the axis variables are introduced in this section, generalising from the 0 correlation assumed in the initial noise cloud and five part cloud examples discussed thus far.

%an extra source of variation in graph is brought in through the cloud construction process. \PD{More details should go here, othervise one may think that you are comparing mappers. }

\subsection{Measures}

A BM graph consists of a series of balls, the number of which is a function of the properties of the underlying data set and the radius selected for the BM algorithm. In assessing the properties of these balls we may measure size, connectivity and colouration as well as the number of balls within the graph. To capture size we consider both the average size as well the minimum and maximum size observed within the data set. Often the smallest ball is just a single point, but the largest balls vary greatly depending on the density of the data set. Between the smallest and largest is then a range, which then mirrors closely the maximum size. In the analysis that follows the range of sizes is therefore not included. To capture connectivity we use both the total number of connections and the average number of connections amongst connected balls. Choice of measuring only amongst connected balls is that it helps us understand the behaviour of the map absent of any outlier ball. An outlier ball is defined as being one which is not connected into any other. Both Figures \ref{fig:egbm1} and \ref{fig:egbm2} demonstrated unconnected balls so we know that for $\epsilon = 2$ there are such zero connection balls. Finally, for the colouration we may also capture minimum, maximum and the range. As previously noted BM requires a random selection of landmark points from the uncovered set and therefore the order of the dataset matters. To account for any variation we run 10,000 repetitions of the code and report the average and 95\% confidence interval there around. 

\subsection{Number of Points}

Connectivity between a pair of landmarks in BM graphs exists when there are points in the intersection of the balls that surround that landmark pair. Where there are more data points it stands to reason that the probability of a point existing in the intersection of two balls is much higher. However, there is also a lower probability that any given point is selected as a landmark. If different landmark selections produce different BM graphs then the statistics will necessarily be slightly different, but the extra density of the areas around the landmark will moderate the difference.  Changing the number of points within a data set will not necessarily extend the domain over which observations are seen. Consequently, we would not expect significant variation in the colouration of balls. More points do mean greater potential for outliers as well as meaning that there will be fewer ``holes'' within the main cloud which do not contain points. A modest increase in the number of balls may be anticipated. Intuitively, more points means bigger balls. Such increases will come both from larger number of points falling within the balls at the centre of the distribution, as well as a similar proportional increase in other parts of the space. Eventually we may expect the end of single points within a ball, even amongst the tails of the respective axes. In this subsection we change the number of data points to show exactly how the number of balls, colourations and connections are affected. 

\begin{figure}
    \begin{center}
        \caption{Number of Points: Noise Cloud}
        \label{fig:nc1}
        \begin{tabular}{c c}
             \includegraphics[width=7cm]{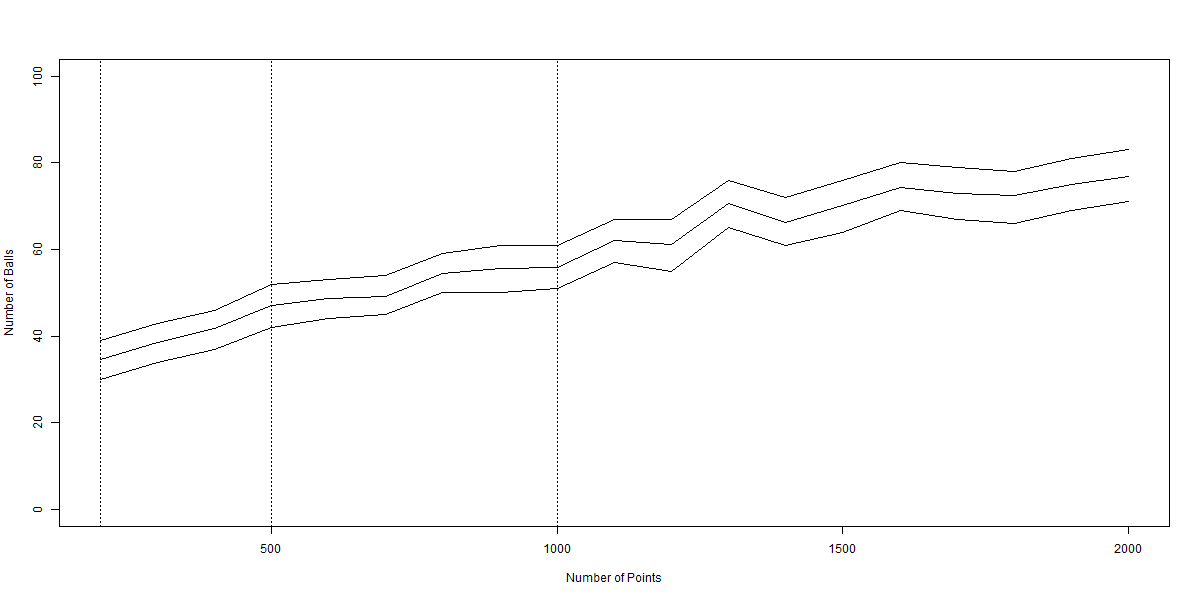} &
             \includegraphics[width=7cm]{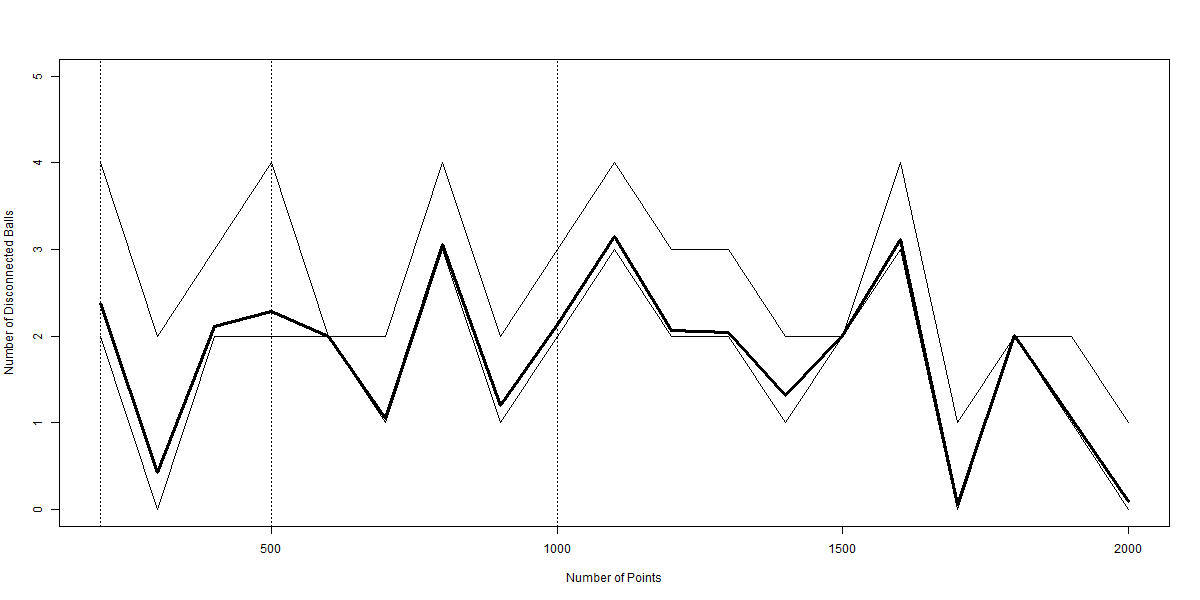} \\
             (a) Number of Balls & (b) Number of Zero Connection Balls \\
             \includegraphics[width=7cm]{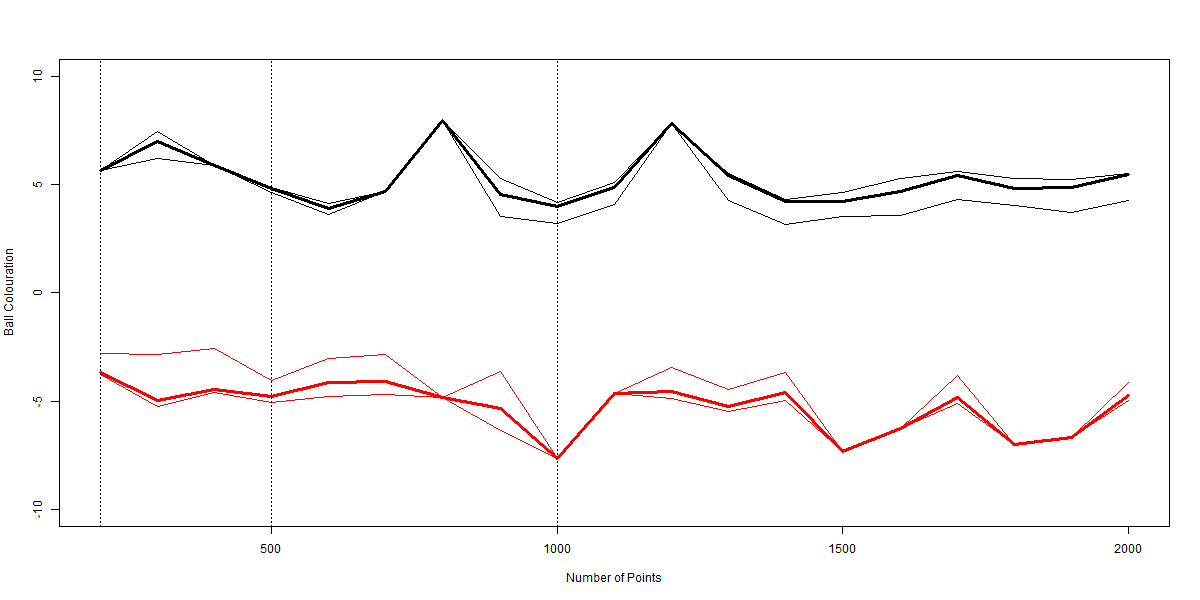} &
             \includegraphics[width=7cm]{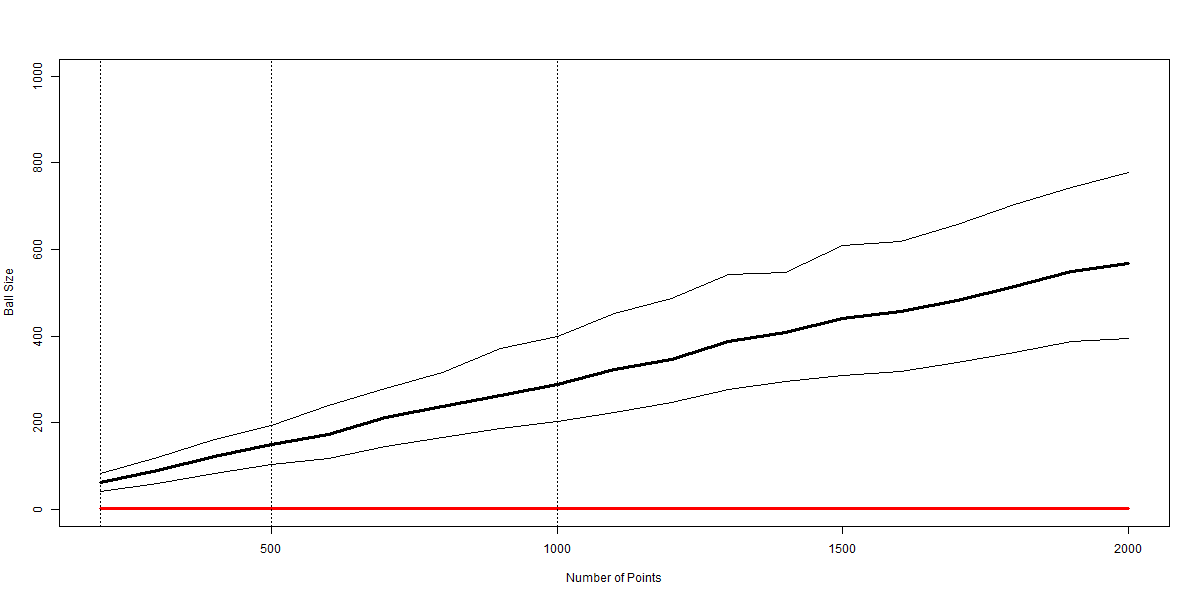} \\
             (c) Colouration & (d) Ball Sizes \\
             \includegraphics[width=7cm]{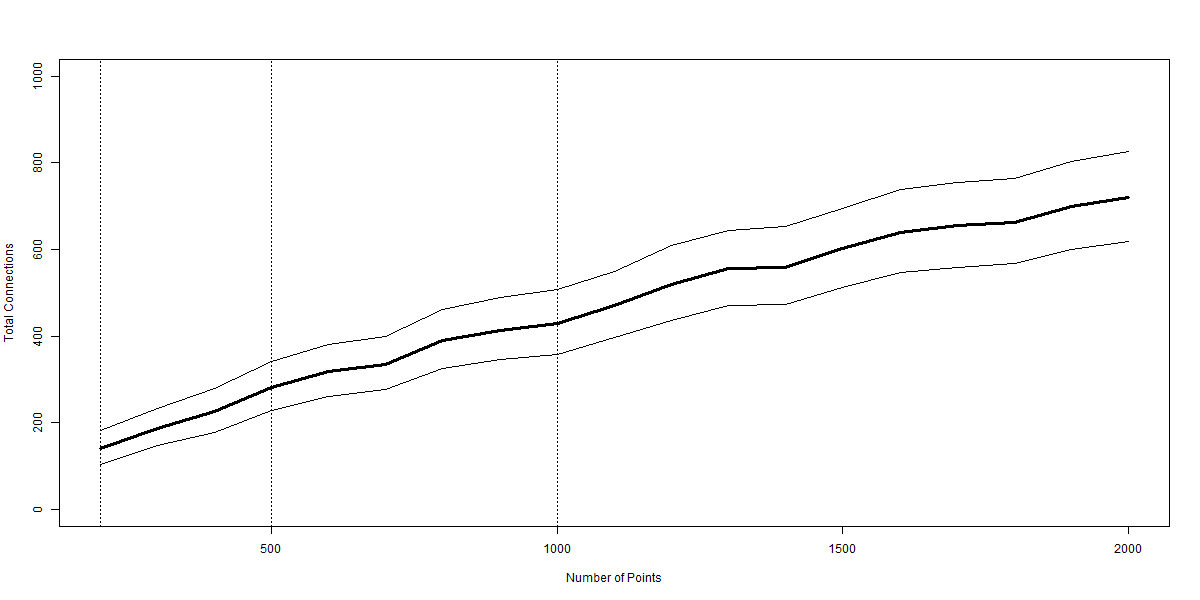} &
             \includegraphics[width=7cm]{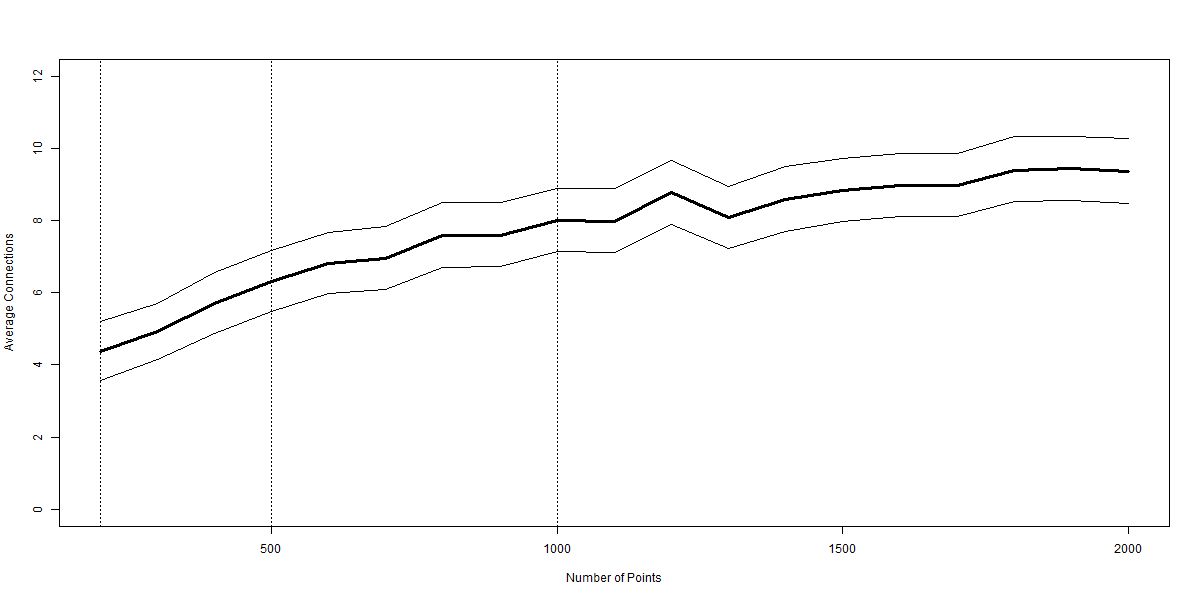}\\
             (e) Total Connections & (f) Average Connections\\
        \end{tabular}
    \end{center}
\raggedright
\footnotesize{Notes: Figures plot the impact of the number of points within the noise cloud on various measures of the BM graph. In each case 10000 repetitions of the BM algorithm \citep{dlotko2019ball} are implemented. A thick line is used to denote the mean from the repetitions, thinner lines denoting the 95\% confidence interval there around. Panel (a) reports the number of balls, and panel (b) the number of balls which have 0 connections to any other balls. Panels (c) and (d) also use red lines to show the maximum and minimum colouration and ball size respectively. Panel (e) reports the total number of connections within the graph, this informs on the points within the overlaps of balls and hence the density of the graphs. Panel (f) plots the average number of connections amongst connected balls. In the case that there are no connected balls then this figure is set to 0. All estimates are generated using the R package \textit{BallMapper} \citep{dlotko2019R}.} 
\end{figure}

\begin{figure}
    \begin{center}
        \caption{Number of Points: Five Part Cloud}
        \label{fig:nc2}
         \begin{tabular}{c c}
             \includegraphics[width=7cm]{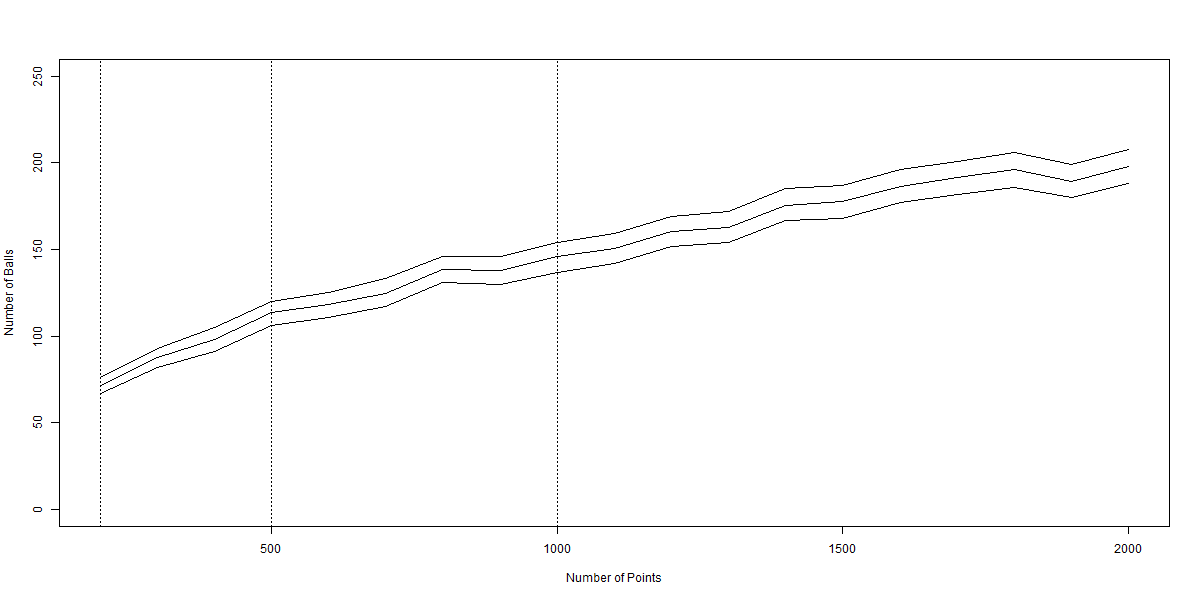} &
             \includegraphics[width=7cm]{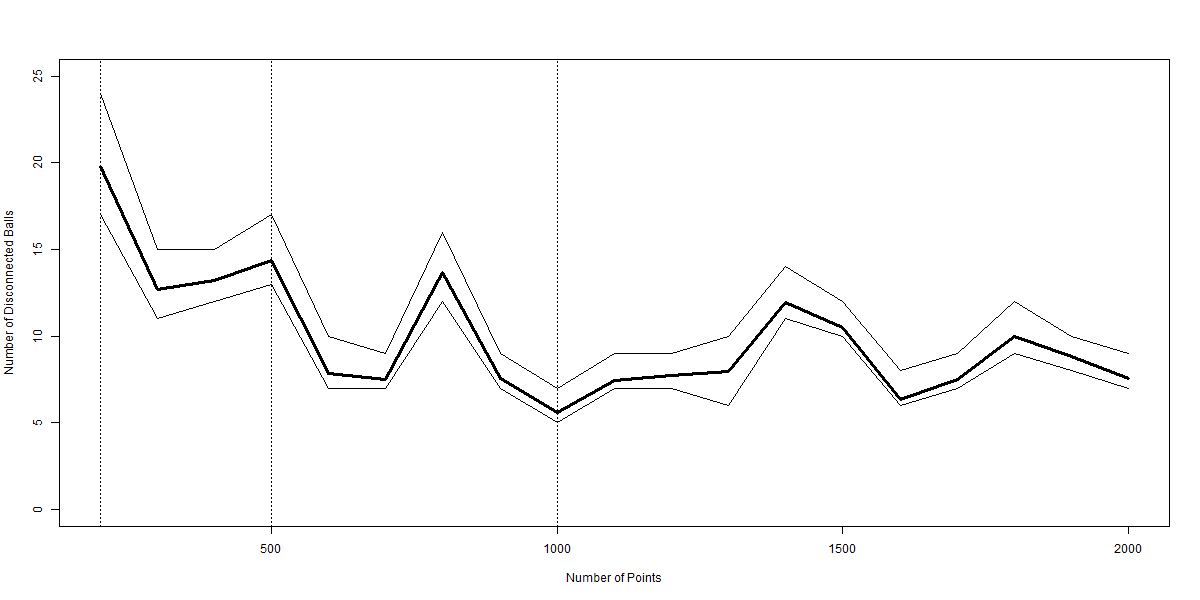} \\
             (a) Number of Balls & (b) Number of Zero Connection Balls \\
             \includegraphics[width=7cm]{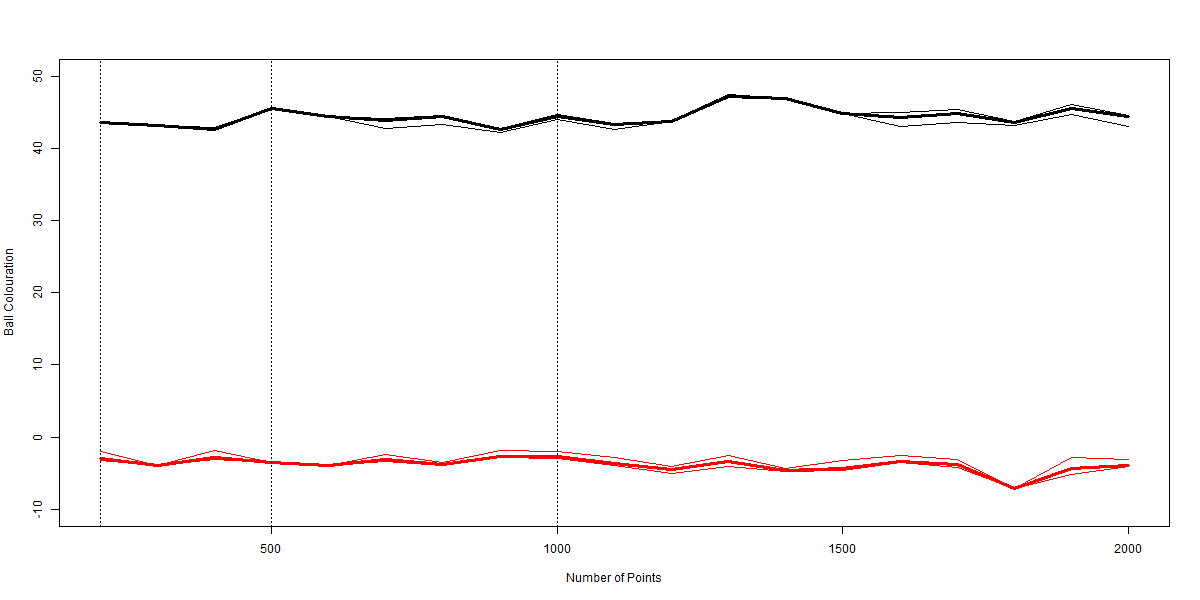} &
             \includegraphics[width=7cm]{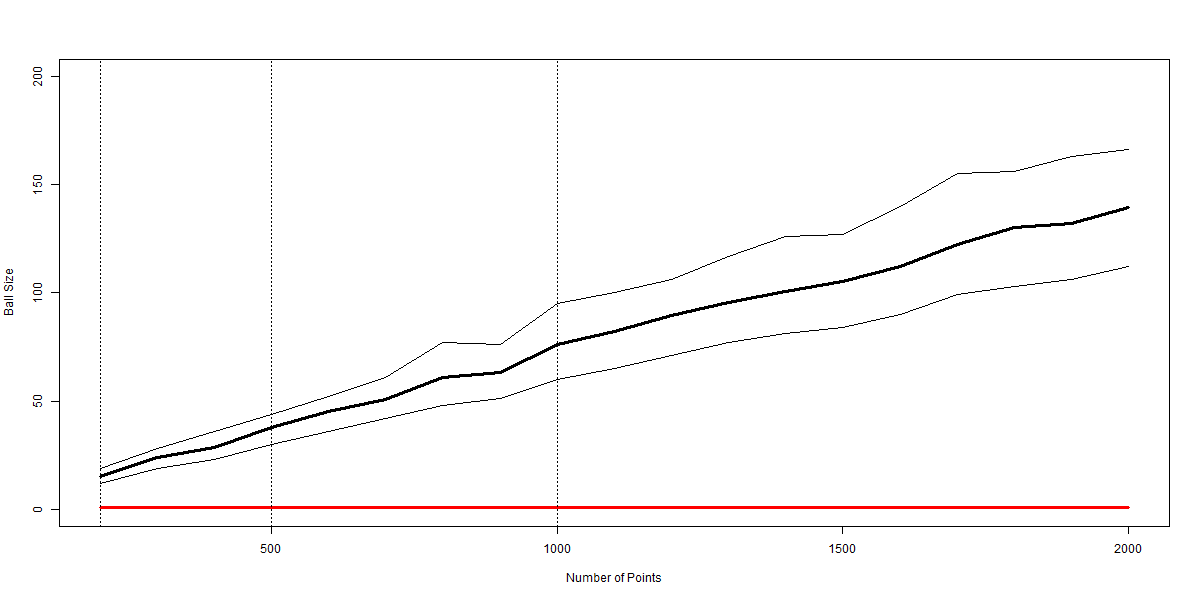} \\
             (c) Colouration & (d) Ball Sizes \\
             \includegraphics[width=7cm]{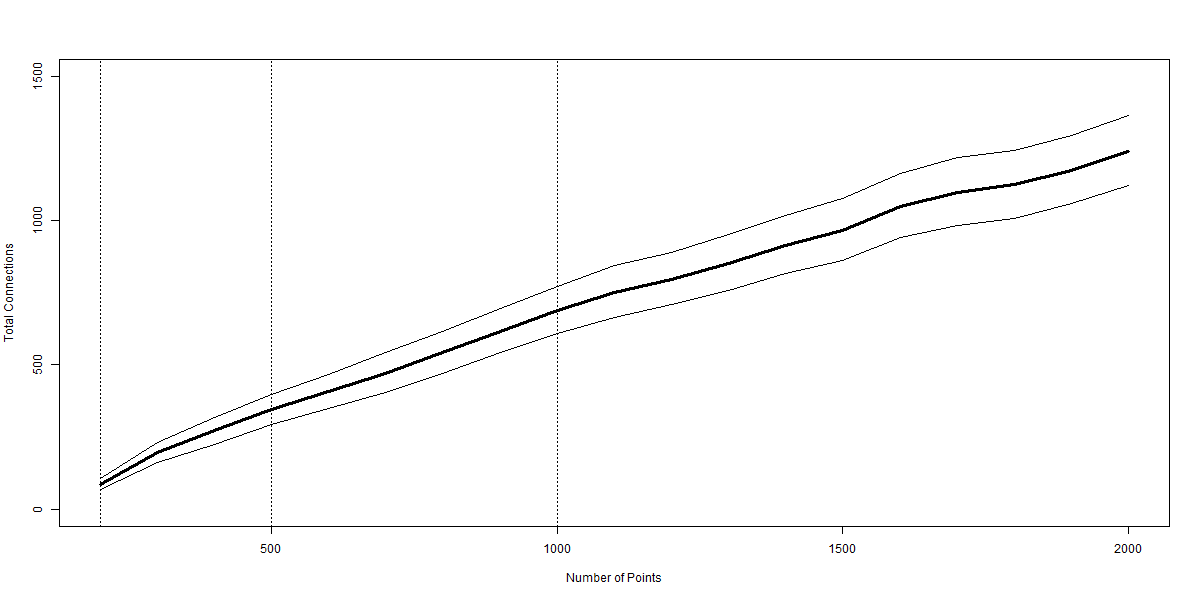} &
             \includegraphics[width=7cm]{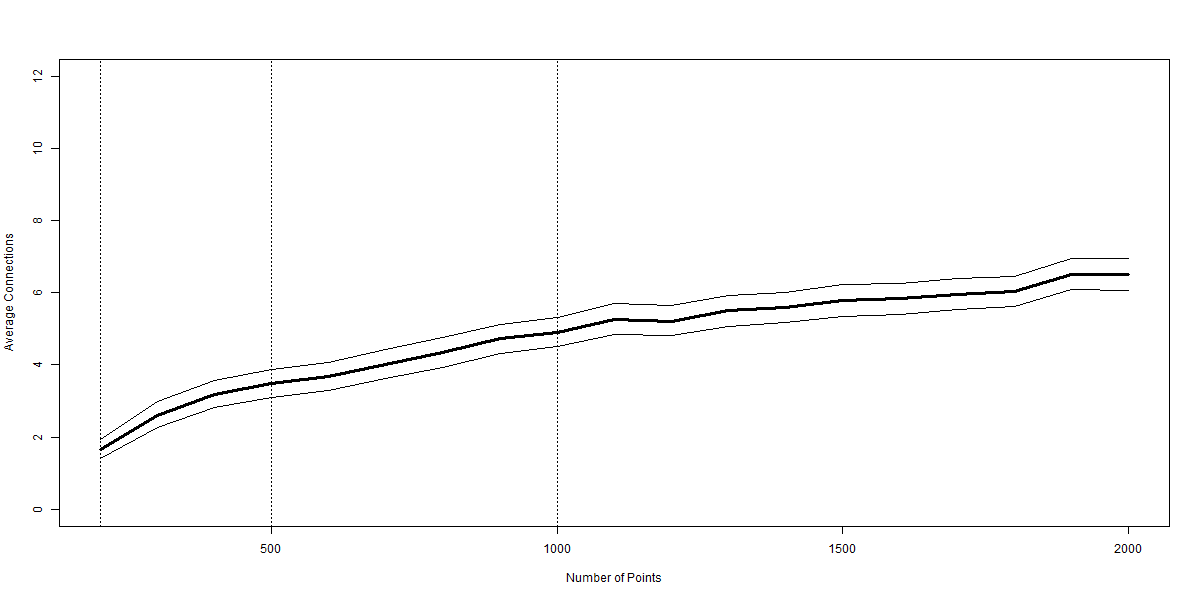}\\
             (e) Total Connections & (f) Average Connections\\
        \end{tabular}
    \end{center}
    \raggedright
\footnotesize{Notes: Figures plot the impact of the number of points within the five point cloud on various measures of the BM graph. In each case 10000 repetitions of the BM algorithm \citep{dlotko2019ball} are implemented. A thick line is used to denote the mean from the repetitions, thinner lines denoting the 95\% confidence interval there around. Panel (a) reports the number of balls, and panel (b) the number of balls which have 0 connections to any other balls. Panels (c) and (d) also use red lines to show the maximum and minimum colouration and ball size respectively. Panel (e) reports the total number of connections within the graph, this informs on the points within the overlaps of balls and hence the density of the graphs. Panel (f) plots the average number of connections amongst connected balls. In the case that there are no connected balls then this figure is set to 0. All estimates are generated using the R package \textit{BallMapper} \citep{dlotko2019R}.} 
\end{figure}

\begin{figure}
    \begin{center}
        \caption{Example Ball Mapper Plots: Number of Points}
        \label{fig:nc3}
        \begin{tabular}{c c c}
             \includegraphics[width=5cm]{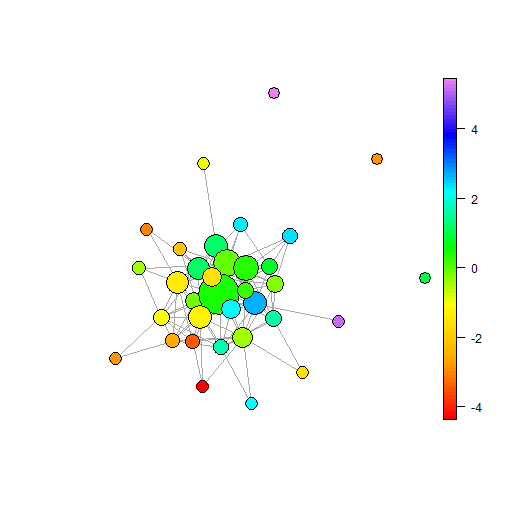} &
             \includegraphics[width=5cm]{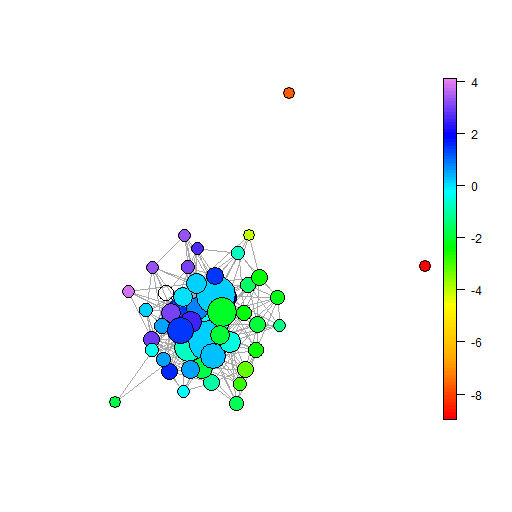} &
             \includegraphics[width=5cm]{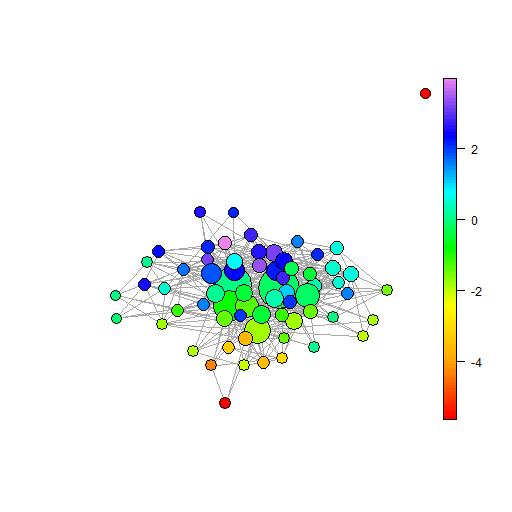} \\
             (a) Noise Cloud ($n=200$) & (b) Noise Cloud ($n=500$) & (c) Noise Cloud ($n=1000$) \\
              \includegraphics[width=5cm]{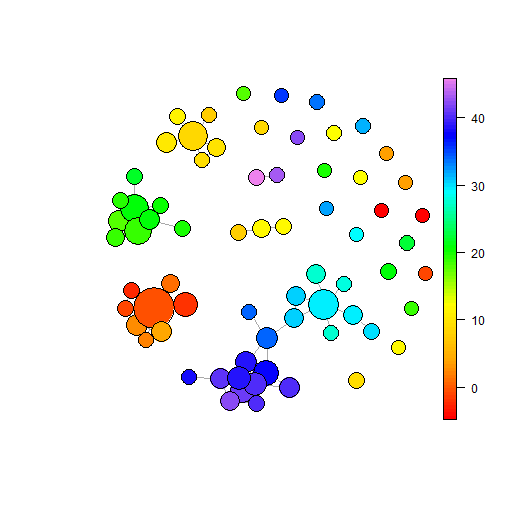} &
             \includegraphics[width=5cm]{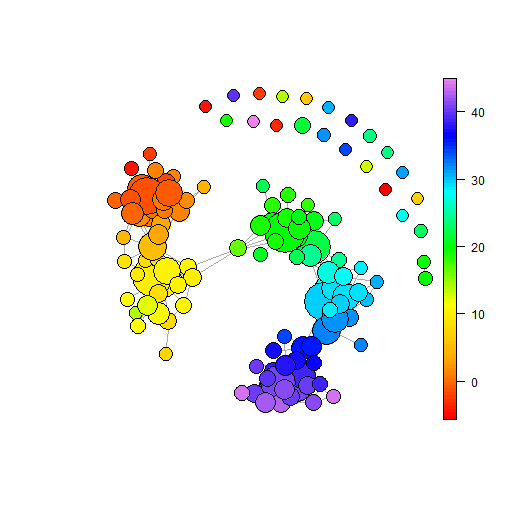} &
             \includegraphics[width=5cm]{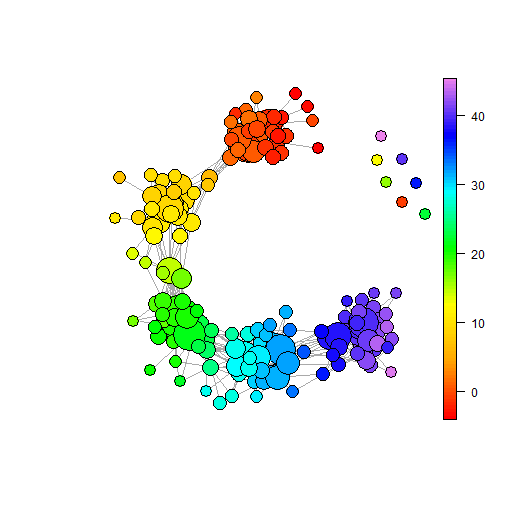} \\
             (d) Five Part Cloud ($n=200$) & (e) Five Part Cloud ($n=500$) & (f) Five Part Cloud ($n=1000$) \\
        \end{tabular}
    \end{center}
\raggedright
\footnotesize{Notes: Figures plot example BM graphs for the stated cloud and point numbers. The noise cloud is constructed using five variables which are independently randomly drawn from standard normal distributions with mean 0 and variance 1. Panels (a) to (c) show this cloud with 200, 500 and 1000 points respectively. The five part cloud is constructed from five sub-clouds where each sub-cloud comprises five variables with a given mean and standard deviation 1. Given means are 0, 2, 4, 6 and 8. All plots generated using the \textit{BallMapper} package in R \citep{dlotko2019R}.}
\end{figure}

In this section we limit our attention to point numbers between 200 and 2000, taking us up to 4 times the number of points as in Section \ref{sec:bmeg}. Such an increase is sufficient to chart the effects of point numbers on the BM graph. Intervals of 100 points are used with selected results reported in Table \ref{tab:nc1}. For brevity only selected values are given and the two clouds are reported within the same table. What we see is that the hypothesised impacts do indeed materialise. However, pattern spotting from these tables is not as simple as if graphs are drawn. Figure \ref{fig:nc1} and \ref{fig:nc2} provide plots of the key measures over the range of data set sizes studied. Within the plots vertical dotted lines are added at 200, 500 and 1000. These are the values used to generate the example plots in Figure \ref{fig:nc3}.

\begin{table}
    \begin{center}
        \caption{Number of Points: Summary}
        \label{tab:nc1}
        \begin{tabular}{l c c c c c l l c c c c c}
        \hline
        \multicolumn{6}{l}{Panel (a): Noise Cloud} && \multicolumn{6}{l}{Panel (b): Five Part Cloud}\\
        Pts & Balls &  $\Delta$Size & $\Delta$Col & Zero & Con && Pts & Balls & $\Delta$Size & $\Delta$Col & Zero & Con\\
        \hline
            200&34.58&59.95&9.33&2.38&4.38&&200&71.47&13.91&46.52&19.77&1.66\\
            &(2.2)&(11.05)&(0.23)&(0.57)&(0.42)&&&(2.21)&(1.55)&(0.32)&(1.85)&(0.13)\\
500&46.97&149.2&9.63&2.29&6.31&&500&113.41&36.65&49.14&14.37&3.48\\
&(2.43)&(26.32)&(0.32)&(0.51)&(0.43)&&&(3.52)&(3.92)&(0)&(1.06)&(0.19)\\
800&54.47&236.67&12.8&3.05&7.58&&800&138.69&59.77&48.18&13.69&4.33\\
&(2.47)&(40.94)&(0)&(0.22)&(0.46)&&&(4.01)&(7.82)&(0.36)&(1.1)&(0.21)\\
1000&55.76&287.96&11.63&2.13&8.01&&1000&145.94&75.33&47.12&5.6&4.91\\
&(2.69)&(53.16)&(0.3)&(0.35)&(0.45)&&&(4.26)&(9.07)&(0.34)&(0.75)&(0.21)\\
1500&70.08&439.14&11.55&2&8.84&&1500&177.78&104.09&49.15&10.51&5.78\\
&(2.93)&(74.42)&(0.31)&(0.07)&(0.45)&&&(4.79)&(11.6)&(0.54)&(0.66)&(0.22)\\
2000&76.88&565.85&10.18&0.09&9.36&&2000&198.17&138.2&48.34&7.53&6.51\\
&(3.12)&(97.23)&(0.4)&(0.29)&(0.45)&&&(5.02)&(14.52)&(0.51)&(0.7)&(0.22)\\
\hline
        \end{tabular}
    \end{center}
\raggedright
\footnotesize{Notes: Pts reports the number of points used to form the point cloud, Balls is the total number of balls within the BM graph, $\Delta$Size is the difference in size between the smallest and largest ball, $\Delta$Col is the difference between the highest and lowest colouration value for any ball within the graph, Zero is the number of balls for which there is no connectivity to any other ball and Con. is the average number of connections per ball amongst those balls that have at least one connection. All figures are the means from the 10000 repetitions at each point number, with figures in parentheses being the standard deviation across all values within the 10000 repetitions. The noise cloud comprises 5 variables each drawn at random from a standard normal distribution of mean 0 and variance 1. The five part cloud comprises 5 sub-clouds each of which contains one fifth of the total number of points. Within the sub-clouds values for each of five variables are drawn at random from a normal distribution of given mean and variance 1. Given means are 0, 2, 4, 6 and 8 for sub-clouds 1 to 5 respectively.}
\end{table}

Figure \ref{fig:nc1} shows the results of 10000 bootstraps of the BM graph with numbers of points from 200 to 2000 in intervals of 100. An increasing number of balls accompanying the increased number of points is clear from panel (a). Whilst the assumed distributions of the variables do not change we may see here that there are more points appearing in the extremes of the distribution. Further as this happens there are also increasing number of points to create connections between former outliers and the main shape. Here it follows that the there will be more connections and a downward pressure on the number of zero connected balls. Panel (b) informs this is the case, with the number of zero connection balls holding constant as the number of points increases. Likewise because the colouration is simply the sum of the $X$ variables, the increase in the number of points means that, as hypothesised, the colouration itself does not change by much. Both the highest (black), and lowest (red), values hold comparatively constant in panel (c). 

As the number of balls has increased so too has the ball size. The largest balls are around 0 on each axis, these being at the centre of the normal distribution where each variable is at its most dense. As more points are added the size of these main balls will inevitably grow. Progressively the confidence interval around the mean ball size becomes larger. Meanwhile the lowest values, shown in red, remain just one observation in every case. Not shown here, but an inevitable consequence of this rise, the range of ball size also increases. Finally panel (e) shows total connections increasing as more points are added, reaffirming the idea that points are appearing in previous gaps in the less dense clouds. Average connections amongst those balls that have at least one connection are also shown to rise. 

A similar message is seen for the five part cloud, but here there is a different pattern to be observed in the colouration owing to the higher range. An effect of the maximum values always being above 40, and the lowest below 0, is that the confidence intervals appear small and the lines more horizontal than in Figure \ref{fig:nc1}. Whilst further variation may arise from alternative distribution assumptions the way that more points facilitate more connections and give rise to higher ball numbers is clear.  

To illustrate the way the increased number of points affects the BM representation, Figure \ref{fig:nc3} contains three columns, one with 200 points, one with 500 points and one with 1000 points. Panels (a) to (c) correspond to the noise cloud, with panels (d) to (f) being the five part cloud. The shape of each is familiar from Section \ref{sec:bmeg} and Figures \ref{fig:egbm1} and \ref{fig:egbm2}. With just 200 points we still see the shapes but the five part cloud in panel (d) lacks the point that connects the third and fourth, green and light blue, sub-clouds meaning that there are now two parts to the main shape. the restoration of the link in the 500 point case of panel (e) is clear. Moving through 500 points to 1000 points there is much more consistency. As noted in the commentary above the presence of more points may induce connectivity between previous outliers and the main shape, but it may equally add more points in the tails of distributions that do not connect. In panel (c) there are more zero connected balls than in panel (b) for this reason. Meanwhile in the case of the five part cloud we note more connectivity forming with fewer outliers in panel (f) than (d).

Consistently the addition of more points to the cloud has shown an ability to obtain a better impression of the underlying shape of the data; this is fully consistent with the standard statistical result that more points on a single variable better capture the distribution thereof. More points induces more connectivity and adds points to existing balls. However, there is limited impact on the number of outliers as there will always be points in the tail. Although not fully uniform, there is limited impact on the colouration either since the points being added to balls necessarily have similar outcome values as their peers. In an applied setting the message remains that more data is desirable.

\subsection{Number of Axes}

Within any modelling design the choice of explanatory variables is key, balancing the desire to consider all possible alternative drivers of the observed outcomes and maintenance of tractability. Within BM graphs the primary impact from increasing the number of variables is that the radius must now be spread over more dimensions. Consider the two dimensional case where the ball is simply a circle of radius $\epsilon$. The maximum distance from the centre of the ball on one axis is $\epsilon$ and that may be achieved only when the other variable is identical. Adding more axes simply means that it would be more exceptional to find two points where the only variation is on one of the axes. In describing the effect of adding more dimensions we therefore consider that the average distance from points to the centre of their ball on any one dimension gets smaller. In turn we can expect more disconnection of balls as the number of axes grows. Averaging should also lead to lower total connections, smaller balls and a larger diversity of colouration for the balls. 

\begin{table}
    \begin{center}
        \caption{Number of Axes: Summary}
        \label{tab:ax1}
        \begin{tabular}{l c c c c c l l c c c c c}
        \hline
        \multicolumn{6}{l}{Panel (a): Noise Cloud ($\epsilon=5$)} && \multicolumn{6}{l}{Panel (b): Five Part Cloud ($\epsilon=10$)}\\
        Axes & Balls &  $\Delta$Size & $\Delta$Col & Zero & Con && Axes & Balls & $\Delta$Size & $\Delta$Col & Zero & Con\\
        \hline
        5&2.61&106.78&0.48&0&0.8&&5&166.32&8.90&47.54&9.58&1.25\\
&(0.69)&(50.12)&(0.36)&(0)&(0.34)&&&(3.75)&(0.92)&(1.77)&(3.02)&(0.06)\\
10&9.38&355.16&3.3&0&4.19&&10&217.8&6.65&92.58&56.85&0.73\\
&(1.59)&(50.56)&(1.06)&(0)&(0.79)&&&(4.88)&(0.84)&(1.43)&(6.98)&(0.03)\\
20&112.38&108.48&14.88&5.44&17.64&&20&274.38&5.32&178.91&135.67&0.67\\
&(4.47)&(28.02)&(1.43)&(0.51)&(1.23)&&&(5.99)&(0.76)&(0.99)&(9.69)&(0.03)\\
30&420.3&7.95&34.43&351.23&1.77&&30&340.5&4&267.5&251.4&0.61\\
&(3.55)&(1.51)&(0)&(2.46)&(0.17)&&&(6.82)&(0.69)&(0.64)&(11.77)&(0.04)\\
50&500&0&39.89&500&0&&50&461.5&1.99&435.6&452.3&0.53\\
&(0.00)&(0.00)&(0.00)&(0.00)&(0.00)&&&(5.18)&(0.45)&(0)&(7.33)&(0.09)\\

\hline
        \end{tabular}
    \end{center}
\raggedright
\footnotesize{Notes: Axes reports the number of axes which define the point cloud, Balls is the total number of balls within the BM graph, $\Delta$Size is the difference in size between the smallest and largest ball, $\Delta$Col is the difference between the highest and lowest colouration value for any ball within the graph, Zero is the number of balls for which there is no connectivity to any other ball and Con. is the average number of connections per ball amongst those balls that have at least one connection. All figures are the means from the 10000 repetitions at each point number, with figures in parentheses being the standard deviation across all values within the 10000 repetitions. The noise cloud comprises the stated number of axis variables each drawn at random from a standard normal distribution of mean 0 and variance 1. The five part cloud comprises 5 sub-clouds each of which contains one fifth of the total number of points. Within the sub-clouds values for each of given number of variables are drawn at random from a normal distribution of given mean and variance 1. Given means are 0, 2, 4, 6 and 8 for sub-clouds 1 to 5 respectively. Owing to the larger spread of characteristics in the five part cloud we use a larger initial radius to assess the effect of axes. In the noise cloud $\epsilon=5$ is used, whilst in the five part cloud we use $\epsilon=10$ as the ball radius.}
\end{table}

Table \ref{tab:nc1} provides a summary of the estimates for the two clouds used as exemplars in this paper. We use larger radii for the balls than in the previous section as the addition of axes will quickly reduce connectivity; a larger $\epsilon$ allows that reduction to be better showcased. The increasing number of balls, and subsequent increase in the number of zero connected balls is clear for both panels. By 50 axes the noise cloud already shows almost 100\% of BM graphs to have 500 individual balls. For the five part cloud the increase is slower, with the average reaching 461.5 at 50 axes. From an early stage in the five part cloud we see the average difference between the smallest and largest balls falling; from a high almost 9 when there are 5 axes the difference is already approaching 5 at 20 axes, and 2 at 50 axes. For the noise cloud the pattern is slightly different since the difference actually rises before falling rapidly to 0 by 50 axes. Average number of connections in the final column shows how adding axes at a given epsilon first increases connection numbers amongst the higher numbers of balls, before this falls down to zero. By contrast the five part cloud the average number of connections does not rise, instead falling steadily as more axes are added. For the five part cloud average connections remain around 0.53, rather than falling to 0 as they do for the noise cloud. In order to see these patterns more clearly we again include graphs.

\begin{figure}
    \begin{center}
        \caption{Number of Axes: Noise Cloud}
        \label{fig:ax1}
         \begin{tabular}{c c}
             \includegraphics[width=7cm]{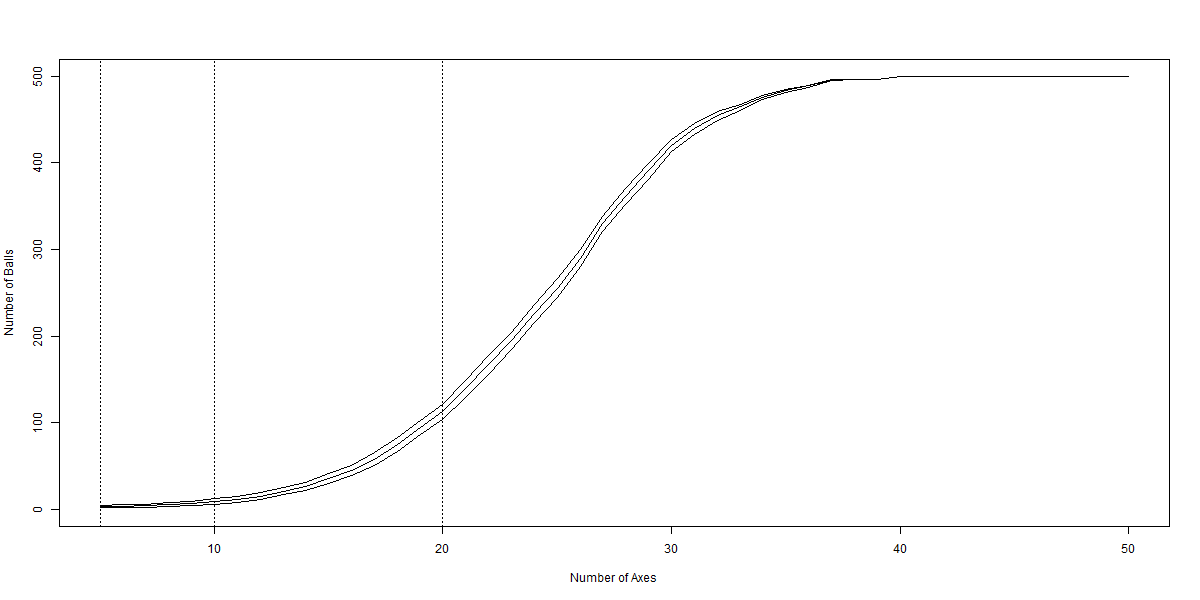} &
             \includegraphics[width=7cm]{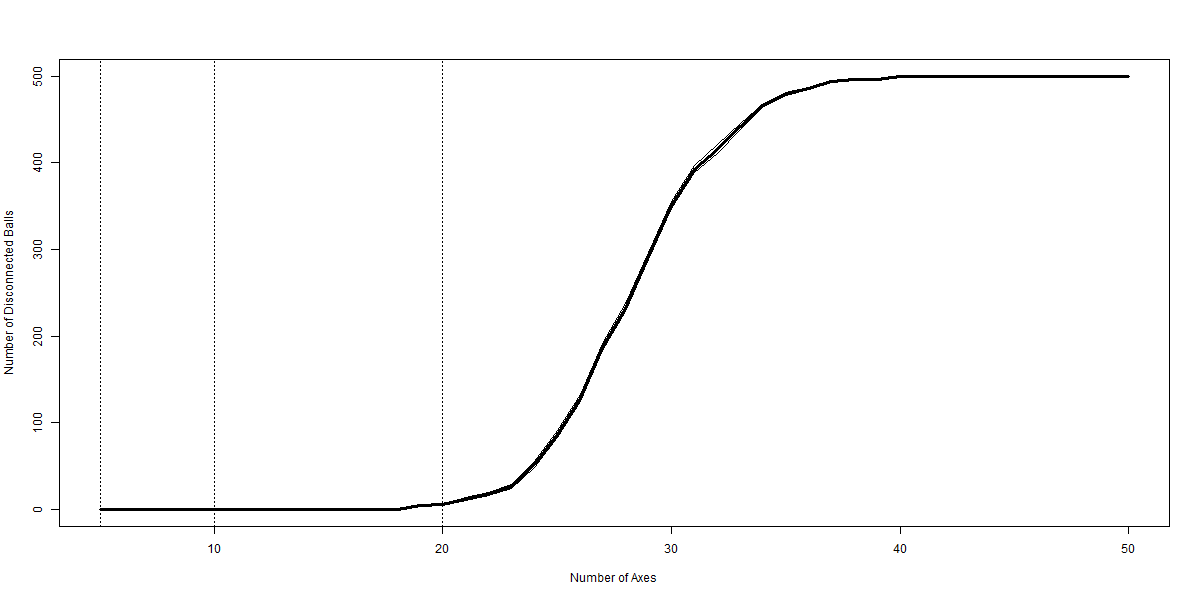} \\
             (a) Number of Balls & (b) Number of Zero Connection Balls \\
             \includegraphics[width=7cm]{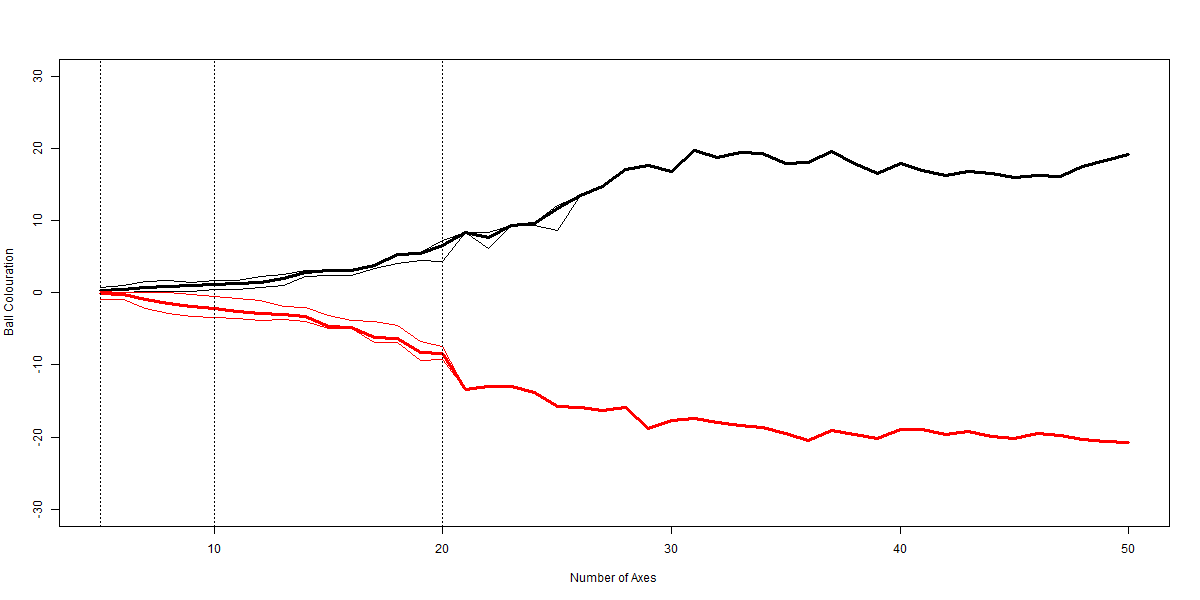} &
             \includegraphics[width=7cm]{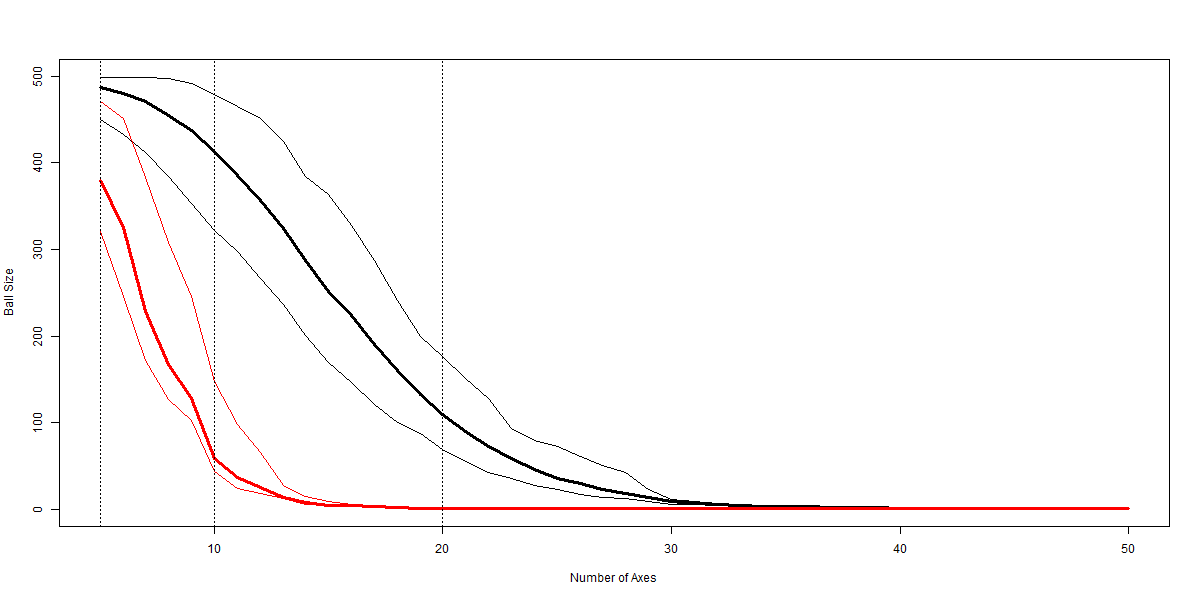} \\
             (c) Colouration & (d) Ball Sizes \\
             \includegraphics[width=7cm]{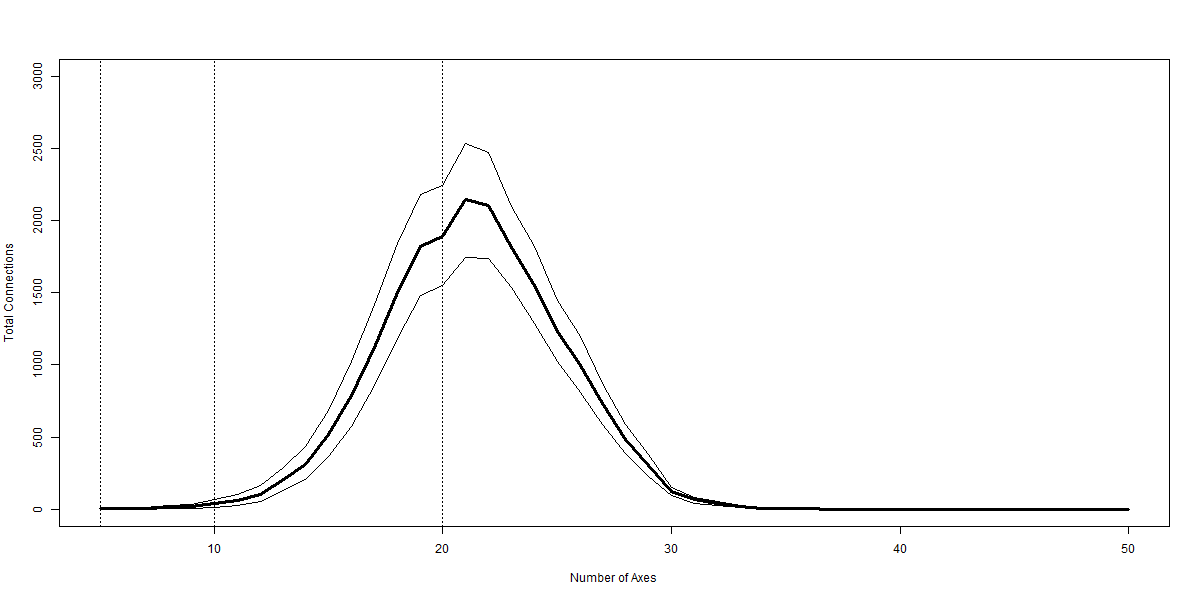} &
             \includegraphics[width=7cm]{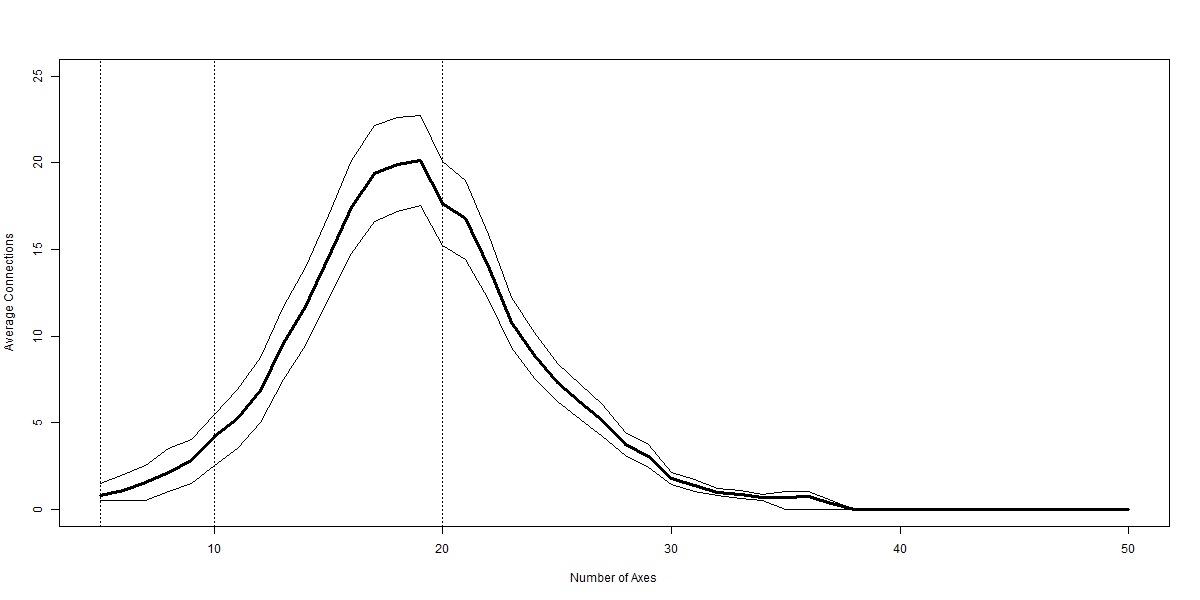}\\
             (e) Total Connections & (f) Average Connections\\
        \end{tabular}
    \end{center}
    \raggedright
\footnotesize{Notes: Figures plot the impact of the number of points axes used in the construction of the noise cloud on various measures of the BM graph. In each case 10000 repetitions of the BM algorithm \citep{dlotko2019ball} are implemented with a ball radius of $\epsilon=5$. A thick line is used to denote the mean from the repetitions, thinner lines denoting the 95\% confidence interval there around. Panel (a) reports the number of balls, and panel (b) the number of balls which have 0 connections to any other balls. Panels (c) and (d) also use red lines to show the maximum and minimum colouration and ball size respectively. Panel (e) reports the total number of connections within the graph, this informs on the points within the overlaps of balls and hence the density of the graphs. Panel (f) plots the average number of connections amongst connected balls. In the case that there are no connected balls then this figure is set to 0. All estimates are generated using the R package \textit{BallMapper} \citep{dlotko2019R}.} 
\end{figure}

Illustrating the evolution of the respective measures over the increasing number of axes, Figure \ref{fig:ax1} shows the increases in the number of balls, and number of balls without any connections to other balls, in panels (a) and (b) respectively. The latter curve is much steeper than the former, still being close to 0 at 20 axes but then hitting 500 around 40 axes. Declining ball sizes are also fully in line with expectation. Total connections and average connections display notable hump shapes, with the connectivity rising as the increasing axis numbers split up the original large balls. Once the largest ball sizes start to approach the smallest, seen from panel (d), the peak of the hump is passed and total connections fall. Average connections in panel (f) can be seen to fall first, the peak being around 18 axes versus 22 for the total connection number. Although we may note some small variations, the confidence intervals that result from the 10000 repetitions are narrow and indicate that the shape of the BM graph is indeed affected by the changes in numbers of axes, rather than any variation at a given number of axes. To see the impact of the number of axes we use 5, 10 and 20 axes for the example plots, these are indicated in Figure \ref{fig:ax1} as vertical dotted lines.

\begin{figure}
    \begin{center}
        \caption{Number of Axes: Five Part Cloud}
        \label{fig:ax2}
         \begin{tabular}{c c}
             \includegraphics[width=7cm]{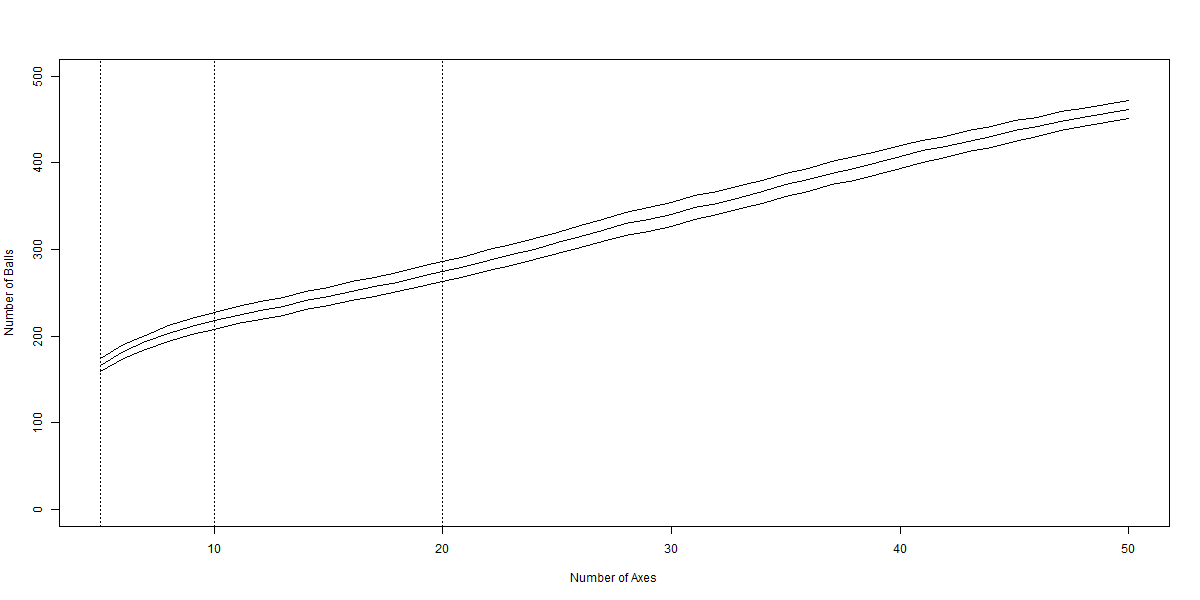} &
             \includegraphics[width=7cm]{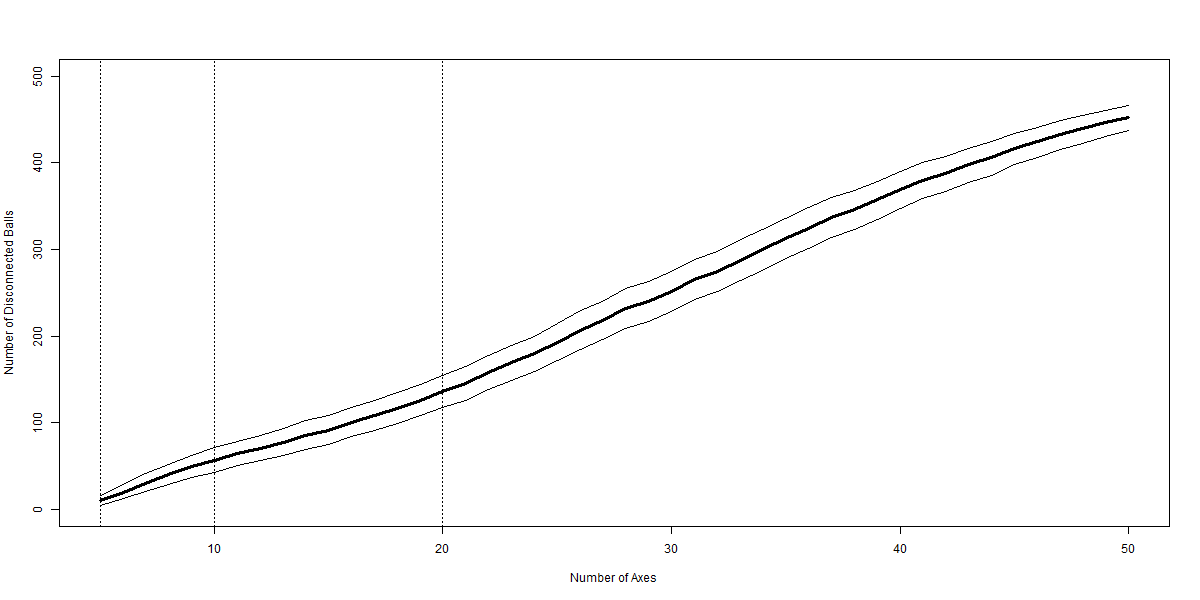} \\
             (a) Number of Balls & (b) Number of Zero Connection Balls \\
             \includegraphics[width=7cm]{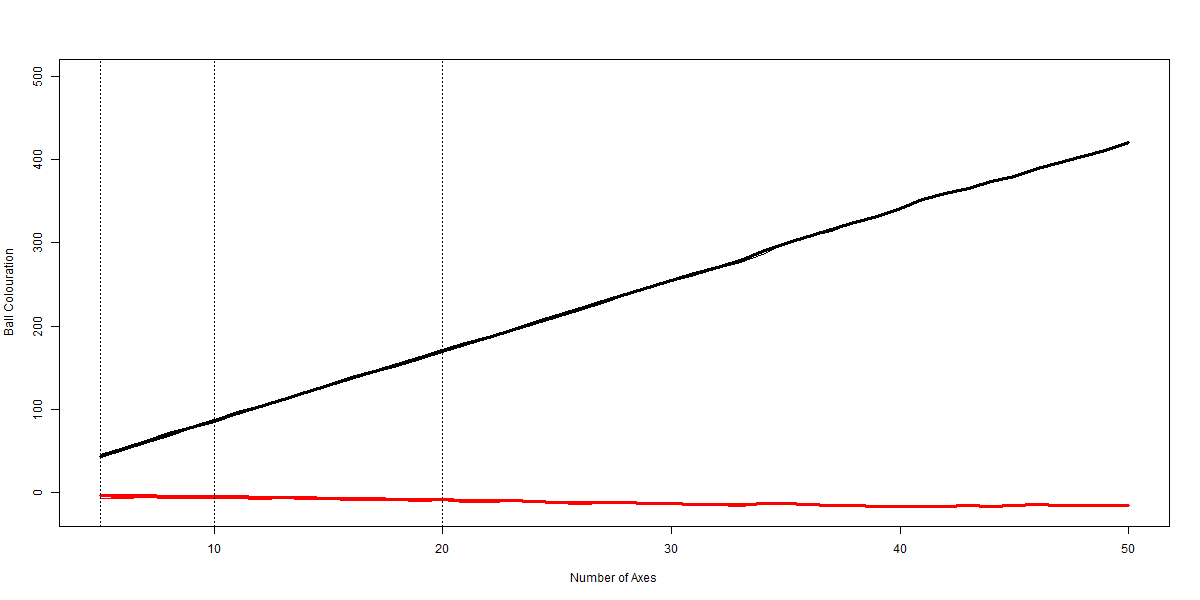} &
             \includegraphics[width=7cm]{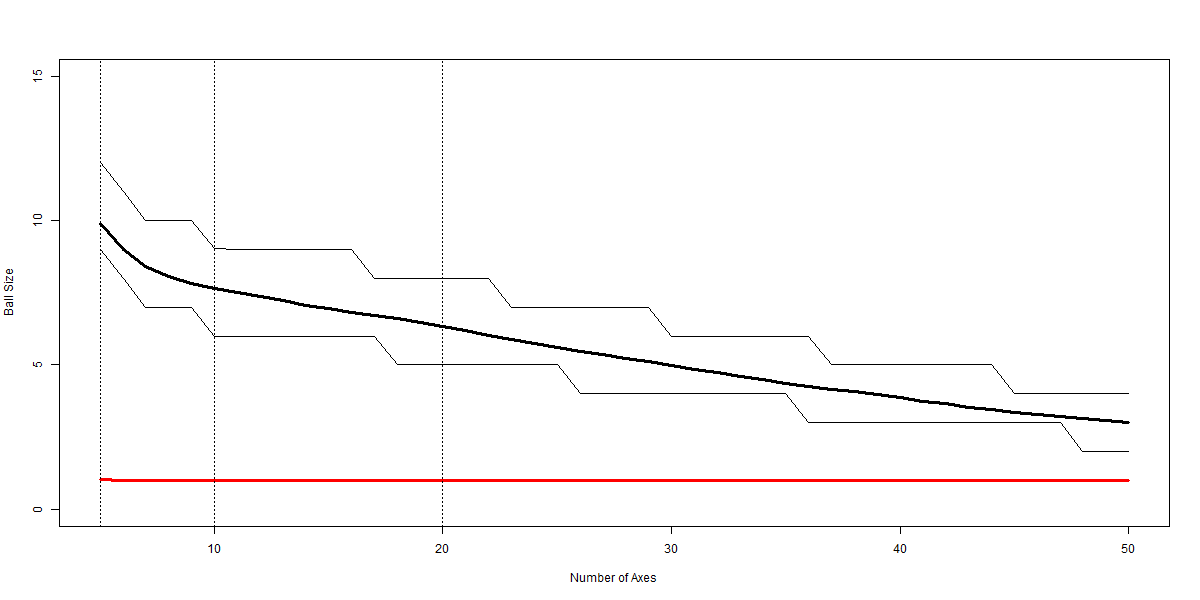} \\
             (c) Colouration & (d) Ball Sizes \\
             \includegraphics[width=7cm]{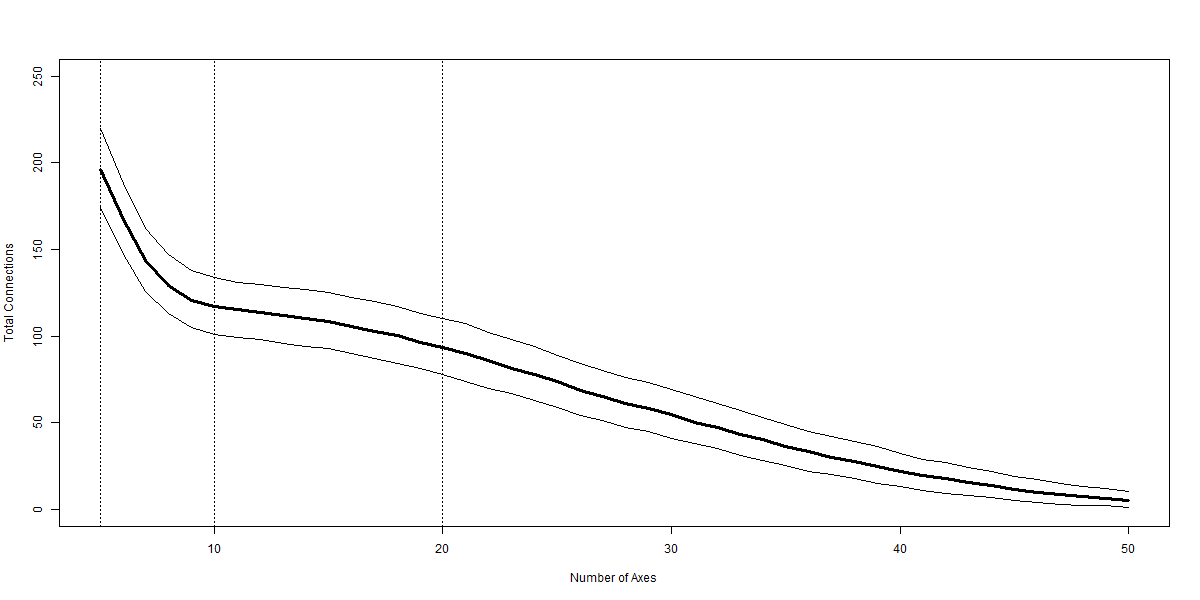}&
             \includegraphics[width=7cm]{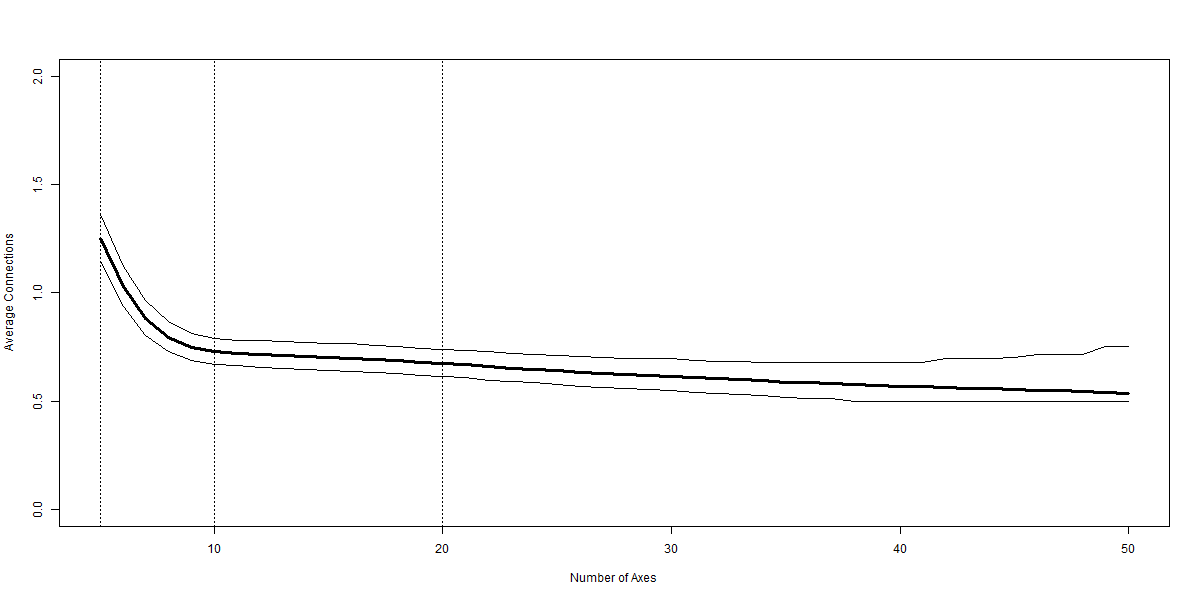}\\
             (e) Total Connections & (f) Average Connections\\
        \end{tabular}
    \end{center}
    \raggedright
\footnotesize{Notes: Figures plot the impact of the number of points axes used in the construction of the five part cloud on various measures of the BM graph. In each case 10000 repetitions of the BM algorithm \citep{dlotko2019ball} are implemented with a ball radius, $\epsilon$, of 10. A thick line is used to denote the mean from the repetitions, thinner lines denoting the 95\% confidence interval there around. Panel (a) reports the number of balls, and panel (b) the number of balls which have 0 connections to any other balls. Panels (c) and (d) also use red lines to show the maximum and minimum colouration and ball size respectively. Panel (e) reports the total number of connections within the graph, this informs on the points within the overlaps of balls and hence the density of the graphs. Panel (f) plots the average number of connections amongst connected balls. In the case that there are no connected balls then this figure is set to 0. All estimates are generated using the R package \textit{BallMapper} \citep{dlotko2019R}.} 
\end{figure}

Results for the five part cloud in Figure \ref{fig:ax2} are fully consistent with this message also. Indeed the numbers of balls and number of zero connection balls, panels (a) and (b) are only differentiated in the time that it takes to reach 500 balls. Both fall just short at 50 axes but already show the flattening of the curve seen around 20 axes in the noise cloud. Colouration varies, a much larger range appearing when the number of axes gets larger. Such is consistent with the way in which the full cloud comprises five sub-clouds with their own assumptions about mean colouration. Maximum ball size in panel (d) falls towards 1 in the same way as it does for the noise cloud, but again we see that has not quite reached 1 at 50 axes. Total connections and average connections, panels (e) and (f), are where the biggest difference is found. Where the noise cloud showed a hump shape the five part cloud only shows falling numbers of connections, this is suggestive that the BM graphs are already past the hump. Further suggestion of this comes in the ball size of panel (d) where the smallest balls are already just 1 point, contrasting with the noise cloud which has 15 axes before the smallest ball size is 1. Choosing different radii for the five part cloud is likely to yield stronger similarity. In the case of the five part cloud we will again give examples with 5, 10 and 20 axes and these are illustrated by the vertical dashed lines on the plot once more. 

\begin{figure}
    \begin{center}
        \caption{Example Ball Mapper Plots: Number of Axes}
        \label{fig:ax3}
        \begin{tabular}{c c c}
             \includegraphics[width=5cm]{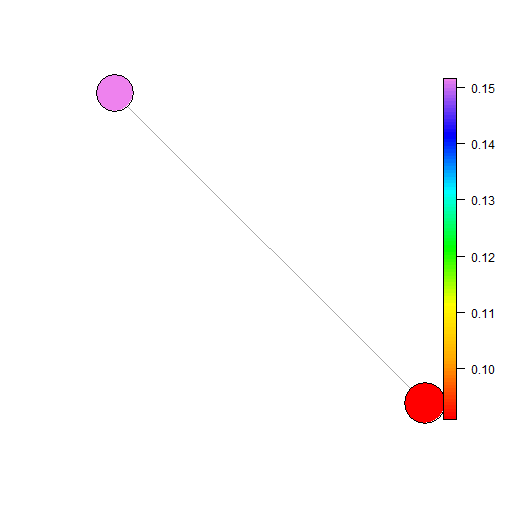} &
             \includegraphics[width=5cm]{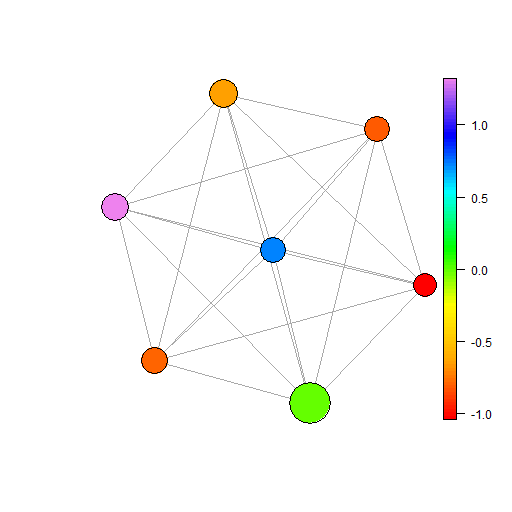} &
             \includegraphics[width=5cm]{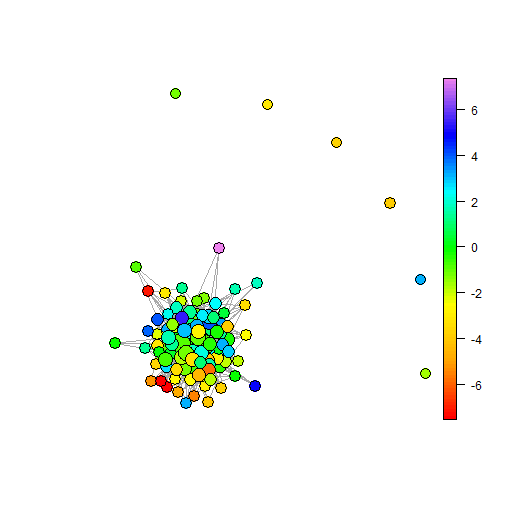} \\
             (a) Noise Cloud: 5 Axes & (b) Noise Cloud: 10 Axes & (c) Noise Cloud: 20 Axes \\
              \includegraphics[width=5cm]{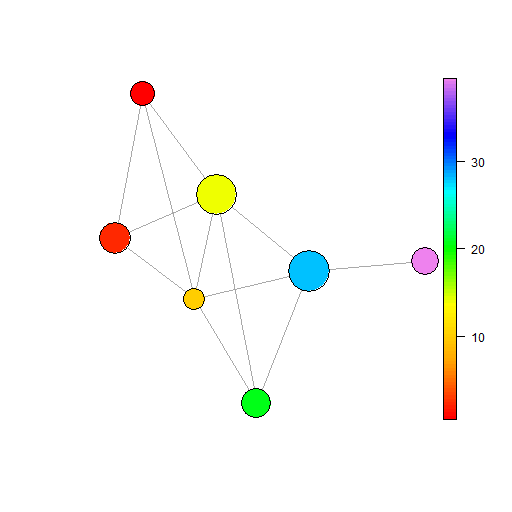} &
             \includegraphics[width=5cm]{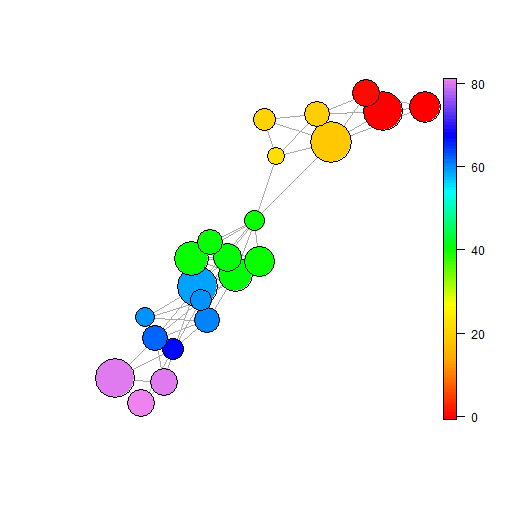} &
             \includegraphics[width=5cm]{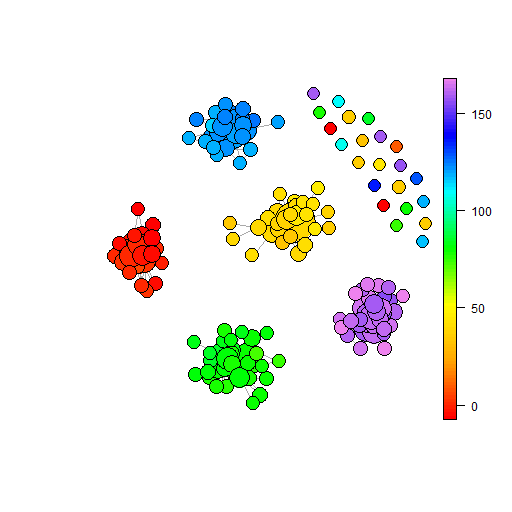} \\
             (d) Five Part Cloud: 5 Axes & (e) Five Part Cloud: 10 Axes & (f) Five Part Cloud: 20 Axes \\
        \end{tabular}
    \end{center}
\raggedright
\footnotesize{Notes: Figures plot example BM graphs for the stated cloud and numbers of axes. The noise cloud is constructed using variables which are independently randomly drawn from standard normal distributions with mean 0 and variance 1. Panels (a) to (c) show this cloud with 200, 500 and 1000 points respectively. The five part cloud is constructed from five sub-clouds where each sub-cloud comprises variables with a given mean and standard deviation 1. Given means are 0, 2, 4, 6 and 8. The number of variables for both clouds is varied in this section, being the number of axes in the cloud. Noise cloud BM graphs use $\epsilon=5$ and five part cloud graphs use $\epsilon=10$. All plots generated using the \textit{BallMapper} package in R \citep{dlotko2019R}.}
\end{figure}

The impact of adding more axes is easily visualised in the two rows of Figure \ref{fig:ax3}. The two balls identified in the first figure contrast with the dense cloud of connected balls in the last figure on each row. Note here that the larger ball radius, $\epsilon=5$, is being used compared to the other examples in this paper where the five axis cloud is plotted with $\epsilon=2$. For the five part cloud we see that the impact of increasing the number of axes is to create a split between the various sub-clouds. Moving from panel (d) to (f) this split is highly apparent. With $\epsilon=5$ panel (e) of Figure \ref{fig:nc3} shows the connection of five sub-clouds on five axes, the increase of $\epsilon$ to panel (d) of Figure \ref{fig:ax3} is discussed in the next section. Panel (e) has five different colourations, although the yellow and orange do merge into the red and there is strong overlap between blue and green. In panel (f) the split on the sub-clouds is achieved by those points which formerly sat in the overlap between the sub-clouds now becoming outliers owing to the number of axes over which the radius must be split. We see outliers with all colourations to confirm this. Each sub-cloud has a similar shape to the noise cloud. 

\subsection{Ball Radius}

Numbers of points, and numbers of axes are typically defined by the available data and are therefore not the choice of the modeller applying BM. For the BM algorithm the single parameter of choice is the radius of the balls that will be used, $\epsilon$. Our final exploration is then the choice of $\epsilon$. Because smaller balls will inevitably contain fewer points it follows that there will be more required to provide a cover for the same data set. Likewise we may expect that there would be lower within ball variation in colour, but that the between ball variation would rise. As $\epsilon$ increases we would expect the ball numbers to fall and the colouration range to narrow. A further consequence of the ball radius increasing is that balls which were previously separate from the main shape come within range; zero connection ball numbers fall. Connections, and hence average connections have counter veiling effects acting upon them. More radius means more balls will connect, but the fact that there will be fewer balls means the total number of connections may also fall.

\begin{table}
    \begin{center}
        \caption{Ball Radius: Summary}
        \label{tab:br1}
        \begin{tabular}{l c c c c c l l c c c c c}
        \hline
        \multicolumn{6}{l}{Panel (a): Noise Cloud} && \multicolumn{6}{l}{Panel (b): Five Part Cloud}\\
        $\epsilon$ & Balls &  $\Delta$Size & $\Delta$Col & Zero & Con && $\epsilon$ & Balls & $\Delta$Size & $\Delta$Col & Zero & Con\\
        \hline
1&245.37&14.59&13.71&134.67&1.98&&1&385.19&4.88&51.11&323.43&0.78\\
&(3.58)&(1.71)&(0)&(2.51)&(0.14)&&&(2.71)&(0.98)&(0.00)&(2.98)&(0.05)\\
2&47.64&142.66&9.81&2.02&6.30&&2&114.42&35.62&50.12&16.05&3.42\\
&(2.54)&(25.65)&(0.82)&(0.14)&(0.42)&&&(3.34)&(4.86)&(0.00)&(0.96)&(0.19)\\
3&14.29&313.34&5.28&0&5.34&&3&37.67&79.97&44.65&0&4.28\\
&(1.47)&(47.61)&(0.8)&(0)&(0.53)&&&(2.44)&(7.03)&(1.23)&(0)&(0.31)\\
4&5.64&321.86&2.23&0&2.32&&4&15.11&120.68&42.59&0&2.87\\
&(0.99)&(54.46)&(0.75)&(0)&(0.5)&&&(1.76)&(18.49)&(1.17)&(0)&(0.38)\\
&&&&&&&5&7.59&122.62&39.17&0&1.71\\
&&&&&&&&(1.27)&(30.83)&(1.87)&(0)&(0.38)\\
&&&&&&&10&2.33&126.07&20.81&0&0.58\\
&&&&&&&&(0.47)&(72.81)&(4.44)&(0)&(0.14)\\

\hline
        \end{tabular}
    \end{center}
\raggedright
\footnotesize{Notes: $\epsilon$ reports the radius of the balls used to form the BM graph, Balls is the total number of balls within the BM graph, $\Delta$Size is the difference in size between the smallest and largest ball, $\Delta$Col is the difference between the highest and lowest colouration value for any ball within the graph, Zero is the number of balls for which there is no connectivity to any other ball and Con. is the average number of connections per ball amongst those balls that have at least one connection. All figures are the means from the 10000 repetitions at each point number, with figures in parentheses being the standard deviation across all values within the 10000 repetitions. Here we report only those epsilon for which 10000 observations were derived. In the case of $\epsilon=5$ the BM representation of the noise cloud only produces graphs with two balls on 8542 occasions. There are no occasions when a BM graph of the noise cloud with $\epsilon=10$. The noise cloud comprises 5 variables each drawn at random from a standard normal distribution of mean 0 and variance 1. The five part cloud comprises 5 sub-clouds each of which contains one fifth of the total number of points. Within the sub-clouds values for each of five variables are drawn at random from a normal distribution of given mean and variance 1. Given means are 0, 2, 4, 6 and 8 for sub-clouds 1 to 5 respectively.}
\end{table}

Summarising the results, Table \ref{tab:br1} informs that the predicted relationships can indeed be seen. It is noteworthy that the maximal radius considered here for the noise cloud is 4. Although we may obtain plots for $\epsilon>4$ there are cases where the algorithm selects an initial point from which all of the points are covered by a single ball. This is no problem in our exposition of the number of axes, but for the purposes of this discussion it is not useful. Hence we restrict attention in the noise cloud to radii of 4 or lower. For the five part cloud we are able to obtain 10000 estimations up to, and including, $\epsilon=11$ so we feature more rows in panel (b). For both examples the rate at which zero connected balls fall to zero is very quick, by $\epsilon=3$ there are no balls that are not connected to another ball. At this point there are still multiple balls and it is only as $\epsilon$ pushes higher that we see the number of balls approaching 1. As noted we do not include cases where there are only one ball, or where the algorithm cannot produce 10000 repetitions with more than one ball. We may again see more by plotting these numbers.

\begin{figure}
    \begin{center}
        \caption{Ball Radius: Noise Cloud}
        \label{fig:br1}
         \begin{tabular}{c c}
             \includegraphics[width=7cm]{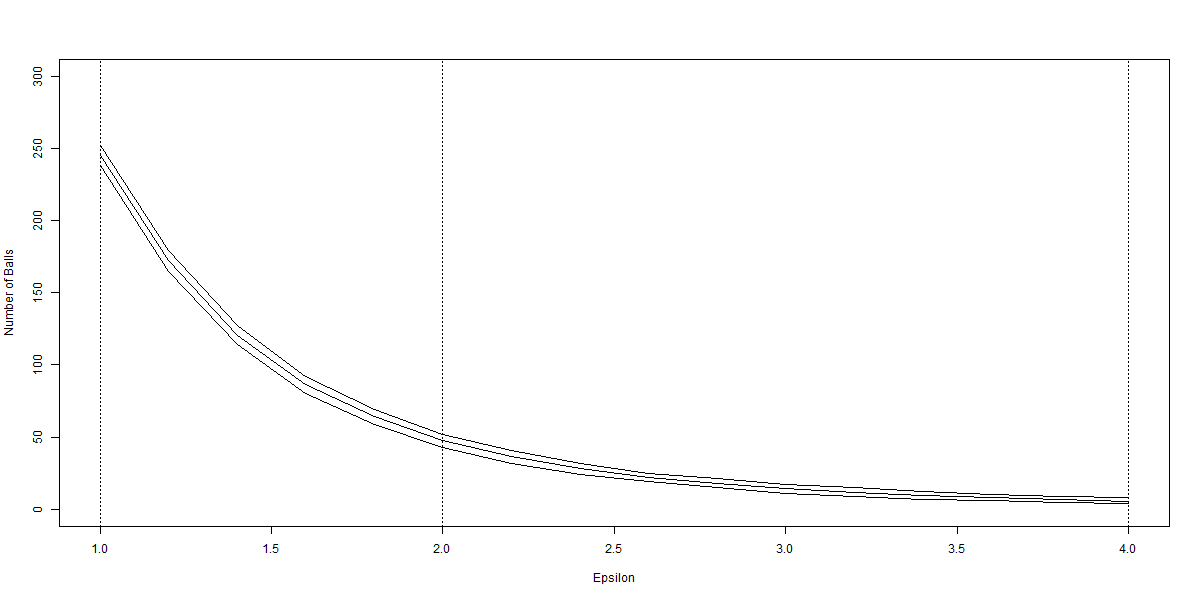} &
             \includegraphics[width=7cm]{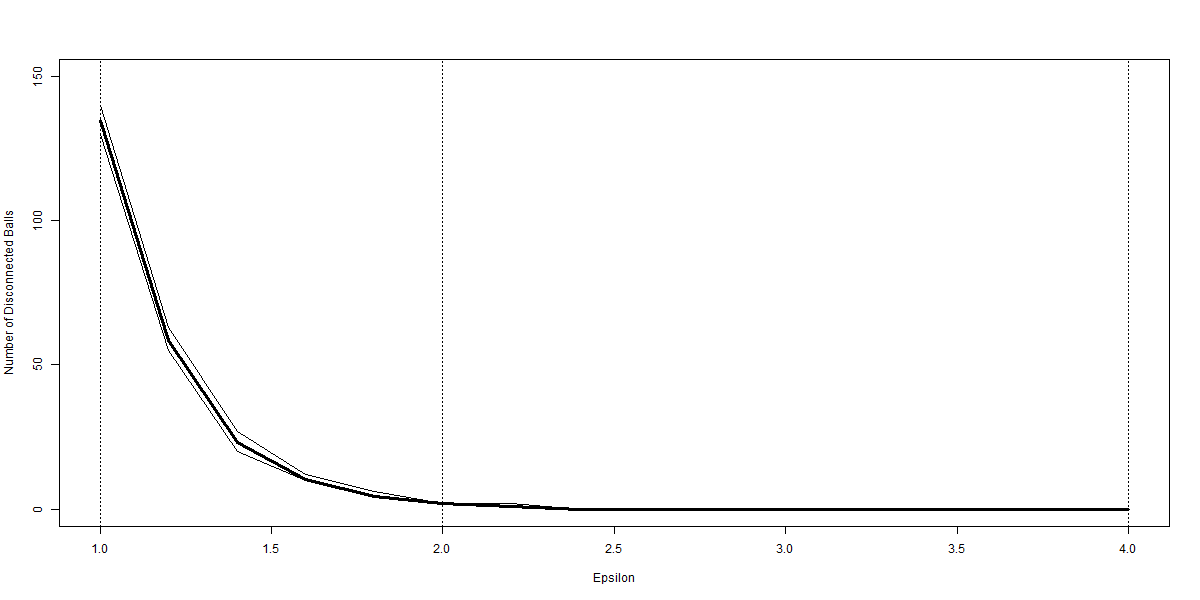} \\
             (a) Number of Balls & (b) Number of Zero Connection Balls \\
             \includegraphics[width=7cm]{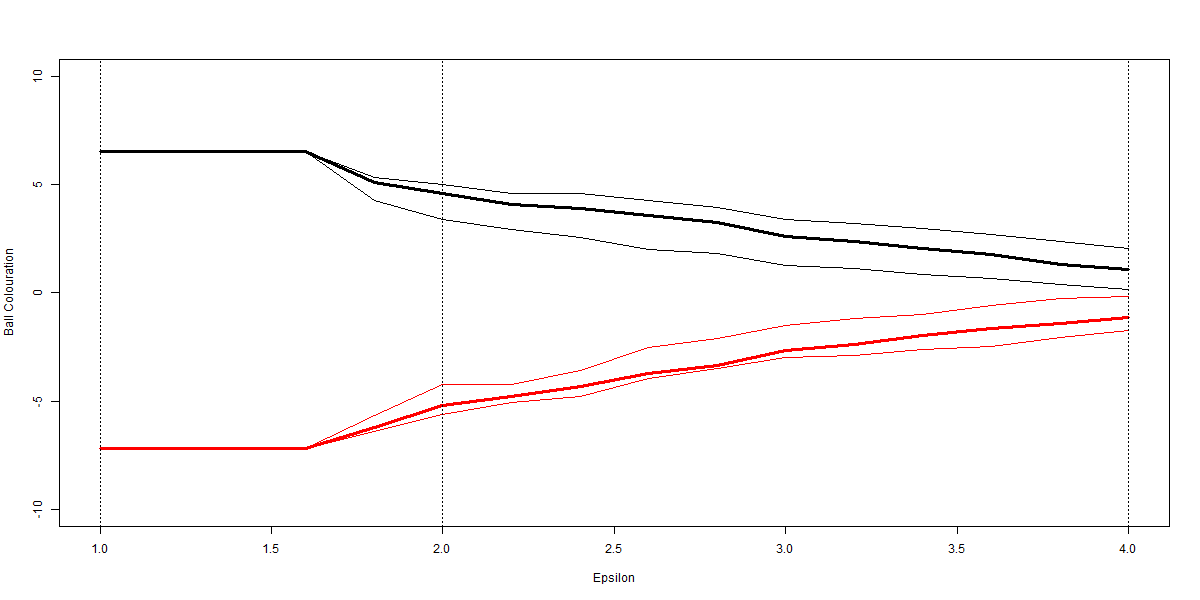} &
             \includegraphics[width=7cm]{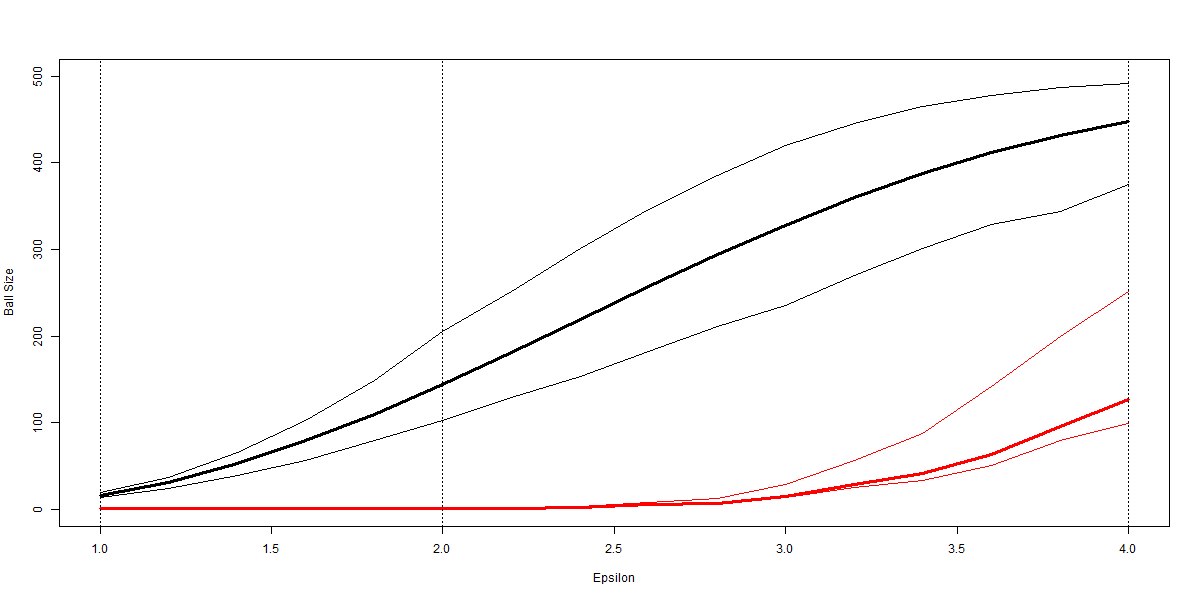} \\
             (c) Colouration & (d) Ball Sizes \\
             \includegraphics[width=7cm]{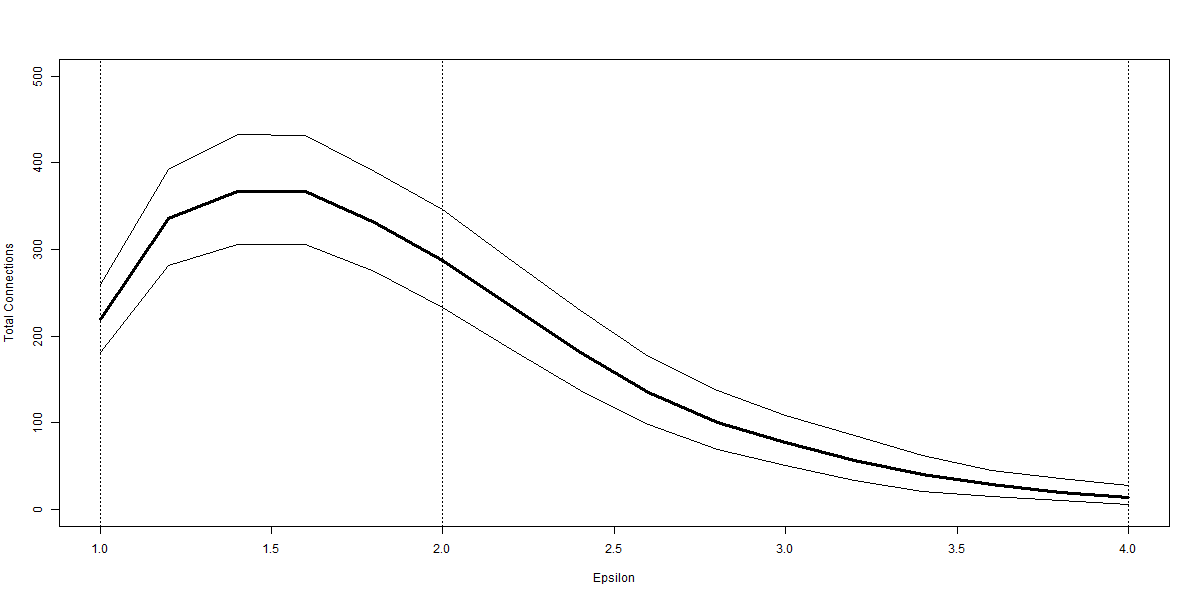} &
             \includegraphics[width=7cm]{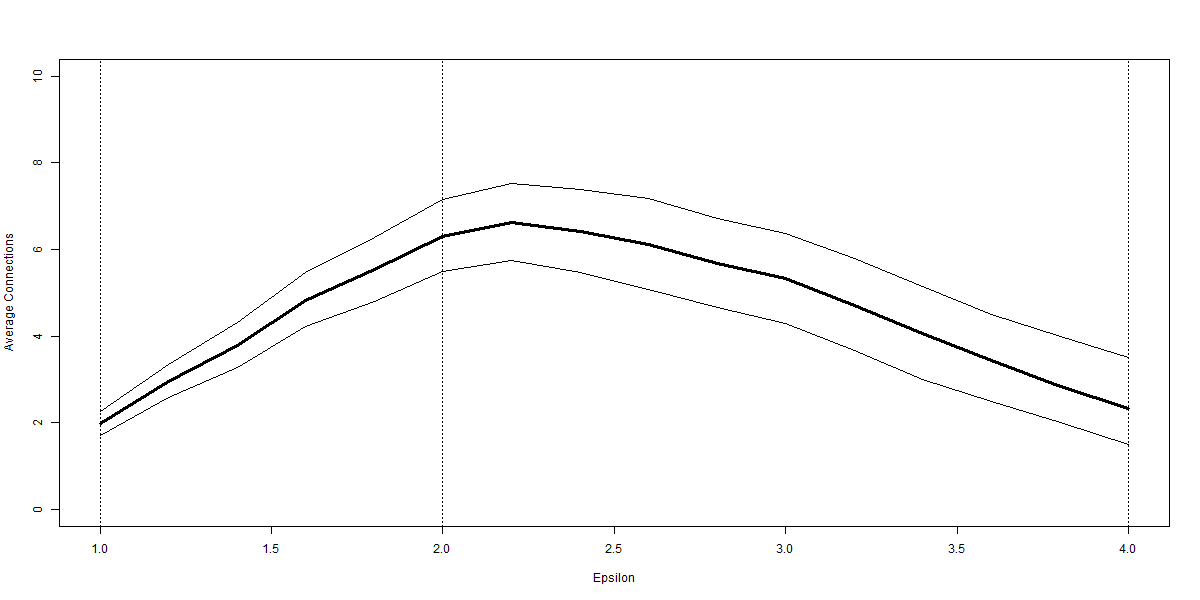}\\
             (e) Total Connections & (f) Average Connections\\
        \end{tabular}
    \end{center}
    \raggedright
\footnotesize{Notes: Figures plot the impact of the radius of the balls, $\epsilon$, used in the construction of the noise cloud on various measures of the BM graph. In each case 10000 repetitions of the BM algorithm \citep{dlotko2019ball} are implemented. A thick line is used to denote the mean from the repetitions, thinner lines denoting the 95\% confidence interval there around. Panel (a) reports the number of balls, and panel (b) the number of balls which have 0 connections to any other balls. Panels (c) and (d) also use red lines to show the maximum and minimum colouration and ball size respectively. Panel (e) reports the total number of connections within the graph, this informs on the points within the overlaps of balls and hence the density of the graphs. Panel (f) plots the average number of connections amongst connected balls. In the case that there are no connected balls then this figure is set to 0. All estimates are generated using the R package \textit{BallMapper} \citep{dlotko2019R}.} 
\end{figure}

\begin{figure}
    \begin{center}
        \caption{Role of Epsilon}
        \label{fig:br3}
        \begin{tabular}{c c c}
        \includegraphics[width=5cm]{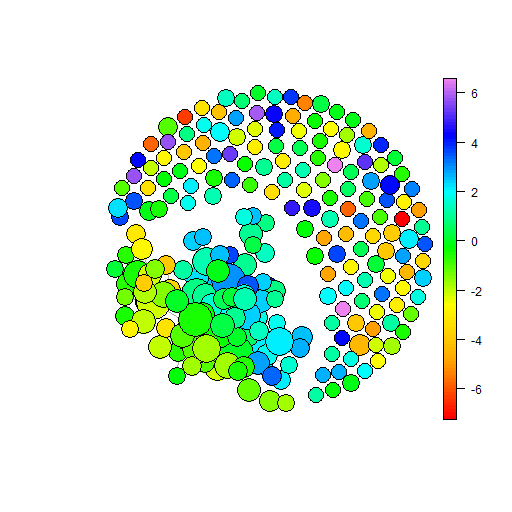}&
    \includegraphics[width=5cm]{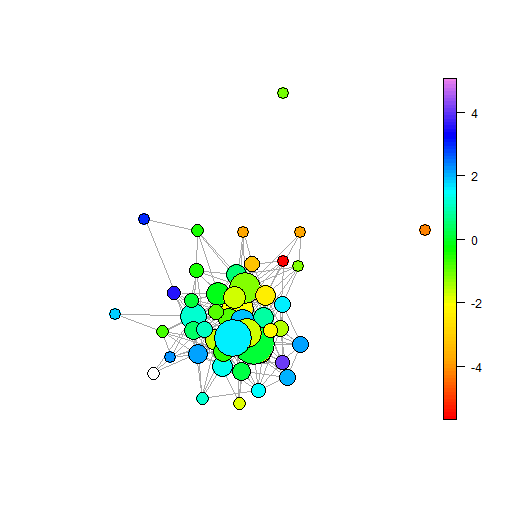}&
    \includegraphics[width=5cm]{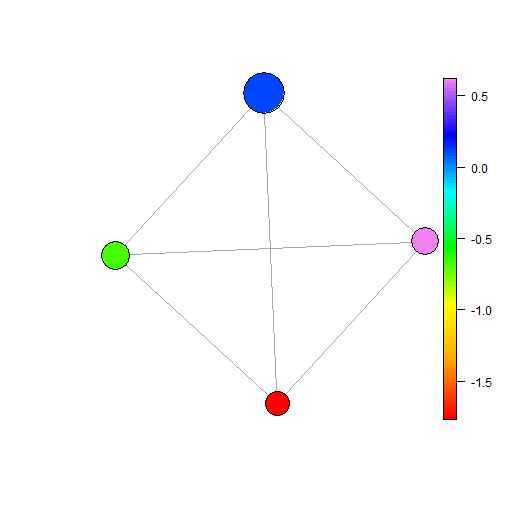}\\
    (a) Noise Cloud: $\epsilon$ = 1 & (b) Noise Cloud: $\epsilon$ = 2 & (c) Noise Cloud: $\epsilon$ = 4\\
    \includegraphics[width=5cm]{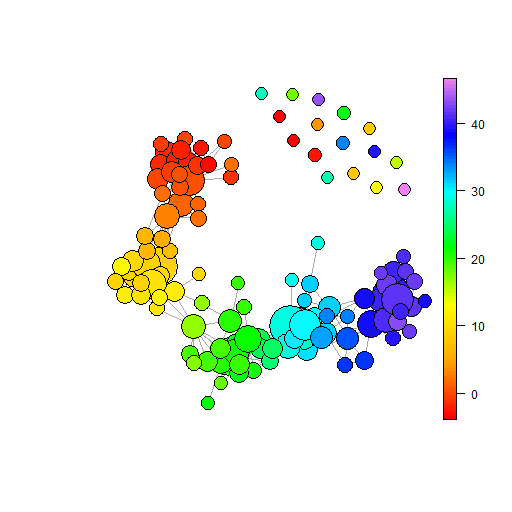}&
    \includegraphics[width=5cm]{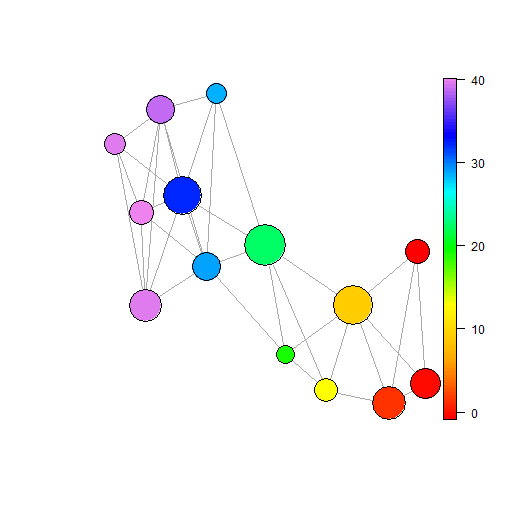}&
    \includegraphics[width=5cm]{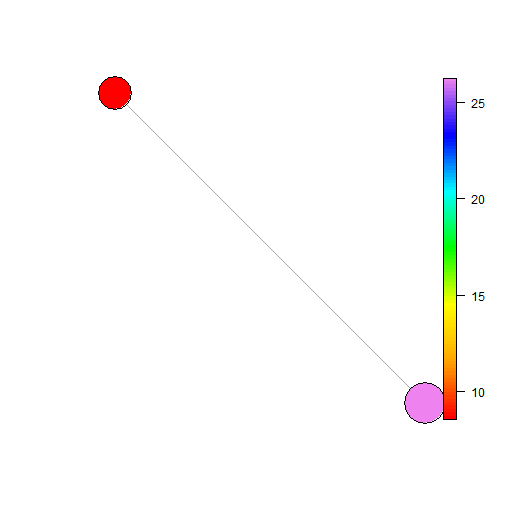}\\
    (d) Five Part Cloud: $\epsilon$ = 2 & (e) Five Part Cloud: $\epsilon$ = 4 & (f) Five Part Cloud: $\epsilon$ = 10 \\
        \end{tabular}
    \end{center}
\raggedright
\footnotesize{Notes: Figures plot example BM graphs for the stated cloud and ball radius. The noise cloud is constructed using five variables which are independently randomly drawn from standard normal distributions with mean 0 and variance 1. Panels (a) to (c) show this cloud with 200, 500 and 1000 points respectively. The five part cloud is constructed from five sub-clouds where each sub-cloud comprises five variables with a given mean and standard deviation 1. Given means are 0, 2, 4, 6 and 8. All plots generated using the \textit{BallMapper} package in R \citep{dlotko2019R}.}
\end{figure}

Figure \ref{fig:br1} plots the ball radius as the horizontal axis, capping the range at 1 and 4. Dotted vertical lines at 1, 2 and 4 show where the example plots in Figure \ref{fig:br3} fit. Panel (a) shows that ball numbers fall smoothly through the range, whilst the number of zero connected balls plotted in panel (b) falls faster. This is as may be hypothesised. Panels (c) and (d) show both the lowest and highest colouration and ball size respectively. What we see as the radius increases is that the homogeneity of the balls increases, particularly the colouration which has a much reduced range of values as $\epsilon$ rises. In panel (d) the biggest ball shows large variation in the mid range of radii, whilst a similar pattern is only evident in the lowest ball size as $\epsilon$ gets close to 4. Patterns in the connection numbers show greatest diversion from the hypothesised effect since we first observe the connections growing before then subsequently contracting. Here we see the balls connecting at low radii before the density of the noise cloud's centre starts to see fewer balls needed to create a cover. In turn these fewer balls require less connections and the total falls. Meanwhile the connecting, and then subsuming, of points on the outside of the cloud will further raise then reduce connections. Given the falling number of balls, and falling number of zero connection balls, the average connections per connected ball hold longer, but the hump shape is still evident in panel (f) of Figure \ref{fig:br1}.  
\begin{figure}
    \begin{center}
        \caption{Ball Radius: Five Part Cloud}
        \label{fig:br2}
         \begin{tabular}{c c}
             \includegraphics[width=7cm]{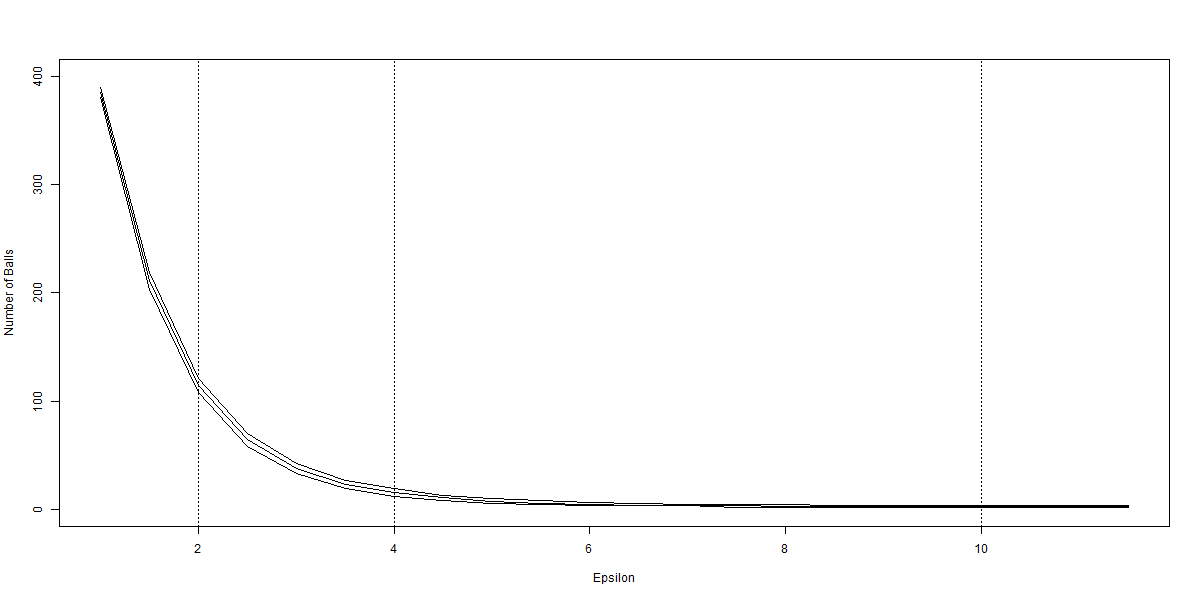} &
             \includegraphics[width=7cm]{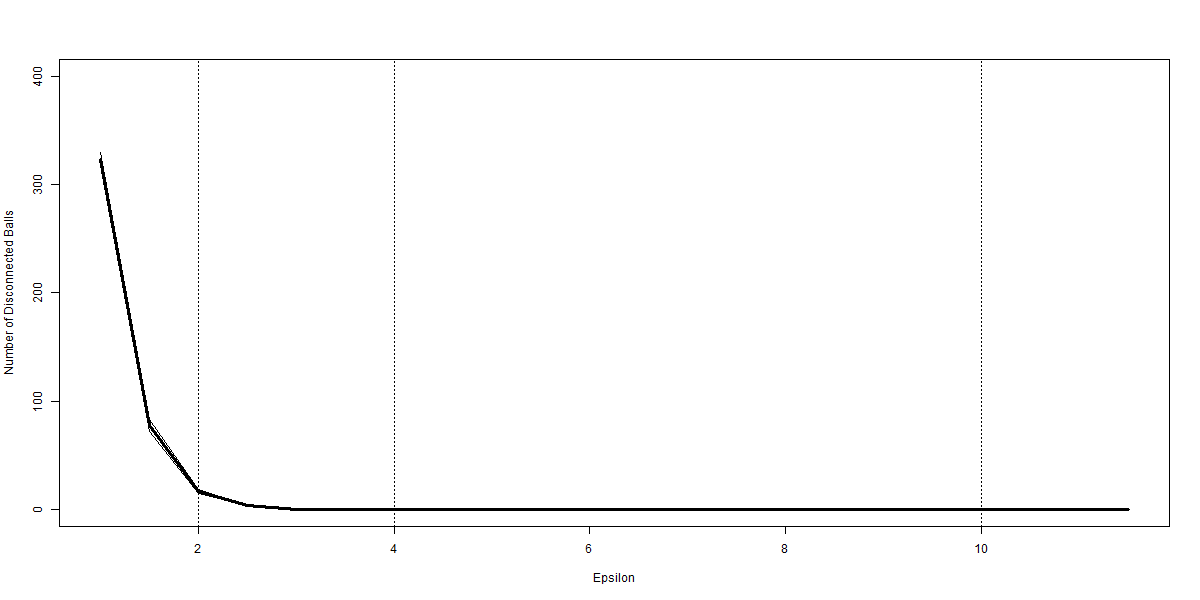} \\
             (a) Number of Balls & (b) Number of Zero Connection Balls \\
             \includegraphics[width=7cm]{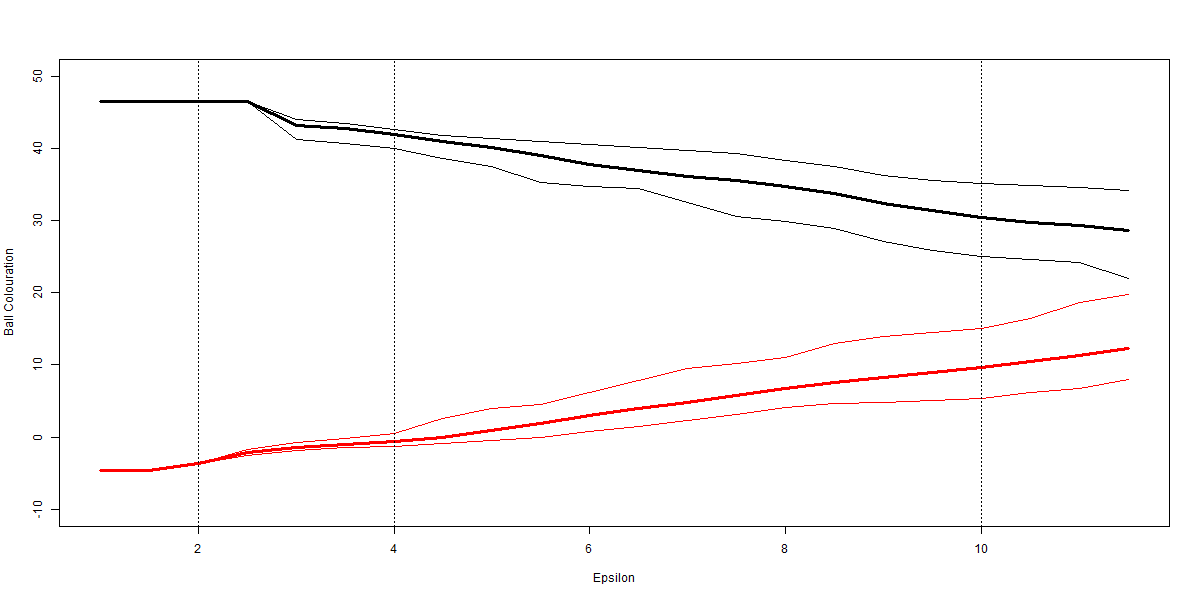} &
             \includegraphics[width=7cm]{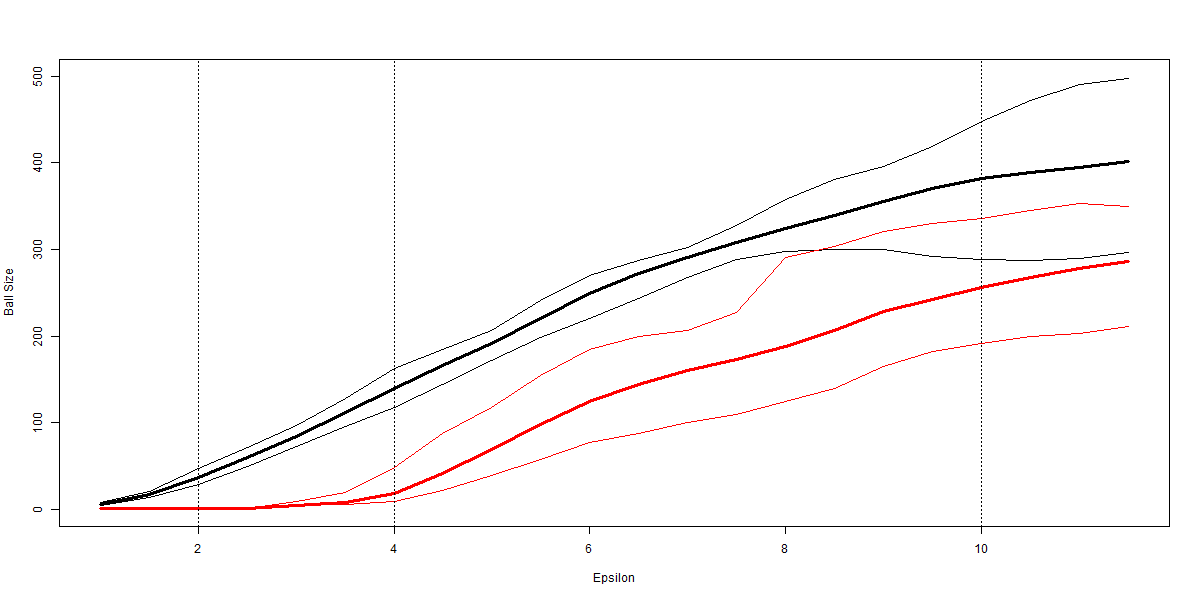} \\
             (c) Colouration & (d) Ball Sizes \\
             \includegraphics[width=7cm]{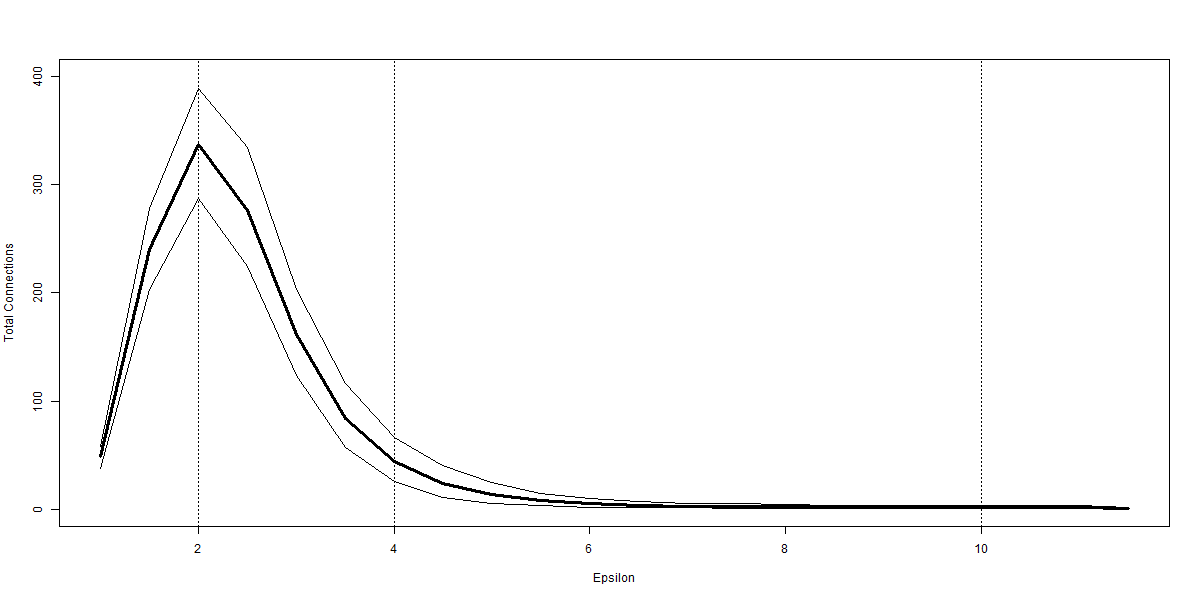} &
             \includegraphics[width=7cm]{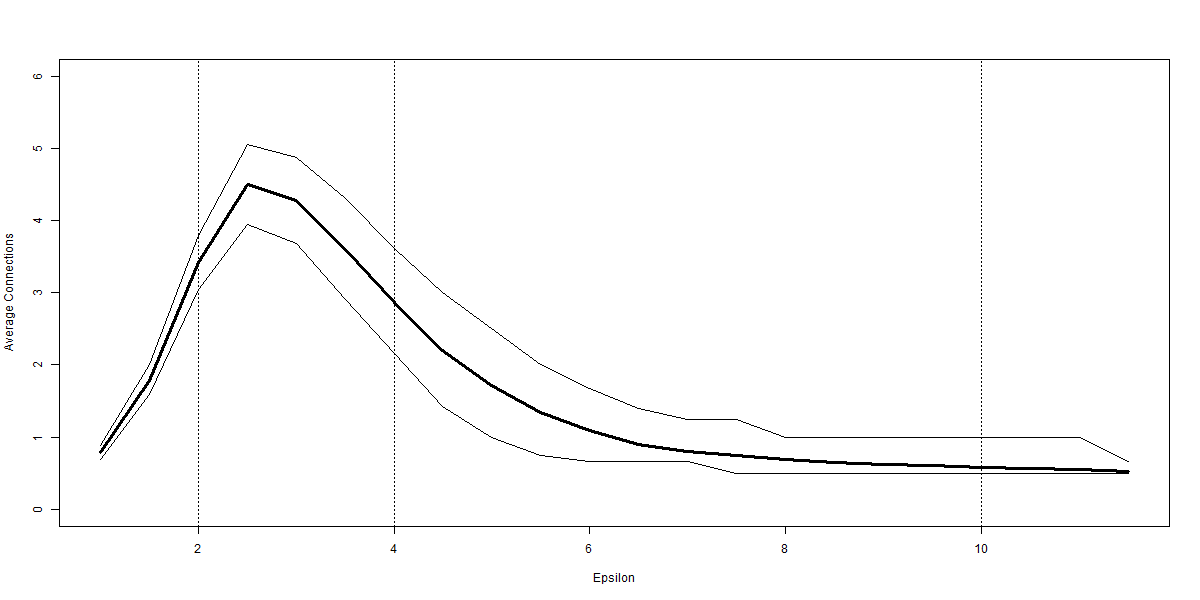}\\
             (e) Total Connections & (f) Average Connections\\
        \end{tabular}
    \end{center}
      \raggedright
\footnotesize{Notes: Figures plot the impact of the radius of the balls, $\epsilon$, used in the construction of the five part cloud on various measures of the BM graph. In each case 10000 repetitions of the BM algorithm \citep{dlotko2019ball} are implemented. A thick line is used to denote the mean from the repetitions, thinner lines denoting the 95\% confidence interval there around. Panel (a) reports the number of balls, and panel (b) the number of balls which have 0 connections to any other balls. Panels (c) and (d) also use red lines to show the maximum and minimum colouration and ball size respectively. Panel (e) reports the total number of connections within the graph, this informs on the points within the overlaps of balls and hence the density of the graphs. Panel (f) plots the average number of connections amongst connected balls. In the case that there are no connected balls then this figure is set to 0. All estimates are generated using the R package \textit{BallMapper} \citep{dlotko2019R}.} 
\end{figure}

Figure \ref{fig:br2} presents similar graphs for the five part cloud, showing again the quick decay of ball numbers and zero connected balls as the radius increases. Once more it is evidenced that the larger balls encompass a greater range of outcome values resulting in a narrowing of the colouration range in panel (c). With the growth of the balls comes not only a large maximum size but also a larger minimum size as well. By the highest radii there is overlap of the 95\% intervals around the mean from the bootstraps. Both the total number of connections in panel (e), and the average number per connected ball in panel (f), display hump shapes as initially the radius increase creates more overlap between balls. When the radius becomes even larger the reduction in ball numbers duly reduces the total number of connections and the number of possible balls a remaining ball may connect to. Here we evidence all of the hypothesised effects from radius increasing, and confirm the humped relationship from the noise cloud.

To see what this means for the BM graphs themselves we may showcase examples from both clouds. In the case of the noise cloud the restricted radius range means we show radii of 1, 2 and 4 in panels (a), (b) and (c) of Figure \ref{fig:br3} respectively. The way that the balls come together to form the main mass is clear, as is the creation of some larger balls in panel (b). When comparing across plots it is not possible to hold the size constant, rather the sizes are on a scale from the largest to smallest. Although we see some similar sized balls in (a) it cannot be stated that these contain as many points as the light blue ball so prominent in (b). Likewise when we move to panel (c) the balls may appear small but they are in fact containing many more points than any in (a) or (b). What we see additionally is the connectivity increasing and then becoming complete in (c). For the five part cloud, panel (d) is the familiar result from the earlier examples of $\epsilon=2$, whilst the simplification to panel (e) produces no outliers. The five different averages remain present, but there is some blurring as a result of the connection through the overlapping points. When the radius reaches 10 and there are only two balls they will naturally be at the opposite ends of the colour spectrum. This reminds on the need to read the scale bar, so doing reveals that the range of colours in panel (f) is much smaller than that in (e) or (d).

Nothing in this section says what the optimal radius is. Indeed there is no way to say that having more balls, or more connectivity is better. In this case having a radius of 2 shows the shape of the clouds well but does not produce a BM graph that is easy to read. Jumping to a radius of 4 makes the graphs simple to discuss but is losing some of the detail of the underlying datasets. The conclusion from this review is thus that including more than one radius would be good practice, with consideration at least given to the impact of different ball radii on results. Section \ref{sec:app} puts these thoughts into practice for two example studies.

\subsection{Distributions}

Assumption of a normal distribution for the artificial data makes natural appeal to the datasets studied within finance. However, the distribution of points is important to the shape of the resulting point cloud in exactly the way it is when visualised on a scatter plot. As a final empirical exercise the role of radius is considered with draws from the random uniform distribution. We contrast the results with the role of radius in a normal distribution as used for the noise cloud of previous subsections. For comparability the variables are all scaled onto [0,1] prior to running the BM algorithm. Results are presented for five dimensional case, in line with the earlier examples of this paper.

\begin{table}
    \begin{center}
        \caption{Summary Statistics for Distribution Comparison Clouds}
        \label{tab:distcss}
        \begin{tabular}{l l c c c c c c c}
        \hline
        Distribution & Variable & Mean & s.d. & Min & q25 & q50 & q75 & Max\\
        \hline
           Normal & $X_1$ & 0.515&0.157&0.000&0.413&0.512&0.615&1.000\\
&$X_2$&0.508&0.171&0.000&0.398&0.507&0.623&1.000\\
&$X_3$&0.527&0.167&0.000&0.416&0.533&0.640&1.000\\
&$X_4$&0.496&0.163&0.000&0.390&0.488&0.604&1.000\\
&$X_5$&0.514&0.170&0.000&0.397&0.523&0.629&1.000\\
Uniform&$X_1$&0.501&0.287&0.000&0.259&0.495&0.734&1.000\\
&$X_2$&0.498&0.291&0.002&0.240&0.504&0.766&1.000\\
&$X_3$&0.518&0.294&0.001&0.252&0.535&0.779&0.998\\
&$X_4$&0.497&0.300&0.001&0.241&0.493&0.767&0.999\\
&$X_5$&0.516&0.299&0.007&0.259&0.516&0.797&0.999\\
\hline
        \end{tabular}
    \end{center}
\footnotesize{Notes: Summary statistics for the two five-dimensional example clouds. Each cloud contains 500 points for which each axis is a random draw from a stated distribution. For the cloud labelled normal are taken from a normal distribution of mean 0 and variance 1, then normalised onto the range $\left[0,1\right]$. Points in the uniform cloud are taken from a uniform distribution on the interval $\left[0,1\right]$. }
\end{table}

\begin{figure}
    \begin{center}
        \caption{Example Uniform and Normal Point Clouds}
        \label{fig:unifnormcloud}
        \includegraphics[width=8cm]{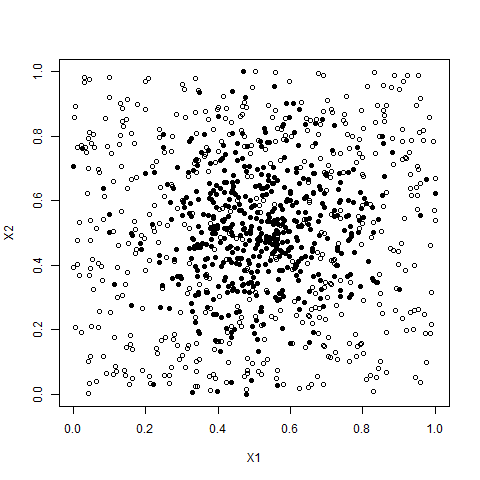}
    \end{center}
\footnotesize{Notes: Scatterplot showing two the co-ordinates of points within five dimension clouds. Each cloud contains 500 points for which each axis is a random draw from a stated distribution. Black points are taken from a normal distribution of mean 0 and variance 1, then normalised onto the range $\left[0,1\right]$. Hollow points are taken from a uniform distribution on the interval $\left[0,1\right]$. All other pairwise combinations between the five axes produce similar results.}
\end{figure}

Table \ref{tab:distcss} provides summary statistics for the two clouds, with Figure \ref{fig:unifnormcloud} showing the key differential between the uniform and normalised noise cloud. Concentration of the noise cloud around the mean, 0.5, is evident, whilst the few points in the tails of the normal distribution also stand out. By contrast the more even spacing of the uniform points is also immediate. Hence the distances between points in the uniform cloud are higher, but the transitions to full connectivity are also quicker because there are no tail observations located far from the main mass of data. In Table \ref{tab:distcss} these effects can be seen in the larger interquartile range within the uniform distribution. Because only 500 points are used the statistics are not equal to their theoretic values, although most are within a short range. Means range from 0.496 to 0.527, and medians likewise are not precisely 0.5. 

\begin{figure}
    \begin{center}
        \caption{Uniform and Normal Distribution Ball Mapper Summaries}
        \label{fig:distss}
        \begin{tabular}{c c}
             \includegraphics[width=7cm]{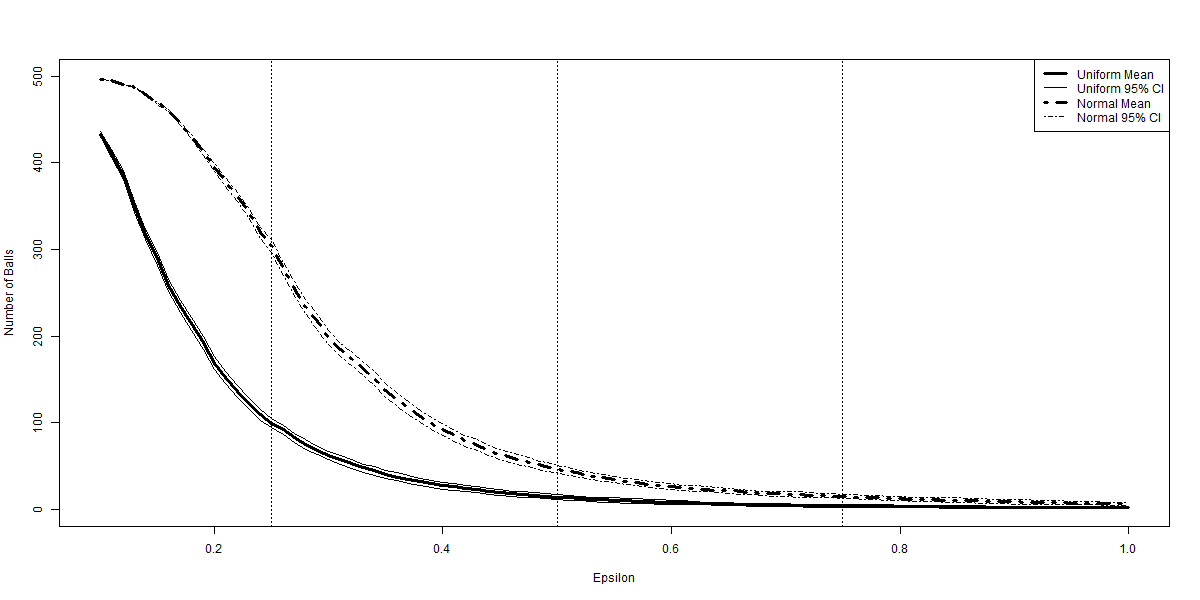}&
             \includegraphics[width=7cm]{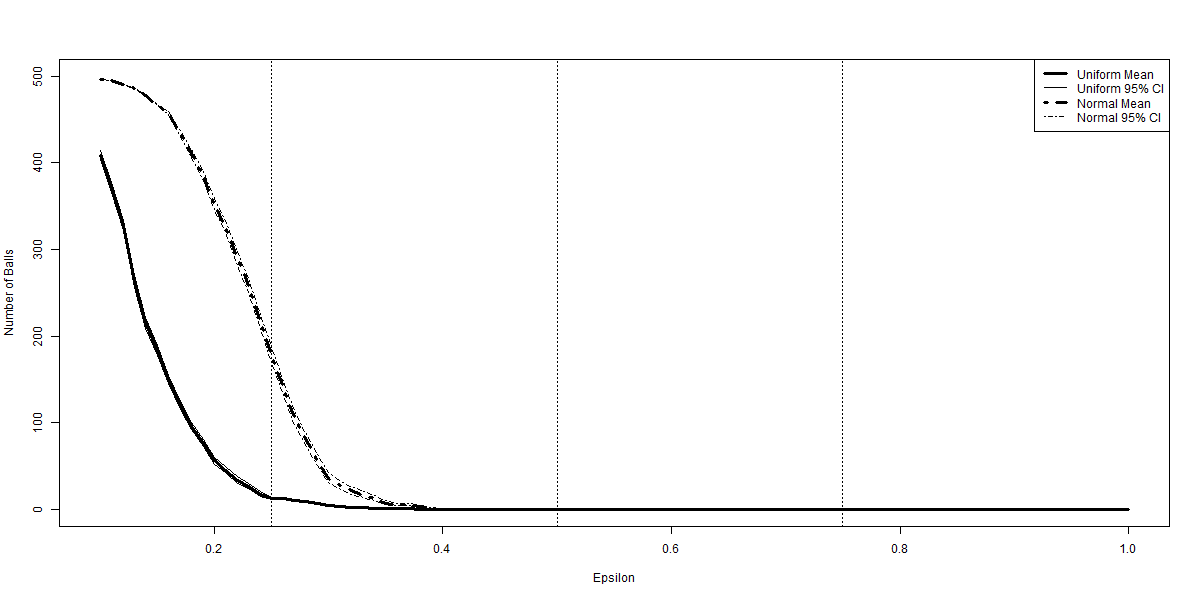}\\
             (a) Number of Balls & (b) Number of Zero Connection Balls \\
             \includegraphics[width=7cm]{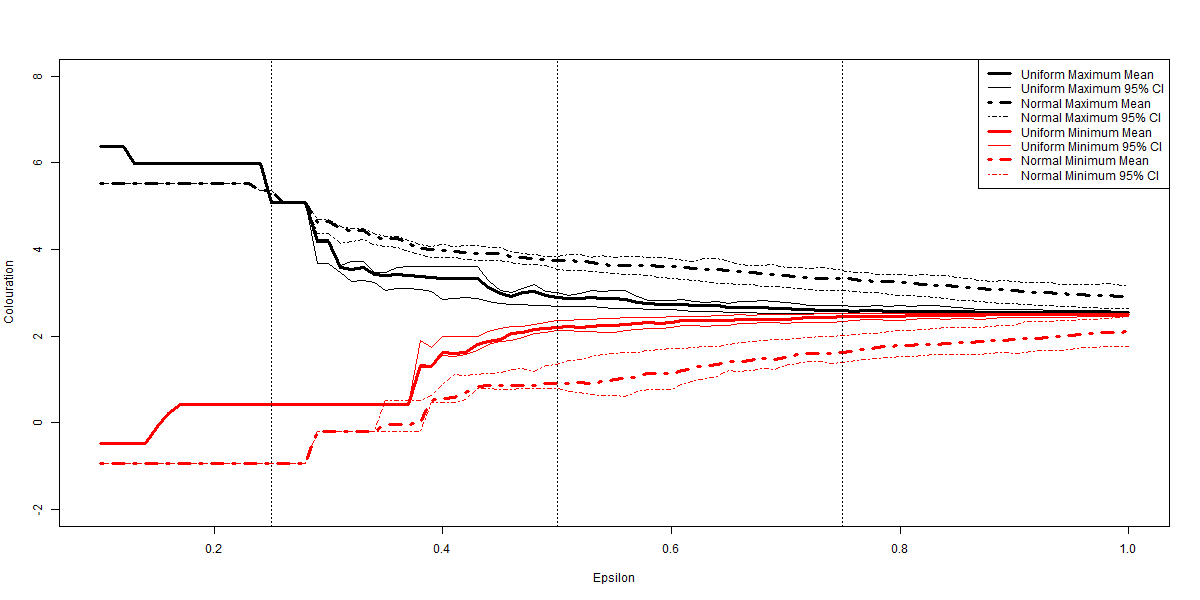}&
             \includegraphics[width=7cm]{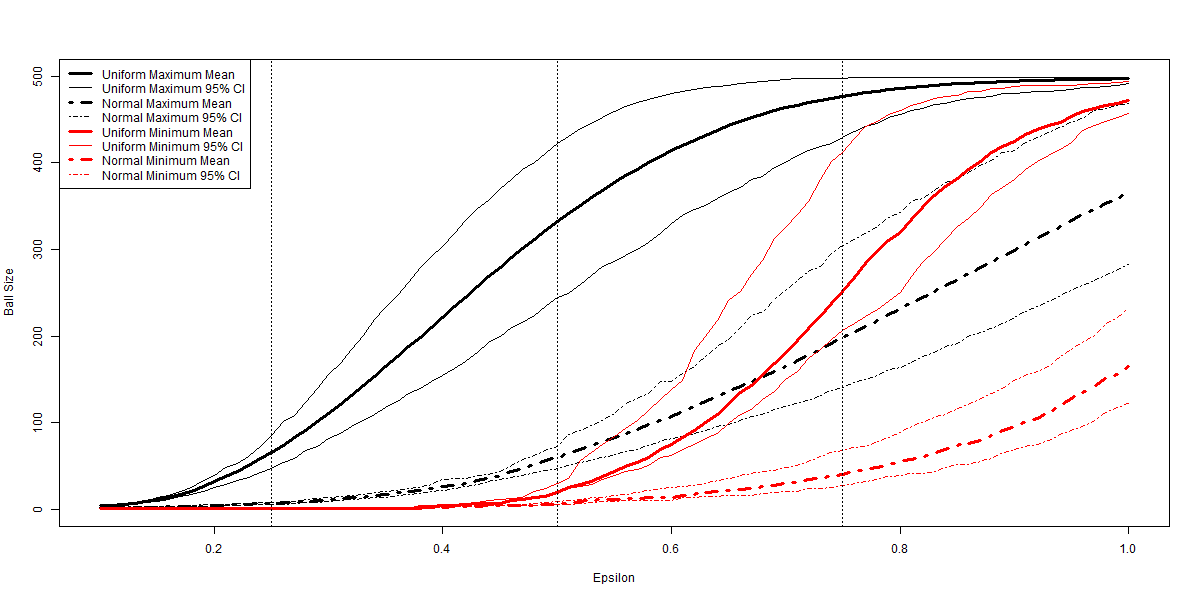}\\
             (c) Colouration & (d) Ball Sizes \\
             \includegraphics[width=7cm]{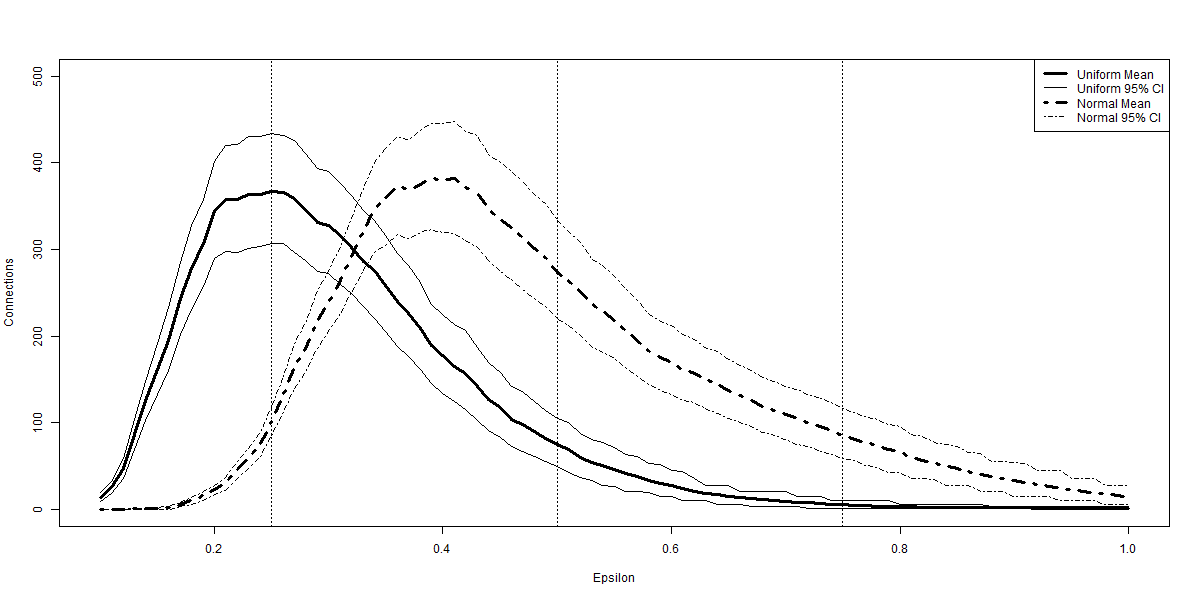} &
             \includegraphics[width=7cm]{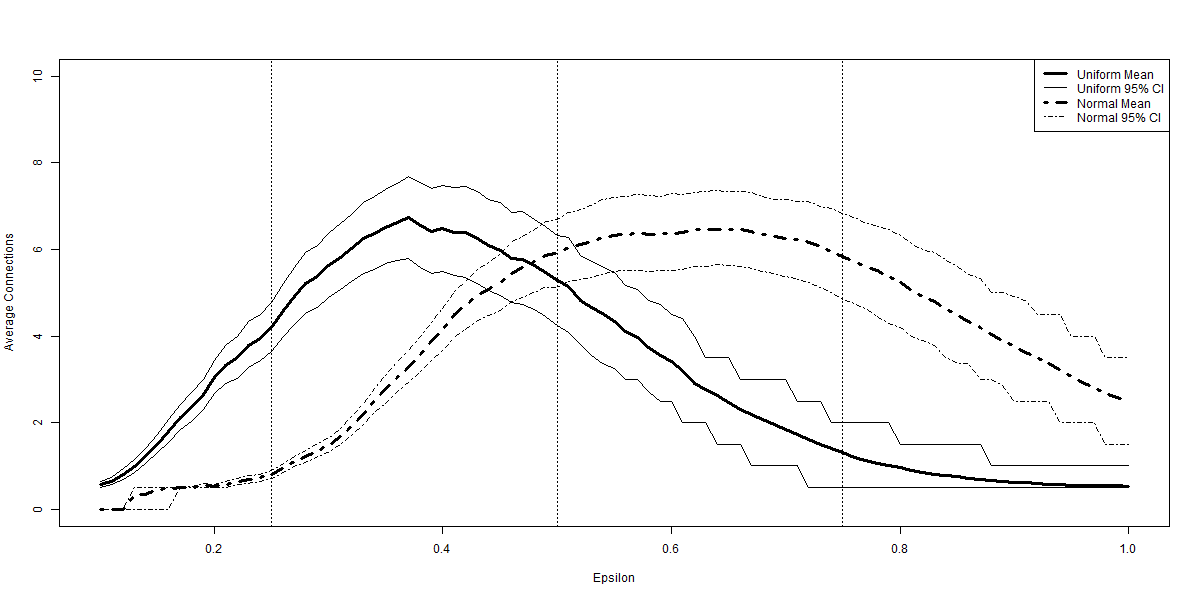} \\
             (e) Total Connections & (f) Average Connections \\
        \end{tabular}
    \end{center}
\footnotesize{Figures plot the impact of the radius of the balls, $\epsilon$, used in the construction of the BM graphs of the normal and uniform clouds. Both clouds contain 500 points. The normal cloud comprises five axes in which the co-ordinates of a point on each axis are drawn randomly from a standard normal distribution of mean 0 and variance 1. After construction of the co-ordinates for all points each of the five axes is normalised onto the interval $\left[ 0,1 \right]$. The uniform cloud comprises five axes in which the co-ordinates of a point on each axis are drawn randomly from a uniform distribution on $\left[ 0,1 \right]$. In each case 10000 repetitions of the BM algorithm \citep{dlotko2019ball} are implemented. A thick line is used to denote the mean from the repetitions, thinner lines denoting the 95\% confidence interval there around. Solid lines represent the normal cloud and dot-dash lines represent the uniform cloud. Panel (a) reports the number of balls, and panel (b) the number of balls which have 0 connections to any other balls. Panels (c) and (d) also use red lines to show the maximum and minimum colouration and ball size respectively. Panel (e) reports the total number of connections within the graph, this informs on the points within the overlaps of balls and hence the density of the graphs. Panel (f) plots the average number of connections amongst connected balls. In the case that there are no connected balls then this figure is set to 0. All estimates are generated using the R package \textit{BallMapper} \citep{dlotko2019R}.}
\end{figure}

Using these two clouds we apply the BM algorithm at radii between 0.1 and 1 in intervals of 0.01. For each radii the algorithm is repeated 10,000 times with random re-ordering of the cloud data between each repetition. The results are summarised in Table \ref{tab:distss} and plotted in Figure \ref{fig:distss}. From the summary statistics it is hypothesised that the number of balls will fall faster for the normal distribution, that the connectivity will be higher in the normal distribution and that the ball sizes will increase quicker for the normal distribution. Because the colouration of each point is derived from the sum of it's co-ordinates we would expect the maximum and minimum values of colouration to be higher in the case of the uniform cloud. 

\begin{table}
    \begin{center}
        \caption{Distribution datasets summary statistics}
        \label{tab:distss}
                \begin{tabular}{l c c c c c l l c c c c c}
        \hline
        \multicolumn{6}{l}{Panel (a): Normal Noise Cloud} && \multicolumn{5}{l}{Panel (b): Uniform Noise Cloud}\\
        $\epsilon$ & Balls &  $\Delta$Size & $\Delta$Col & Zero & Con && Balls & $\Delta$Size & $\Delta$Col & Zero & Con\\
        \hline
            0.1&433.27&3.1&6.87&408.77&0.57&&496&1&6.45&496&0\\
0&(1.73)&(0.55)&(0)&(2.52)&(0.04)&&(0)&(0)&(0)&(0)&(0)\\
0.25&99.45&65.03&4.66&12.13&4.20&&304.15&6.04&6.3&178.76&0.8\\
0&(2.99)&(10.79)&(0)&(1.02)&(0.29)&&(3.76)&(0.52)&(0)&(4.55)&(0.05)\\
0.5&14.07&312.22&0.69&0&5.28&&46.27&53.99&2.83&0&5.91\\
0&(1.46)&(47.08)&(0.14)&(0)&(0.53)&&(2.44)&(7.65)&(0.19)&(0.02)&(0.4)\\
0.75&3.63&226.31&0.14&0&1.31&&14.53&157.3&1.7&0&5.85\\
0&(0.91)&(62.86)&(0.07)&(0)&(0.45)&&(1.36)&(40.96)&(0.23)&(0)&(0.5)\\
0.9&2.24&69.67&0.06&0&0.62&&8.58&203.26&1.14&0&3.75\\
0&(0.47)&(37.84)&(0.04)&(0)&(0.23)&&(1.15)&(57.48)&(0.27)&(0)&(0.56)\\
\hline
        \end{tabular}
    \end{center}
\raggedright
\footnotesize{Notes: Summary statistics for normal and uniform noise clouds with 500 points. The normal cloud comprises five axes in which the co-ordinates of a point on each axis are drawn randomly from a standard normal distribution of mean 0 and variance 1. After construction of the co-ordinates for all points each of the five axes is normalised onto the interval $\left[ 0,1 \right]$. The uniform cloud comprises five axes in which the co-ordinates of a point on each axis are drawn randomly from a uniform distribution on $\left[ 0,1 \right]$. $\epsilon$ reports the radius of the balls used to form the BM graph, Balls is the total number of balls within the BM graph, $\Delta$Size is the difference in size between the smallest and largest ball, $\Delta$Col is the difference between the highest and lowest colouration value for any ball within the graph, Zero is the number of balls for which there is no connectivity to any other ball and Con. is the average number of connections per ball amongst those balls that have at least one connection. All figures are the means from the 10000 repetitions at each point number, with figures in parentheses being the standard deviation across all values within the 10000 repetitions. Here we report only those epsilon for which 10000 observations were derived.}
\end{table}

Evidence form Table \ref{tab:distss} and Figure \ref{fig:distss} shows that the hypothesised relationships do indeed appear. The number of balls falls as the radius increases, as it does in all cases. However, the rate of fall is much faster for the normal distribution. Panel (a) of Figure \ref{fig:distss} shows that at a radius of 0.01 there are still almost 500 individual balls, whilst the normal distribution has already seen multiple points in one ball. This message is reinforced in the range of sizes where the normal distribution has 3.10 but the uniform distribution produces a maximum range of just 1. The message on connection also comes through clearly, beginning with the much more rapid fall in the number of zero connection balls in the normal cloud. In panel (b) of Figure \ref{fig:distss} we see both distributions reaching 0, but the faster progress of the normal distribution case to that point is evident. Closer inspection reveals that although the normal cloud does fall fast the transition to 0 is actually slower as it takes higher radii to capture those last few outliers. Panels (e) and (f) support the faster construction of the connections in the normal cloud, both cases having their peak value below the corresponding peaks of the uniform cloud. Panel (c) helps us understand the impact on colouration, with the normal distribution illustrated as solid lines. We see that the highest and lowest colourations converge much quicker than the corresponding uniform lines. By the highest radius of 1, the uniform maximum and minimum colouration are still not within each others confidence interval. 

When dealing with data in practice, the distributions of the points are given and need not be either uniform or normal. The message from this section is that if data is spread out then BM is able to better show what is happening than when the data becomes focused on a very narrow part of the space. Where data is normally distributed the solution is to introduce smaller radii in the first instance and focus on the messaging within the densest part of the space. A second plot with higher radius can then comment on the connectivity of the sparser regions near the tails. In many instances the fact that the balls are not connected to the main mass will be sufficient for inference. Within the finance literature it is common to perform winzorisation on a dataset to reduce outliers. In such cases the values of each of the axis variables that are above (below) a stated quantile are replaced by that stated quantile. For BM winzorisation will mean that there are multiple points with identical co-ordinates and may result in larger balls at the ends of the distributions. This should be born in mind by any researcher applying winzorisation prior to the BM algorithm.

\begin{figure}
    \begin{center}
        \caption{Uniform and Normal Distribution Ball Mapper Examples}
        \label{fig:disteg}
        \begin{tabular}{c c c}
            \includegraphics[width=5cm]{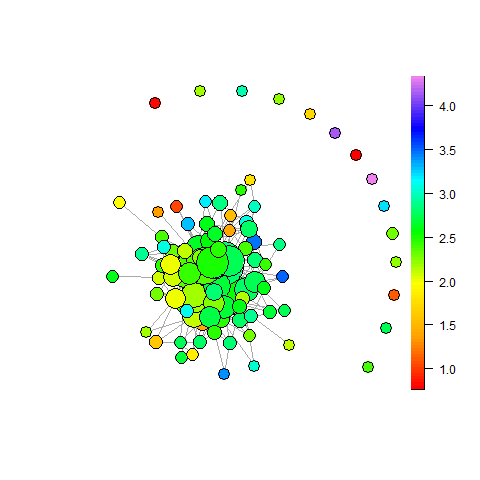}&
             \includegraphics[width=5cm]{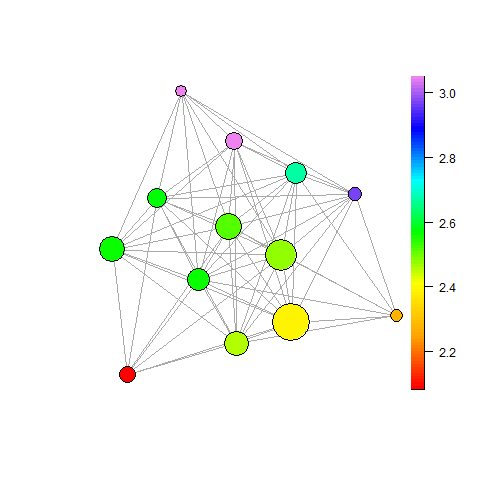}&
             \includegraphics[width=5cm]{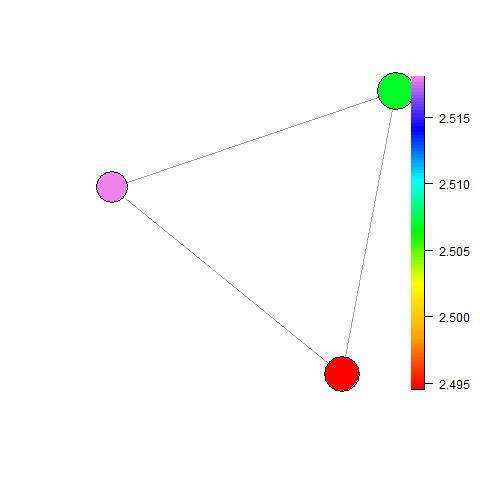}\\
             (a) Normal $\epsilon=0.25$ & (b) Normal $\epsilon=0.50$ & (c) Normal $\epsilon=0.75$ \\
             \includegraphics[width=5cm]{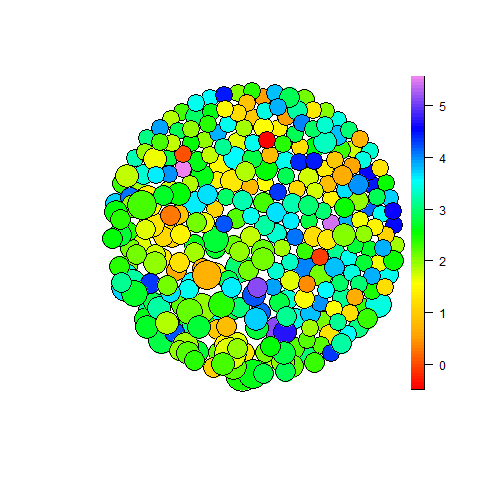}&
             \includegraphics[width=5cm]{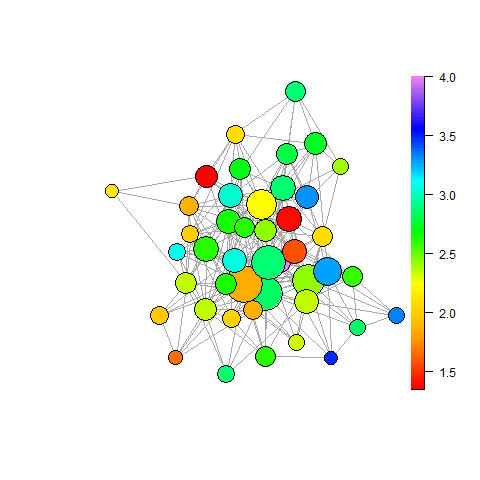}&
             \includegraphics[width=5cm]{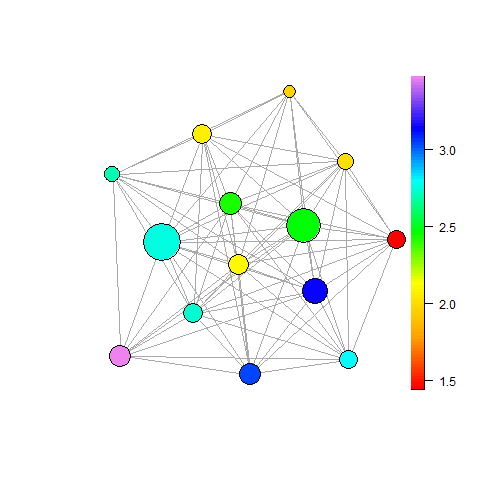}\\
             (d) Uniform $\epsilon=0.25$ & (e) Uniform $\epsilon=0.50$ & (f) Uniform $\epsilon=0.75$ \\
        \end{tabular}
    \end{center}
\footnotesize{Notes: Figures show example ball mapper diagrams for three radii, $\epsilon$, on each of two point clouds. Normal refers to a five-dimensional cloud in which each co-ordinate is drawn randomly from a standard normal distribution. All five axes are then scaled onto $[0,1]$ for comparability with the uniform cloud. The uniform cloud is also five-dimensional with the co-ordinate of a point on each axis drawn randomly from a uniform distribution on $[0,1]$. Both clouds contain 500 points. Colouration of the BM plots is based upon the sum of the co-ordinates of each point. All estimates are generated using the R package \textit{BallMapper} \citep{dlotko2019R}.}
\end{figure}

Figure \ref{fig:disteg} provides BM plots of the normal and uniform noise clouds used within this section at three radii. Through these plots the density of the normal distribution near the centre of the space comes through clearly, as does the rapid reduction in the number of balls when the radius increases. At the radius of 0.25 the uniform cloud in panel (d) of Figure \ref{fig:disteg} does not show any value, whilst the normal cloud is already emerging as a dense mass with some outliers in panel (a). Continuing to the radius of 0.5, we see that the normal distribution produces larger balls that are nearer the middle value and then smaller balls whose colouration is at the ends of the range. The colour bar associated with panel (b) covers a much shorter range, approximately 2.35 to 3.45, compared to the range of 1.4 to 4.1 on the uniform plot in panel (e). Although we see some larger balls, these cover many different colouration levels so do not depict a central mass. Likewise we note that the BM algorithm scales ball size according to the range of number of points within the ball; the number of points within a ball in panel (b) will be much higher than one which is given the same size on the plot in panel (e). Finally by panel (c) the normal distribution is producing just three balls and each is similar in colouration. The uniform case by contrast still has many balls and a colouration range of 1.5 to 3.4, far wider than the normal produced at the radius of 0.50. 

A natural question emerging from the comparison of normal and uniform distribution clouds in Figure \ref{fig:disteg} concerns the similarity of panels (b) and (f). These are produced by the normal cloud with radius 0.50 and the uniform cloud with radius 0.75 respectively. Closer inspection reveals that the larger balls in panel (b) have similar colouration values, whilst those in panel (f) do not. We see all of the balls in (b) around the 2.4, 2.5 level, whilst in panel (f) the two largest balls have colourations of 2.4 and 2.7. There is a large blue ball with colouration 3.0 in panel (f) but no large ball in panel (b) comes close to this. To see BM as a way of distinguishing between distributions it is necessary to apply similar radii, but even with different radii the message can be taken from colouration. Were we to colour by the co-ordinates on any one of the axes then the differences between (b) and (f) would appear strongly again. 

This section serves as a brief overview of the impact of distributions on BM plots. It reminds that our data will behave differently according to the underlying distribution of each of the axis variables. Extension may be made to look at how skewness and kurtosis change the pictures. Assuming colouration continues to be the sum of the co-ordinates of a point, the impact of such is perhaps intuitive since skewness creates a long tail and leaves the colouration of the largest balls nearer to one end of the overall range. Kurtosis meanwhile concentrates the peak and means that the tails are longer on both sides. For application in Finance the distribution is important, but it is the other factors that have been reviewed here, such as numbers of points, axes and the ball radius that are the important application decisions.

\section{Alternative Colouration}
\label{sec:colour}

To this point all graphs have been plotted with the sum of the axis values as the colouration variable. The average value obtained across all of the points within a given ball providing the colouration value. Such colouration is of use when the outcome variable is continuous, but we may also consider other options. A first discussion is made of cases where there is a binary outcome. The mean of a dummy variable for a specific characteristic represents the proportion of observations possessing that particular characteristic. For example a dummy variable for firm failure in the subsequent year has an average equal to the proportion of all firms in the present period data which fail in the subsequent year. Within a ball colouration by the mean of a dummy is the proportion of observations in the specific part of the joint distribution having the characteristic. Secondly, BM can colour according to a user defined function instead of the mean. Because BM knows which points are within a ball it becomes a straightforward exercise to code by functions such as the standard deviation, range, minimum, maximum or modal values amongst the identified set of points. For illustration here we consider the standard deviation of outcomes amongst ball members as the colouration function.

\subsection{Binary Outcomes}

\begin{figure}
    \begin{center}
        \caption{Proportion of Points with $X_3>0$}
        \label{fig:binary}
        \begin{tabular}{c c}
             \includegraphics[width=7cm]{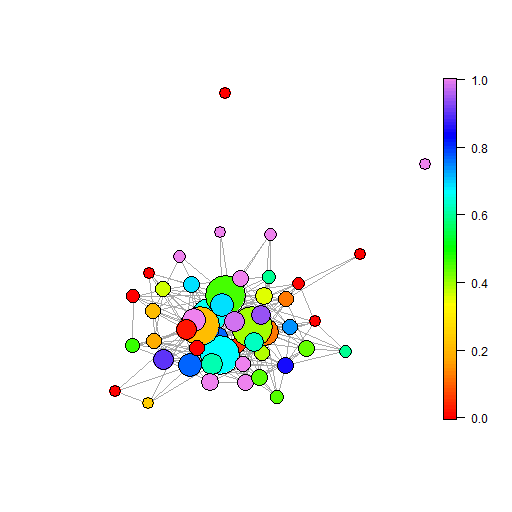} & \includegraphics[width=7cm]{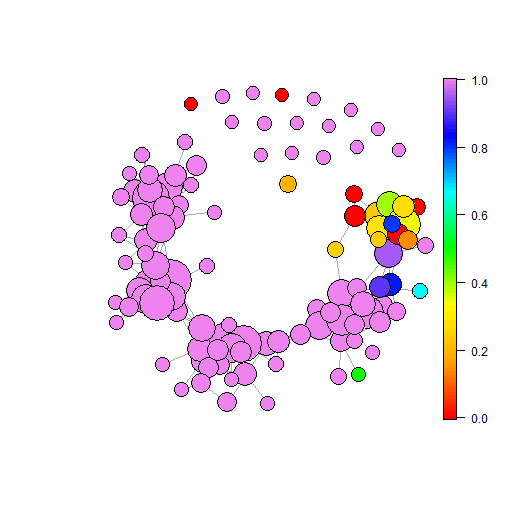}  \\
             (a) Noise Cloud & (b) Five Part Cloud 
        \end{tabular}
    \end{center}
\raggedright
\footnotesize{Notes: Both clouds are constructed from 500 points with co-ordinates on five axes. Each co-ordinate is drawn at random from a normal distribution with given mean and variance 1. For the noise cloud all 500 points are drawn with mean 0, whilst for the five part cloud there are five draws for each variable, one hundred from a distribution with mean 0, 100 with mean 2, 100 mean with mean 4, 100 mean 6 and finally 100 with mean 8. Colouration of these balls is according to the proportion of points within the ball that have $X_3>0$. BM graphs are created using the R package \textit{BallMapper} \citep{dlotko2019R} with ball radius $\epsilon=2$.}
\end{figure}

In the case of a binary variable the average value is the proportion of points within the ball that have the given characteristic. For example we may employ the binary variable for $X_3<0$, which was used in Section \ref{sec:rep} to show how scatter plots can use different shape points to showcase further information. In this case let us instead create a dummy variable in our dataset for $X_3>0$ and then use that variable in the colouration of the BM plot. Figure \ref{fig:binary} presents the resulting BM graph.

Panel (a) shows how the lack of correlation between the axes of the noise cloud means that there is a reasonable spread of $X_3>0$ across the full range of values of other variables. Despite this BM is able to concentrate the higher values within the overall cloud, those balls in the blues and purples being proximate in most instances. Where we do see value from BM is in the colouration of the five part cloud in panel (b). Here the 0 mean cloud is where we see cases of $X_3<0$; the colouration is not purple. However we can also see how there are some balls in the next block that do not have all of their member observations with $X_3>0$. There is a green ball to the bottom right of the plot that shows such well. Recalling that BM preserves knowledge of which points are in which ball we may query to find out more about either this green ball or the outliers that do not have coloration 1. 

In most instances where a binary variable is used the inference would be for something stronger than just the value of one axis. In the two applications for Section \ref{sec:app} the outcome variables are binary, being firm failure and the stock market moving in an upward direction. Note that for these two example clouds there is no meaningful dummy that is not related directly to one of the axes. 

\subsection{Alternative Colouration Functions}

\begin{figure}
    \begin{center}
        \caption{Standard Deviation of Outcome: Five Part Cloud}
        \label{fig:sd5}
        \begin{tabular}{c c}
             \includegraphics[width=7cm]{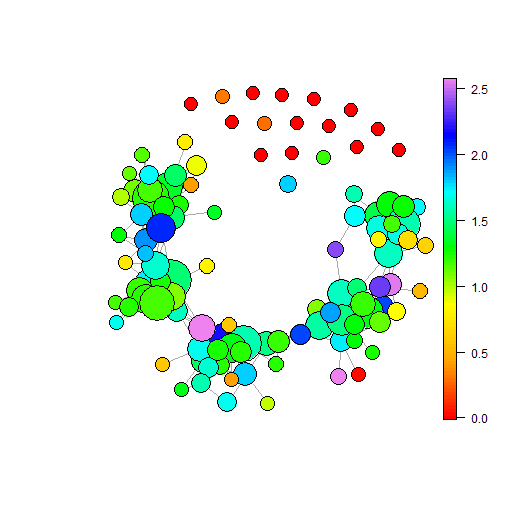} & \includegraphics[width=7cm]{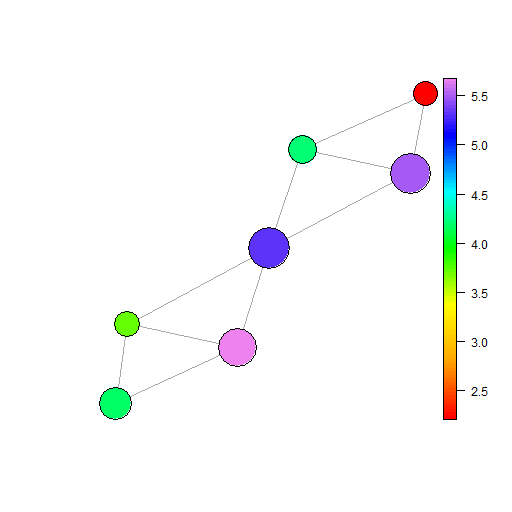}  \\
             (a) Five Part Cloud ($\epsilon=2$) & (b) Five Part Cloud ($\epsilon=5$)
        \end{tabular}
    \end{center}
\raggedright
\footnotesize{Notes: Both clouds are constructed from 500 points with co-ordinates on five axes. Each co-ordinate is drawn at random from a normal distribution with given mean and variance 1. There are five draws for each variable, one hundred from a distribution with mean 0, 100 with mean 2, 100 mean with mean 4, 100 mean 6 and finally 100 with mean 8. Colouration of these balls is according to the standard deviation of the value of $Y$ within each ball. For point $i$, $Y_i$ is the sum of the five axis coordinates. BM graphs are created using the R package \textit{BallMapper} \citep{dlotko2019R} with ball radius $\epsilon=2$ and $\epsilon=5$ as indicated.}
\end{figure}

Understanding average values of an outcome variable in a space is akin to plotting the distribution across one variable, the size of the ball corresponding to the density. There is inherent value in seeing the multidimensional surface created by the axis variables. However, we do not need to restrict ourselves to the average value within a ball. One obvious area of interest is the amount of variability of the outcome variable within the ball. To capture such we may consider the standard deviation of the outcome variable as an alternative colouring function. The results are presented in Figure \ref{fig:sd5}

Within the core of each sub-cloud cluster the main large balls of Figure \ref{fig:sd5} are green in colour, showing the variation we might expect from having so many points. In our case the outcome variable is the sum of all inputs so it is no surprise that the lowest standard deviations are in smaller balls. When the outcome is a noisy function of the axis variables, or there is limited dependence, then we may expect high standard deviations within balls. High within variation may therefore be used as a further summary of the data and hypothesised relationships. For the five point cloud, Figure \ref{fig:sd5} shows that when the radius increases we find balls that cover more than one sub-cloud and inherently this gives them a larger standard deviation. At the extreme, there are balls in the middle of the shape that have the potential to cover three sub-clouds and these will have the highest standard deviation. Here again the evidence from panel (b) of Figure \ref{fig:sd5} supports the position with the highest colouration being towards the centre of the shape.

\section{Reading Ball Mapper Graphs}
\label{sec:label}

To this point the BM graphs have been left without annotation, allowing the viewer to understand the messages about the shape of the underlying data that the BM representation conveys. However, to ensure effective use of the plots it is helpful to label each ball. Once labelled the plots must be drawn bigger to facilitate easier reading of the numbers within the balls, though smaller coloured by axis plots can be added since the layout of all balls will be identical\footnote{Layout is determined by a plotting seed which is set to 123 by default. The user may change this seed to slightly alter the abstraction performed by R in creating the visualisation. An identical seed will then ensure an identical plot regardless of the colouration variable chosen.}. We may then take the BM graph and map it back to the underlying data set by sequentially numbering the points in the data and using the functionality of the \textit{BallMapper} package \citep{dlotko2019R} to produce a mapping of point numbers to ball numbers. Where points appear in multiple balls, because they were in the intersection, that is also identified through this process.

\begin{figure}
    \begin{center}
        \caption{Reading Ball Mapper Graphs: Five Part Cloud}
        \label{fig:sd}
        \includegraphics[width=12cm]{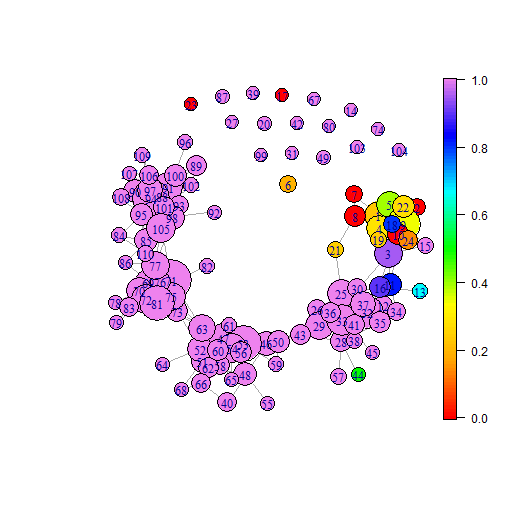}
    \end{center}
\end{figure}

For this section let us return to the colouration rule for binary outcomes introduced in Section \ref{sec:colour}. The colouration function is simply the proportion of points within the ball which satisfy $X_3>0$. Most balls in panel (a) of Figure \ref{fig:sd} are therefore purple, corresponding to the fact that all of their points have sufficiently high $X_3$. In the area corresponding to the first of the five sub-clouds we see the variety of colouration as here the mean of $X_3$ is around 0 and hence we may expect around half of the observations in the sub-cloud would have $X_3<0$. Using the labels we see balls 7 and 8 have high proportions, whilst ball 3 contains almost no points with $X_3<0$.

An interesting case here is ball 44. This is a ball which is coloured green, but is part of the second sub-cloud. There are many balls for which $X_3>0$ between ball 44 and the other non-purple balls. Using the underlying data we may identify which points are in ball 44 and see why this different colouration occurs. Ball 44 contains just two points, one of which has $X_3<0$ and hence the colouration is 0.5; the co-ordinates on all other axes are very similar and hence the ball connects through the point with $X_3>0$ to the main shape. Overall the impact of the different $X_3$ co-ordinates is mitigated somewhat by the other axes and the standard deviation of points within this ball is lower than might be expected, being 0.277. In Figure \ref{fig:sd5} panel (a) it may be seen that ball 44 has a red colour to indicate low values. 

Here we have looked at the value of an input variable to give the colouration. However we may imagine situations in which the differing colouration of a ball, like number 44, would be of relevance. If it is assumed that the outcome should be a continuous function of the axis variables then BM can show simply the cases where that does not hold. The ability to then drill into the data, as demonstrated here, can then facilitate the deeper investigation of what is happening in that part of the characteristic space. We turn to two examples now in which identification using the numbering of BM graphs is used as part of the data analysis.

\section{Applications}
\label{sec:app}

\subsection{Firm Failure Prediction}

Our first example follows \cite{qiu2020refining}, taking a point cloud from the five financial ratios in the \cite{altman1968financial} z-score model. In \cite{qiu2020refining} we demonstrated that firm failures are constrained within a small subset of the financial ratio space. Here we demonstrate the robustness of that result under the metrics for analysing BM graphs developed within this paper. Our demonstration focuses on 2015 data, being the most recent example in \cite{qiu2020refining}.

Five variables form the explanatory set. First is the ratio of working capital to total assets, this is denoted as $X_1$. Secondly the ratio of retained earnings to total assets is incorporated. In what follows the retained earnings to assets ratio is $X_2$. Both of these give an impression of financial health and the ability to generate future income. Third is the ratio of earnings before income and tax to total assets. This is the return on assets and measures the productivity of the firm. High profitability means that the firm should be safe from bankruptcy. $X_3$ is the return on assets. Fourth is the ratio of the market value of equity to total liabilities; higher values mean the market values the firm sufficiently relative the amount it owes others. We denote this ratio as $X_4$. Finally, the total sales to assets ratio informs how productively the firm is making use of its assets to generate revenue. $X_5$ is used for the sales to assets ratio. Later versions of \cite{altman1968financial} have used more ratios, but here we stick with these original set\footnote{See the recent review paper by \cite{altman2017financial} for more.}. Failure in this case is defined as the firm being removed from Compustat citing bankruptcy as the reason.

\begin{table}
    \begin{center}
        \caption{Summary of Firm Failure Prediction Data}
        \label{tab:alt0}
        \begin{tabular}{l c c c c l l c c c c c}
             \hline
             \multicolumn{5}{l}{Panel (a): Summary Statistics}&&\multicolumn{5}{l}{Panel (b): Correlations}\\
             & Mean & s.d. & Min & Max &&  $X_1$ & $X_2$ & $X_3$ & $X_4$ & $X_5$ \\
             \hline
             $X_1$&0.170&0.214&-0.471&0.748 && 1 &&&&\\
            $X_2$&-0.370&1.228&-7.664&0.694&& -0.094 & 1 &&&\\
            $X_3$&-0.013&0.191&-1.005&0.308&& -0.043 & 0.544 & 1 &&\\
            $X_4$&3.072&4.225&0.094&31.92&&0.411&-0.109&-0.051&1&\\
            $X_5$&0.840&0.645&0.000&3.520&&0.152&0.041&0.211&-0.061&1\\
            $Fail$&0.00&0.017&0&1 &&&&&&\\
             \hline
        \end{tabular}
    \end{center}
\raggedright
\footnotesize{Notes: Summary statistics and pairwise correlations for the firm failure dataset. Following \cite{altman1968financial} $X_1$ is the ratio of working capital to total assets, $X_2$ is the ratio of retained earnings to total assets, $X_3$ is the return on assets, $X_4$ is the ratio of market value of equity to total liabilities, and $X_5$ is the total sales to total assets ratio. Fail is a dummy which takes the value 1 if the firm fails in the subsequent year. Here all financial ratios are calculated for 2015, with the failure dummy representing failure in 2016. Panel (a) reports the mean, standard deviation minimum and maximum for each variable. Panel (b) reports the pairwise correlations for the five axis variables used in the construction of the point cloud.$n=3605$}
\end{table}

From the summary statistics in Table \ref{tab:alt0} it may be seen that $X_4$, the market value of equity to total liabilities ratio has an average value far higher than the others. There is also variation in the standard deviations and ranges of observed values that mean we may confirm these variables to be on different scales. For this reason variables are normalised onto the range $[0,1]$ before the BM is run. This is the approach adopted in \cite{qiu2020refining}. When performing BM analysis it is imperative that the scales of variables are similar. Having normalised the range of potential axis values is much smaller and so the range of ball radii to consider must include very small values. In the analysis we use $\epsilon=0.1$ as the minimum and $\epsilon=1$ as the maximum. Correlations in panel (b) show that there is more relationship between the variables than there is in the noise cloud, but the majority of correlations are still very low.  

\begin{table}
    \begin{center}
        \caption{Firm Failure Prediction: Summary}
        \label{tab:alt1}
        \begin{tabular}{l c c c c c }
        \hline

        Radius ($\epsilon$) & Balls &  $\Delta$Size & $\Delta$Col & Zero & Con \\
        \hline
       0.1&926.41&403.97&1&558.97&3.56\\
&(6.67)&(49.73)&(0)&(4.32)&(0.14)\\
0.2&259.55&1191.74&0.16&64.89&4.01\\
&(4.77)&(129.5)&(0.04)&(2.8)&(0.2)\\
0.4&44.39&2455.53&0.02&2.03&5.29\\
&(2.18)&(197.3)&(0.01)&(0.18)&(0.39)\\
0.5&24.51&2849.32&0.01&0&5.05\\
&(1.7)&(171.95)&(0.01)&(0)&(0.42)\\
0.8&6.79&3254.41&0&0&2.81\\
&(1.01)&(123.12)&(0)&(0)&(0.47)\\
1&3.64&2675.85&0&0&1.32\\
&(0.75)&(663.64)&(0)&(0)&(0.37)\\

\hline
        \end{tabular}
    \end{center}
\raggedright
\footnotesize{Notes: Radius ($\epsilon$) reports the radius of ball that is used to construct the BM graph, Balls is the total number of balls within the BM graph, $\Delta$Size is the difference in size between the smallest and largest ball, $\Delta$Col is the difference between the highest and lowest colouration value for any ball within the graph, Zero is the number of balls for which there is no connectivity to any other ball and Con. is the average number of connections per ball amongst those balls that have at least one connection. All figures are the means from the 10000 repetitions at each point number, with figures in parentheses being the standard deviation across all values within the 10000 repetitions. The cloud used comprises five axes, being the ratio working capital to total assets, the ratio of retained earnings to total assets, the ratio of earnings before interest and taxation to total assets, the ratio of total market valuation to total liabilities and finally the ratio of sales to total assets. These are the five ratios suggested in \cite{altman1968financial}.}
\end{table}

Once again we vary the ball radius and take 10000 BM graphs for each radius considered. Consequently we may take summary statistics on the results from these repetitions to inform about the behaviour of the firm failure cloud. Table \ref{tab:alt1} demonstrates the same properties over increasing $\epsilon$ as the artificial cases, the ball number falling rapidly and the size increasing accordingly. Connectivity follows the inverted u-shape as well, the average connectivity amongst connected balls peaking at 5.29 when $\epsilon=0.4$. It may be noted that the number of zero connection balls has almost reached zero at this stage. Given the application of this model to identifying failing firms the difference in colour being 1 at $\epsilon=0.1$ confirms there are some balls in which all firms fail. However, the average number of balls being 926 means that this is not practical from a visualisation perspective. In the examples that follow we use $\epsilon=0.2$, but this already shows a maximum colour difference of just 0.16. First though we may apply the metrics to produce graphs which summarise the 10000 repetitions of the BM algorithm.

\begin{figure}
    \begin{center}
        \caption{Ball Radius: Firm Failure Example}
        \label{fig:alt1}
         \begin{tabular}{c c}
             \includegraphics[width=7cm]{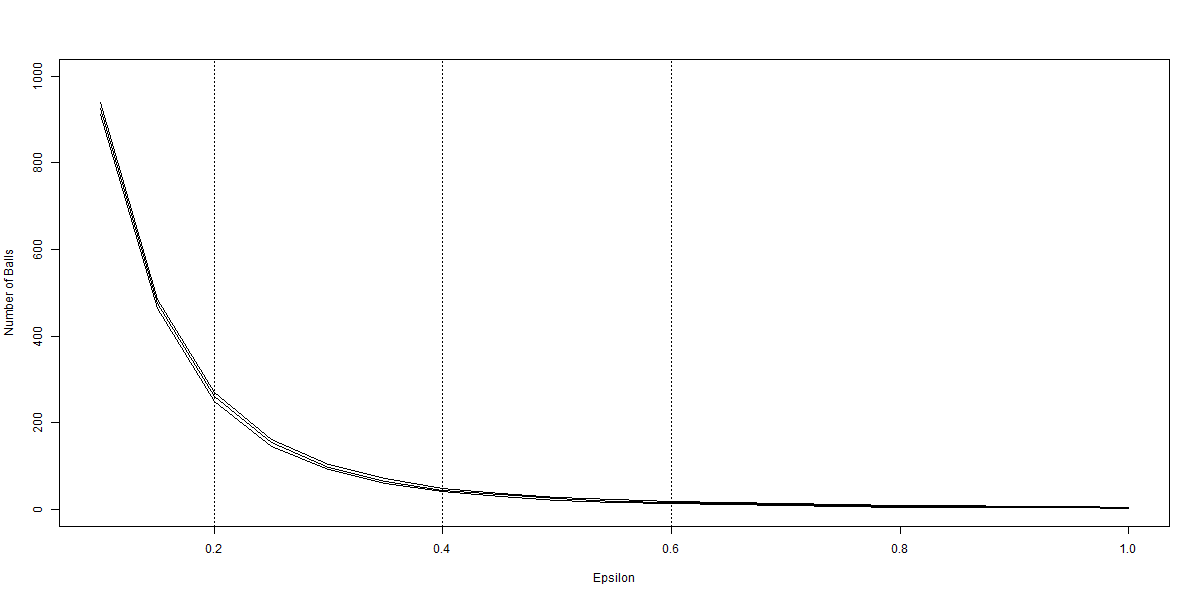} &
             \includegraphics[width=7cm]{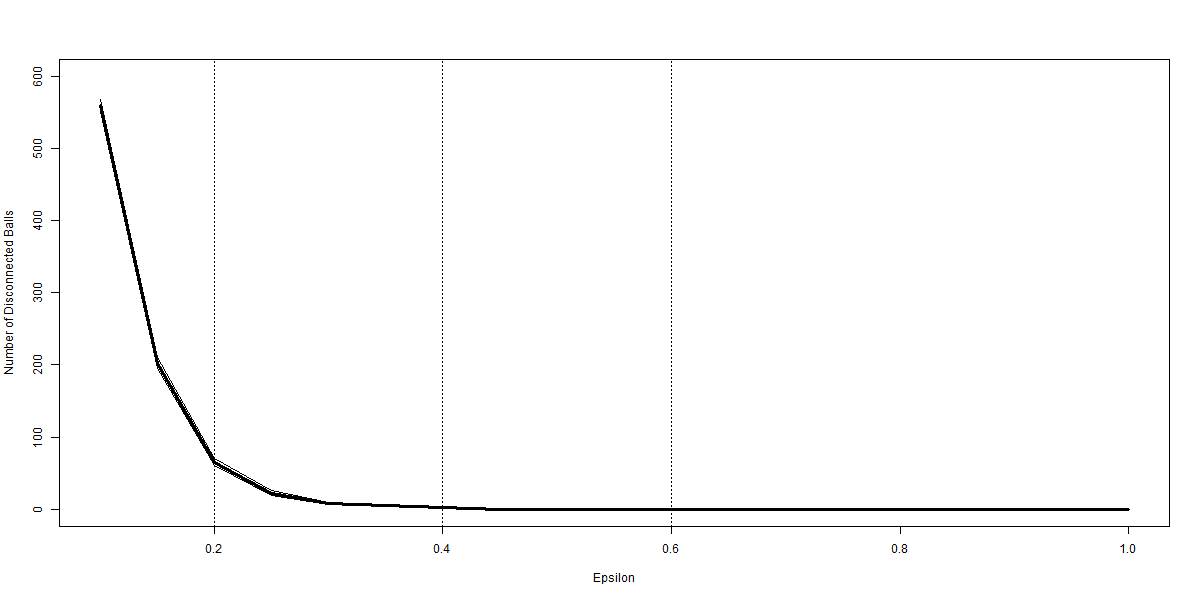} \\
             (a) Number of Balls & (b) Number of Zero Connection Balls \\
             \includegraphics[width=7cm]{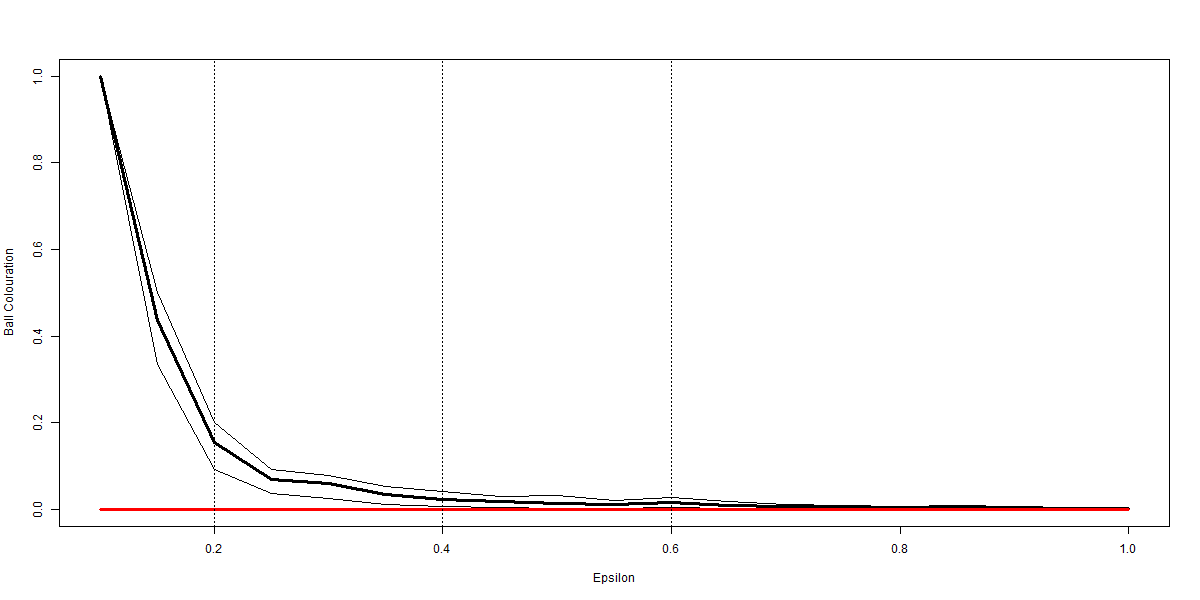} &
             \includegraphics[width=7cm]{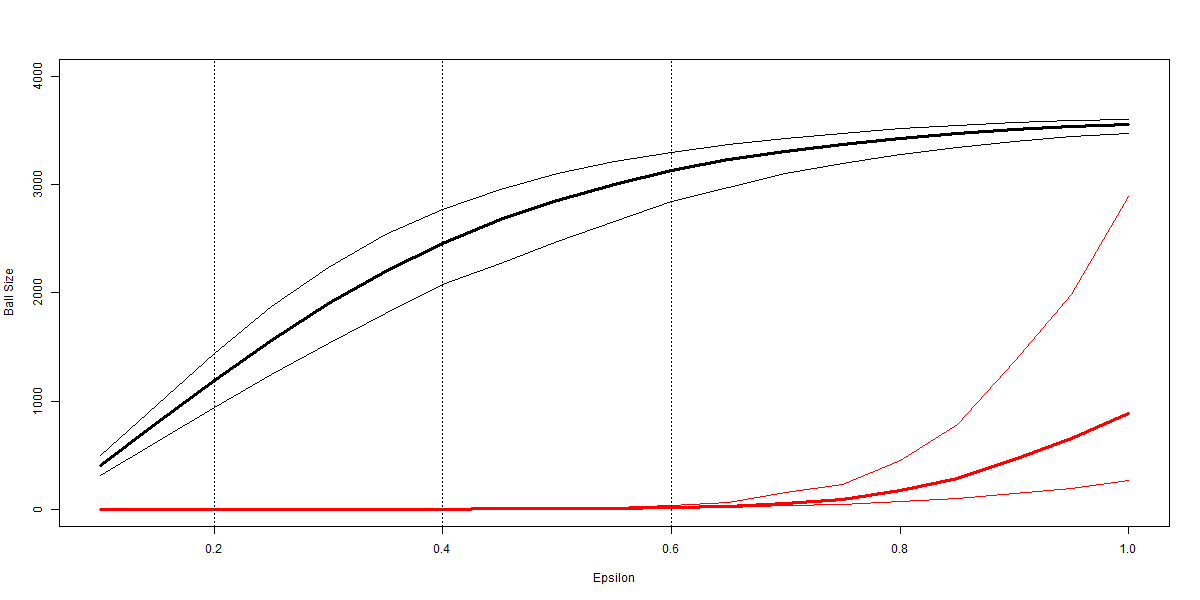} \\
             (c) Colouration & (d) Ball Sizes \\
             \includegraphics[width=7cm]{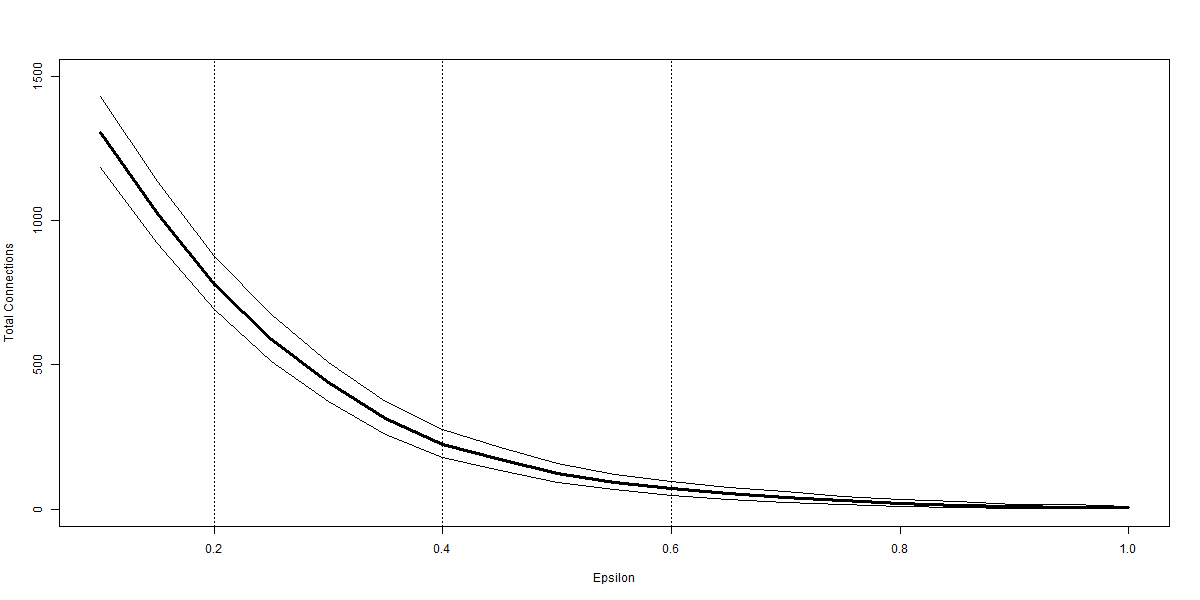} &
             \includegraphics[width=7cm]{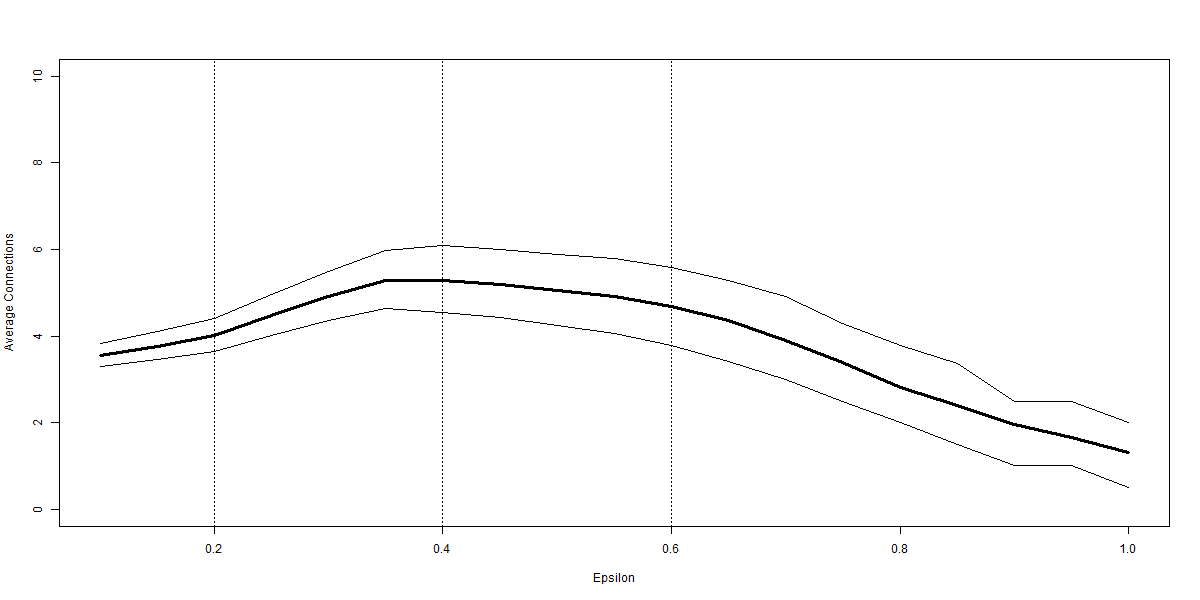}\\
             (e) Total Connections & (f) Average Connections\\
        \end{tabular}
    \end{center}
      \raggedright
\footnotesize{Notes: Figures plot the impact of the radius of the balls, $\epsilon$, used in the construction of the BM graph for the credit scoring example. In each case 10000 repetitions of the BM algorithm \citep{dlotko2019ball} are implemented. A thick line is used to denote the mean from the repetitions, thinner lines denoting the 95\% confidence interval there around. Panel (a) reports the number of balls, and panel (b) the number of balls which have 0 connections to any other balls. Panels (c) and (d) also use red lines to show the maximum and minimum colouration and ball size respectively. Panel (e) reports the total number of connections within the graph, this informs on the points within the overlaps of balls and hence the density of the graphs. Panel (f) plots the average number of connections amongst connected balls. In the case that there are no connected balls then this figure is set to 0. The cloud used comprises five axes, being the ratio working capital to total assets, the ratio of retained earnings to total assets, the ratio of earnings before interest and taxation to total assets, the ratio of total market valuation to total liabilities and finally the ratio of sales to total assets. These are the five ratios suggested in \cite{altman1968financial}. Ratios are normalised onto the interval $[0,1]$ prior to implementation of the BM algorithm. All estimates are generated using the R package \textit{BallMapper} \citep{dlotko2019R}.} 
\end{figure}

\begin{figure}
    \begin{center}
        \caption{Credit Scoring Example Ball Mapper Graphs}
        \label{fig:alt2}
        \begin{tabular}{c c}
             \multicolumn{2}{c}{\includegraphics[width=10cm]{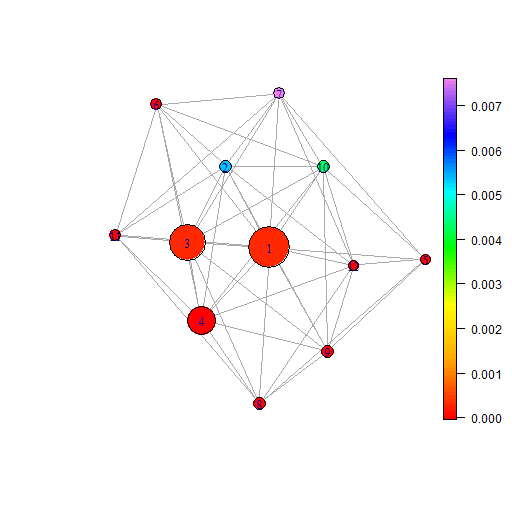}}\\
             \multicolumn{2}{c}{(a) Epsilon = 0.6} \\
             \includegraphics[width=7cm]{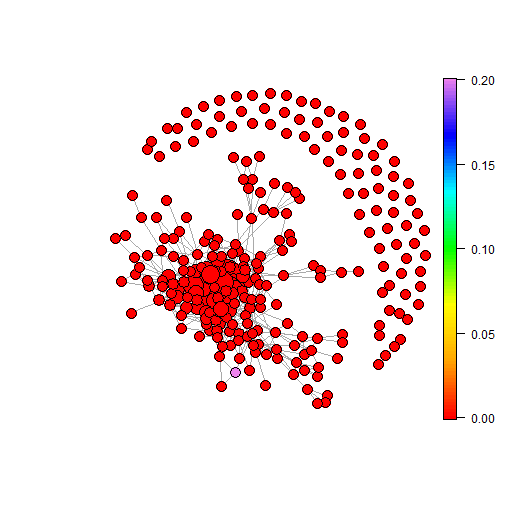} & \includegraphics[width=7cm]{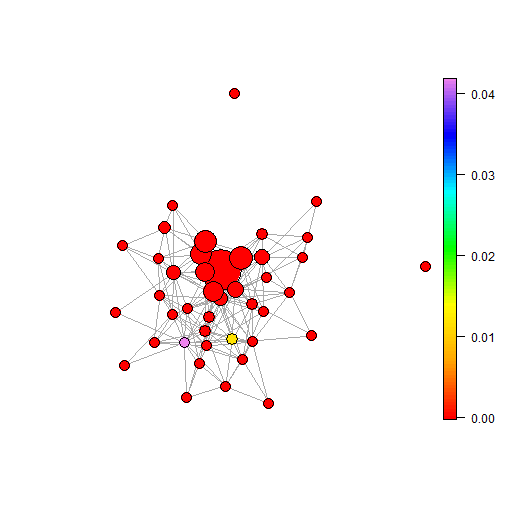} \\
             (b) Epsilon = 0.2 & (c) Epsilon = 0.4
        \end{tabular}
    \end{center}
\raggedright
\footnotesize{Notes: Figures plot example BM graphs for the stated ball radius $\epsilon$. The cloud used comprises five axes, being the ratio working capital to total assets, the ratio of retained earnings to total assets, the ratio of earnings before interest and taxation to total assets, the ratio of total market valuation to total liabilities and finally the ratio of sales to total assets. These are the five ratios suggested in \cite{altman1968financial}. All variables are normalised onto the interval $[0,1]$ prior to implementation of the BM algorithm. Plots are generated using the R package \textit{BallMapper} \citep{dlotko2019R}.} 
\end{figure}

Each of the six panels in Figure \ref{fig:alt1} resembles the shapes from the artificial examples, especially the role of increasing ball radius in reducing ball numbers. However, the colouration picture is different owing to the application. Where the artificial examples took colour from the sum of the axes in this case it is the proportion of firms within a ball that fail. As there are always balls with no failures the minimum colouration stays at 0, whilst the maximum colouration falls quickly. The nature of the data set means that there are small balls for much of the considered range, only above $\epsilon=0.7$ do we see larger values for the minimum ball size in panel (d). The total number of connections is falling throughout the range, which does differ from the artificial cases. There are many potential causes for this pattern, bu all confirm the cloud for this section does not share the properties of the noise cloud. On all plots we add vertical lines at $\epsilon=0.2$, $\epsilon=0.4$ and $\epsilon=0.6$ to show the radii for which example plots are developed. These example plots follow in Figure \ref{fig:alt2}.

Our example graphs are shown without the ball number in panels (b) and (c) of Figure \ref{fig:alt2} for clarity. Given the very low proportion of firm failures it is immediate that the majority of the balls have 0 failed firms and are coloured red. A result identified in \cite{qiu2020refining} is that the failed firms occupy a very small subspace of the whole cloud, and this is repeated here in those few balls that do have a proportion of failing firms. In the larger panel (a) failures may be found in balls 2, 7 and 10, which are connected at the top of the shape. Those points in the overlap of 7 and 6, 11, and 5 are not failures as the colouration of those latter three balls is all 0. A concern for the identification of the fail region from $\epsilon=0.6$ is that there are some failing firms in balls 1 and 3, their colour is not quite as red as the others. The proportion is below 0.0005 but that may still be seen as conveying some risk to lenders. Of more use is the $\epsilon=0.4$ that is considered in \cite{qiu2020refining}. Reducing the radius can produce a lot more balls as panel (b) shows. Here again the separation of the failing firms is strong, including one ball where 20\% of the constituent firms failed. However looking at the BM graph it is hard to fully visualise the message from the data. 

Within the \cite{altman1968financial} example the ability of BM to identify regions of the characteristic space in which firms fail is of high value. This information may then inform further modelling, or simply act as a useful guide to help lenders identify potentially safe firms from within the distress zone that \cite{altman1968financial} suggests. 

\subsection{Stock Market Direction Forecasts}

For the second example we consider the use of financial variables to forecast one month ahead stock movements. Our example follows the discussion in \cite{nyberg2011forecasting} and \cite{nyberg2013predicting} and makes use of the 3 month treasury bill rate, 10 year constant maturity treasury bond rate and the first differences thereof. We denote these variables as 3M, 10Y, $\Delta$3M and $\Delta$Y10. Further use is made of the term spread (term), the dividends to price (DP) and earnings to price (EP) ratios and the volatility (Vol) of the market returns\footnote{Following \cite{christoffersen2006financial} this is calculated as the sum of the squared daily returns for the period.} As such we have a 8 dimensional point cloud. Colouration is then a binary indicator which takes the value one if the return on the S\&P 500 exceeds the three month treasury bill rate in the subsequent month. Our sample runs from January 1960 to September 2020, with the market return exceeding the risk free rate in 57\% of these periods.

\begin{sidewaystable}
    
   % \begin{small}
        \caption{Summary of Stock Market Direction Prediction Data}
        \label{tab:ny01}
        \begin{center}
   \begin{tabular}{l c c c c l l c c c c c c c c}
            \hline
            \multicolumn{5}{l}{Panel (a): Summary Statistics} && \multicolumn{8}{l}{Panel (b): Correlations} \\
            & Mean & s.d. & Min & Max && 3M & $\Delta$3M & 10Y & $\Delta$10Y & Term & DP & EP & Vol \\
            \hline
             3M&4.579&3.221&0.01&16.3 && 1 &&&&&&&\\
             $\Delta$3M &-0.004&0.431&-4.62&2.61 &&0.076&1 &&&&&&\\
             10Y&6.09&2.952&0.62&15.32 &&0.923&0.015&1 &&&&&\\
             $\Delta$10Y&-0.005&0.282&-1.76&1.61&&0.089&0.58&0.069&1&&&&\\
             Term&1.511&1.236&-2.65&4.42&&-0.4&-0.161&-0.018&-0.067&1&&&\\
             DP&0.339&0.133&0.135&0.733&&0.733&-0.035&0.78&0.015&-0.048&1&&\\
             EP&0.734&0.308&0.139&1.709&&0.737&-0.03&0.721&0.016&-0.198&0.882&1&\\
             Vol & 0.002 & 0.005 & 0.000 & 0.081 &&-0.096&-0.137&-0.074&-0.112&0.074&-0.036&-0.048&1\\
             Up &0.567&0.496&0&1&&&&&&&&&\\
             \hline 
        \end{tabular}
    
    \end{center}
  %  \end{small}
\raggedright
\footnotesize{Notes: Summary statistics and pairwise correlations for the firm failure data set. Data used in constructing the point cloud includes the 3 month treasury bill rate, 10 year constant maturity treasury bond rate and the first differences thereof. We denote these variables as 3M, 10Y, $\Delta$3M and $\Delta$Y10. Further use is made of the term spread (term), the dividends to price (DP) and earnings to price (EP) ratios and the volatility (Vol) of the market returns. Panel (a) reports the mean, standard deviation, minimum and maximum for each variable. Panel (b) provides the pairwise correlations between the variables. $n=693$}
\end{sidewaystable}

Summary statistics in Table \ref{tab:ny01} help understand the data that we are considering here. Panel (a) gives the mean, variance and range information for each of the variables. Meanwhile panel (b) shows that the correlations between some of the variables in the cloud is high. Dividends to price and earnings to price are correlated at 0.882, whilst the three month and ten year interest rates have a correlation of 0.923. Such high correlation is problematic for the inclusion of all variables within a regression model but does not create a challenge for BM. These high correlations are similar to those for the five part artificial cloud considered earlier. We may take differences in mean value and ranges from panel (a) of Table \ref{tab:ny01} that make it optimal to first normalise the data before applying the BM algorithm. 

\begin{table}
    \begin{center}
        \caption{Stock Direction Forecasting: Summary}
        \label{tab:ny1}
        \begin{tabular}{l c c c c c }
        \hline

        Radius ($\epsilon$) & Balls &  $\Delta$Size & $\Delta$Col & Zero & Con \\
        \hline
        0.1&254.6&32.7&1&126.8&1.21\\
         &(3.67)&(1.74)&(0)&(3.12)&(0.06)\\
        0.2&87.51&84.66&1&28.34&2.47\\
        &(2.65)&(8.91)&(0)&(1.2)&(0.19)\\
        0.4&23.32&317.73&1&4.24&2.36\\
        &(1.68)&(41.33)&(0)&(0.45)&(0.29)\\
        0.5&15.27&434.18&1&2.17&2.30\\
        &(1.33)&(54.88)&(0)&(0.38)&(0.29)\\
        0.8&5.62&594.7&0.27&0&1.39\\
        &(0.75)&(39.95)&(0.07)&(0.00)&(0.22)\\

\hline
        \end{tabular}
    \end{center}
\raggedright
\footnotesize{Notes: Axes reports the number of axes which define the point cloud, Balls is the total number of balls within the BM graph, $\Delta$Size is the difference in size between the smallest and largest ball, $\Delta$Col is the difference between the highest and lowest colouration value for any ball within the graph, Zero is the number of balls for which there is no connectivity to any other ball and Con. is the average number of connections per ball amongst those balls that have at least one connection. All figures are the means from the 10000 repetitions at each point number, with figures in parentheses being the standard deviation across all values within the 10000 repetitions. The data cloud follows the discussion in \cite{nyberg2011forecasting} and \cite{nyberg2013predicting} and makes use of the 3 month treasury bill rate, 10 year constant maturity treasury bond rate and the first differences thereof. Further use is made of the term spread, the dividends to price and earnings to price ratios and the volatility of the market returns.}
\end{table}

BM is run on the normalised data for intervals of $\epsilon$ of 0.05 over the range 0.1 to 0.8. For each radius 10000 BM graphs are constructed and the summary statistics reported in Table \ref{tab:ny1}. Measures are those developed in Section \ref{sec:meas}. As in all other cases the increasing of the radius causes a reduction in the number of balls and an increase in the size of balls. In Table \ref{tab:ny1} we note an increase in the difference between the largest and smallest ball as captured in $\Delta$Size. A positive excess return to the S\&P 500 index is a binary variable and therefore a ball with all increases has a colouration of 1 and a ball with only decreases has a colouration of 0. As the radius increases the difference between the highest and lowest colouration moves from close to 1 to much closer to 0. This may be seen in the move from $\epsilon=0.5$ to $\epsilon=0.8$. Intuitively the number of disconnected balls falls again. Because the number of balls is falling we see a reduction in the average number of connections per connected ball. 

\begin{figure}
    \begin{center}
        \caption{Ball Radius: Stock Market Direction Forecasting}
        \label{fig:ny1}
         \begin{tabular}{c c}
             \includegraphics[width=7cm]{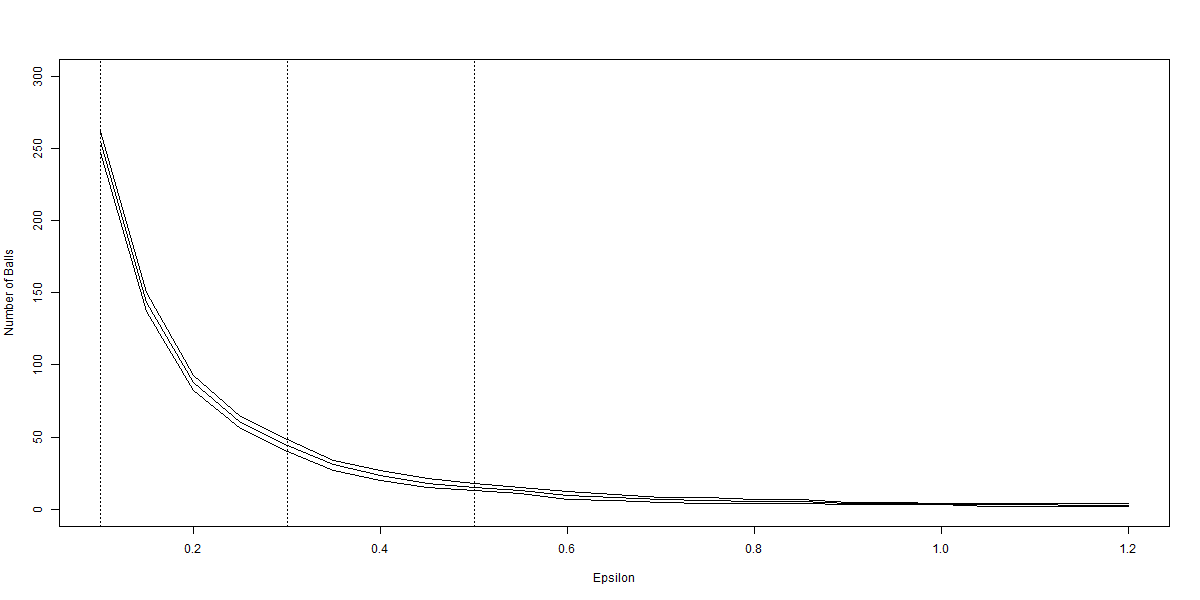} &
             \includegraphics[width=7cm]{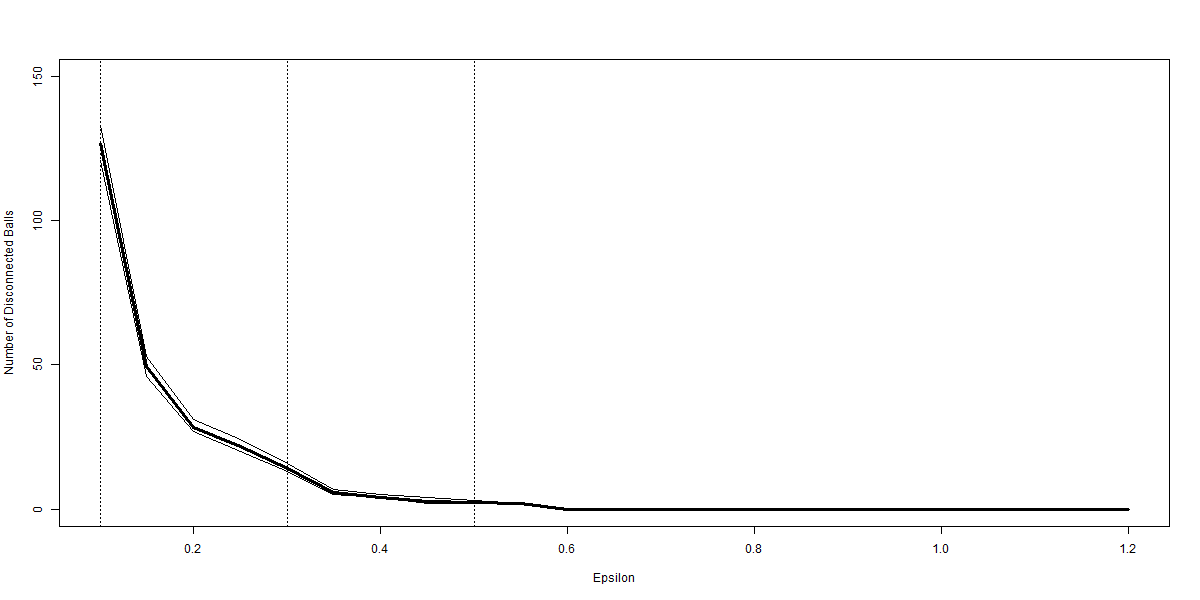} \\
             (a) Number of Balls & (b) Number of Zero Connection Balls \\
             \includegraphics[width=7cm]{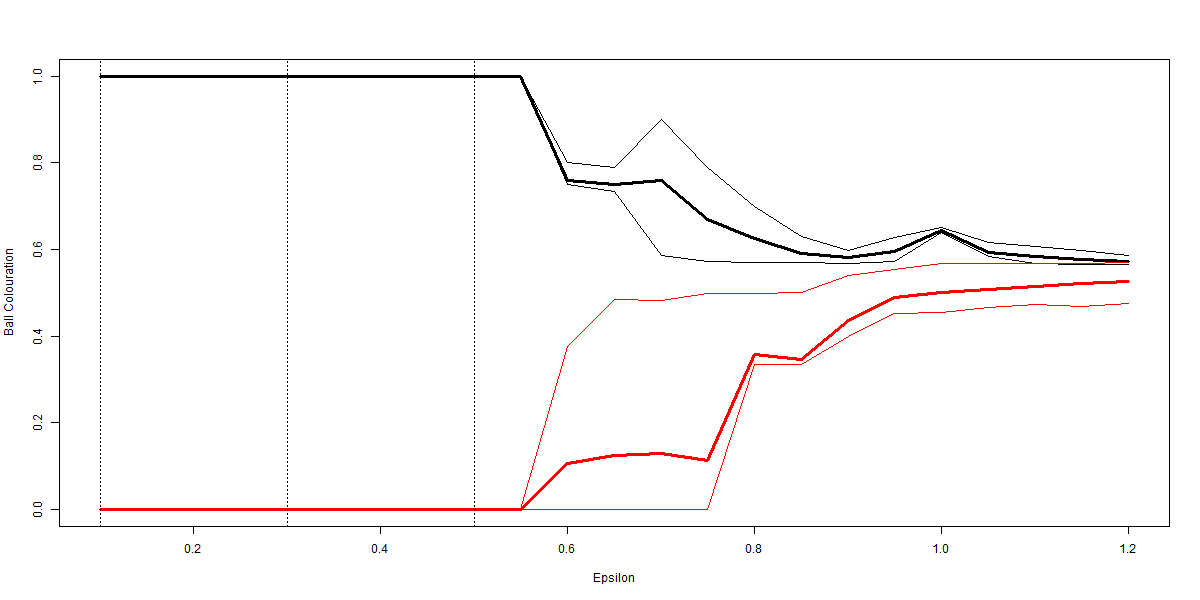} &
             \includegraphics[width=7cm]{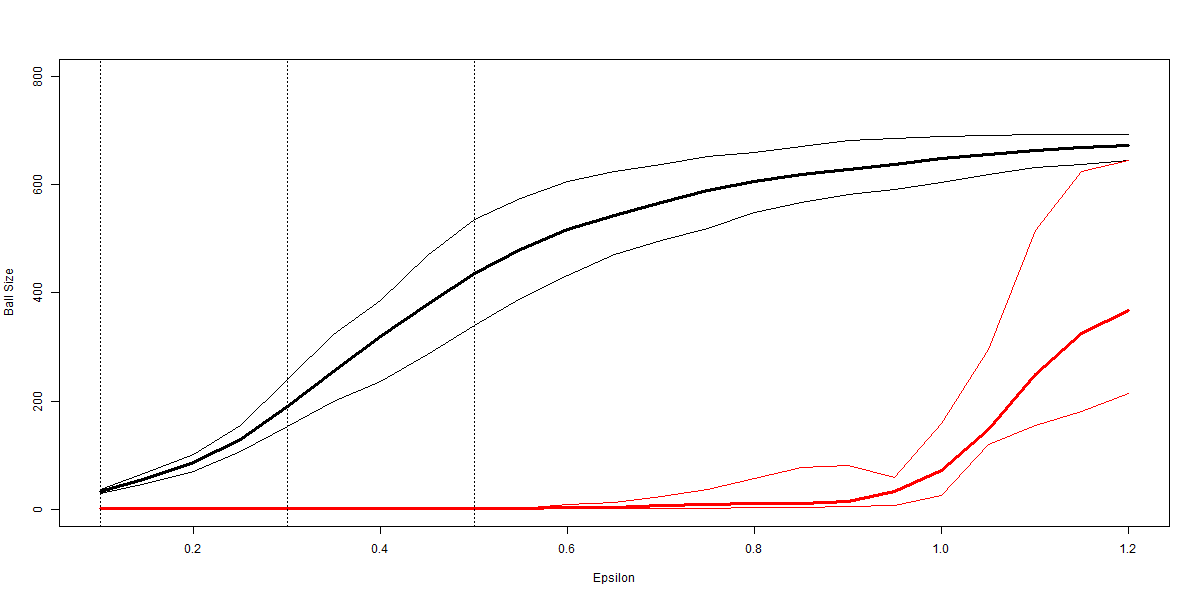} \\
             (c) Colouration & (d) Ball Sizes \\
             \includegraphics[width=7cm]{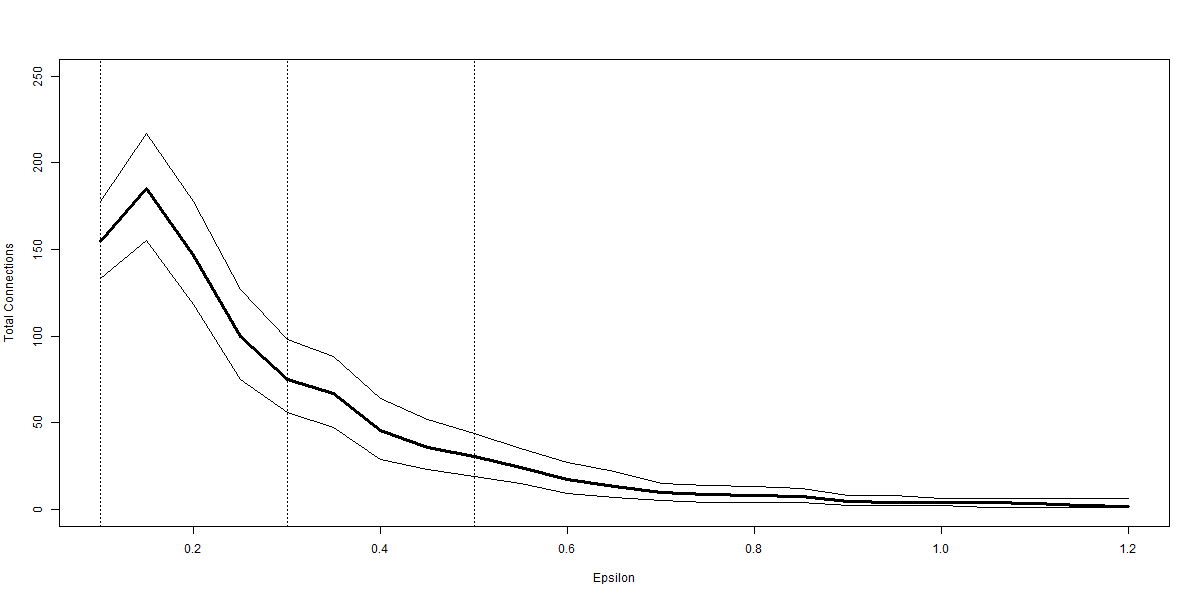} &
             \includegraphics[width=7cm]{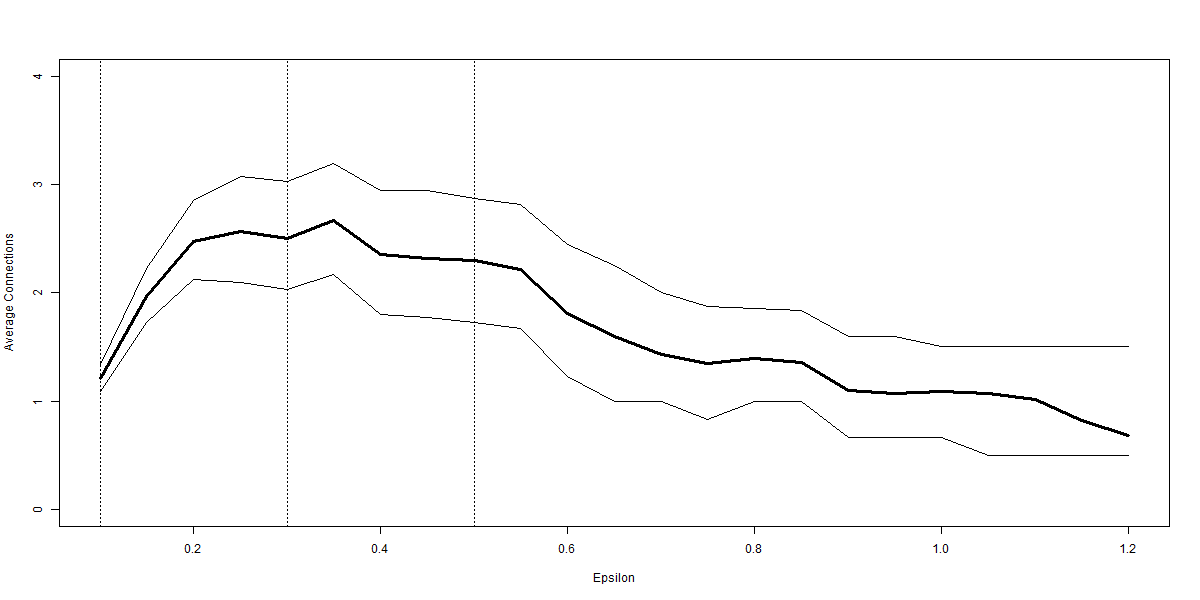}\\
             (e) Total Connections & (f) Average Connections\\
        \end{tabular}
    \end{center}
    \raggedright
    \footnotesize{Figures plot the impact of the radius of the balls, $\epsilon$, used in the construction of the BM graph for the credit scoring example. In each case 10000 repetitions of the BM algorithm \citep{dlotko2019ball} are implemented. A thick line is used to denote the mean from the repetitions, thinner lines denoting the 95\% confidence interval there around. Panel (a) reports the number of balls, and panel (b) the number of balls which have 0 connections to any other balls. Panels (c) and (d) also use red lines to show the maximum and minimum colouration and ball size respectively. Panel (e) reports the total number of connections within the graph, this informs on the points within the overlaps of balls and hence the density of the graphs. Panel (f) plots the average number of connections amongst connected balls. In the case that there are no connected balls then this figure is set to 0. The data cloud follows the discussion in \cite{nyberg2011forecasting} and \cite{nyberg2013predicting} and makes use of the 3 month treasury bill rate, 10 year constant maturity treasury bond rate and the first differences thereof. Further use is made of the term spread, the dividends to price and earnings to price ratios and the volatility of the market returns. All variables are normalised onto the interval $[0,1]$ prior to implementation of the BM algorithm.}
\end{figure}

Where Table \ref{tab:ny1} is useful in clarifying the broad picture, the graphical representations developed in Section \ref{sec:meas} can provide a more direct visualisation of the robustness of the BM graph. Figure \ref{fig:ny1} reveals many consistencies with the artificial examples as the number of balls falls rapidly with even a small increase in the ball radius. This reduction comes with a reduction in the number of zero connection balls, the fall in numbers being faster than the overall ball number. The hump shape of panels (e) and (f) in Figures \ref{fig:br1} and \ref{fig:br2} can be seen at the very low radii, but the effect is not as pronounced. Recalling that the range of possible outcomes is $[0,1]$, the horizontal line on the maximum in panel (c) of Figure \ref{fig:ny1} reveals that many of the balls in which the market always increases in the following month do not connect into other balls until $\epsilon$ is sufficiently large. Results for the minimum colouration remain at 0 until a similar radius. We note that the rise in the minimum colouration is large compared to the fall in the maximum. Panel (d) completes the picture by showing how the largest balls grow steadily, whilst there remain balls with just one region in until a high radius. This, and the results on colouration, remind that there are some periods where the market characteristics are very different from normal times.

\begin{figure}
    \begin{center}
        \caption{Stock Market Direction Forecasting Example Ball Mapper Graphs}
        \label{fig:ny2}
        \begin{tabular}{c c}
             \multicolumn{2}{c}{\includegraphics[width=10cm]{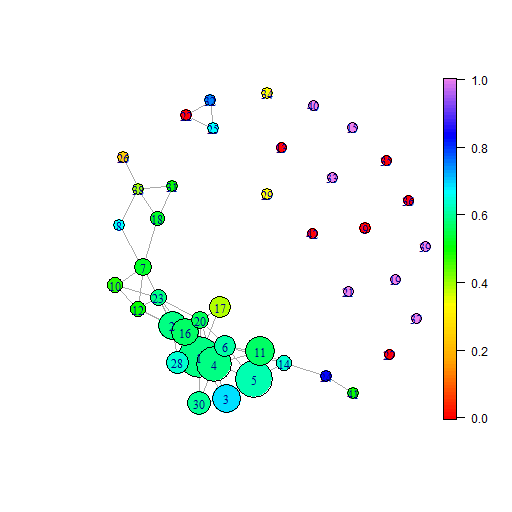}}\\
             \multicolumn{2}{c}{(a) Epsilon = 0.3} \\
             \includegraphics[width=7cm]{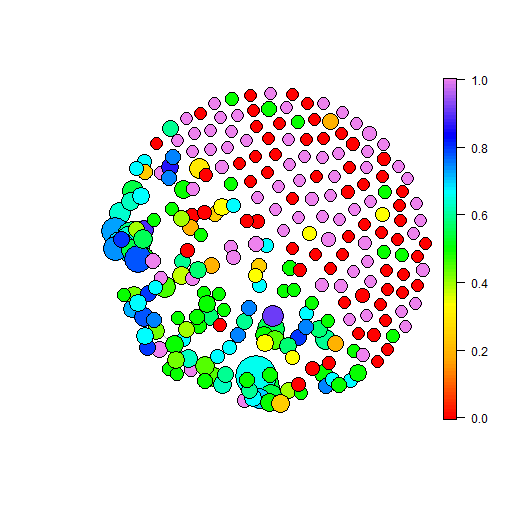} & \includegraphics[width=7cm]{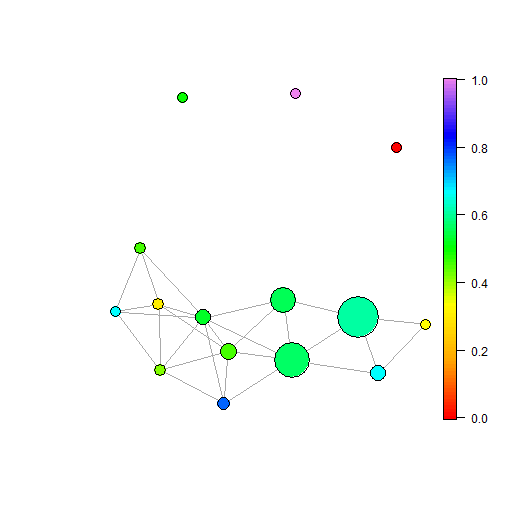} \\
             (b) Epsilon = 0.1 & (c) Epsilon = 0.5
        \end{tabular}
    \end{center}
\raggedright
\footnotesize{Notes: Plots provide three examples of BM graphs for the stock market direction prediction example. Colouration is based on the proportion of points within a ball for which the subsequent month sees an increase in the S\&P 500 index. The data cloud follows the discussion in \cite{nyberg2011forecasting} and \cite{nyberg2013predicting} and makes use of the 3 month treasury bill rate, 10 year constant maturity treasury bond rate and the first differences thereof. Further use is made of the term spread, the dividends to price and earnings to price ratios and the volatility of the market returns. All variables are normalised on the interval $[0,1]$.}
\end{figure}

Through the example plots we can see that the increases in the S\&P 500 index are more concentrated at one end of the main shape, this comes through strongly in all three panels. Panel (a), with $\epsilon=0.3$, shows the most obvious grading of the colour scheme across the space. High correlations identified in Table \ref{tab:ny01} manifest in the narrow shape of the central collection of balls. At the high end of this we see the proportion of rises in the subsequent month being 70\% or higher. Indeed given that across the whole sample 56\% of months have increases. It is also informative how many balls at the lower end of the shape have proportions below 40\%. The lowest colouration is around 0.2. Hence there is no ball where the market never rises, and indeed there is none where the market always rises. However, the information from the data shows that these predictors combine to give a very good signal of the market movement. When the radius is made larger we see that the biggest ball is the red with a low proportion of increases. This connects into blue balls with 70\% of cases having a subsequent rise. The overlap balls that create the edge contribute greatly to the greater colouration relative to the lowest balls at smaller $\epsilon$. 

Considering the examples in Figure \ref{fig:ny2} prompts natural interest in the characteristics in that upper part of the cloud. As a next stage we may plot the same cloud but colour according to the eight axes that make up the cloud. We plot these without ball numbers, but the correspondence to Figure \ref{fig:ny2} is immediate.

%\section{Hypothesis Testing}

%Using multiple repetitions at the same radius we are able to understand the proportion of graphs in which certain features arise.

%Examples might include:

 %   \item Number of times the Brexit voting proportion is higher in a fewer number of balls than the number for which remain is higher - test the conclusion that the Brexit vote was more concentrated. 
  %  \item All points begin as their own ball (unless their co-ordinates are identical) and therefore every combination found within a ball must appear for some radius. Can identify the point at which combinations form.
   % \item May imagine a situation where one constituency is a remain voter but becomes combined with a brexit voting constituency resulting in an average vote for leave. If this ball expands to also include more remain then the story is not as interesting as if it expands to include more leave. 

\section{Summary}
\label{sec:discuss}

Topological data analysis ball mapper (BM) offers a consistent and robust way to visualise multiple dimensional data sets in an abstract two-dimensional form. It's strength lies in the way these representations embed information on connectivity, data density and correlation. Moreover the scope to visualise outcome functions across the space adds a further dimension of value. Within Finance there are many examples where multiple explanatory factors are linked to an outcome, be that the modelling of return movements on market characteristics, or the consideration of corporate financial performance across accounting data. This paper has provided a comprehensive overview of the functionality of BM, the impacts of key properties of data sets on BM plots and some examples of additional ways in which BM graphs may be understood. Two applications have been discussed to show BM in action.

BM is just one of the ways in which topological data analysis is being used to create covers of data sets for further analysis. Alternative forms lack the stability and ease of interpretation of the BM algorithm; these being key features for financial data. There are natural parallels with clustering, but critically BM is looking to represent as best as possible with evenly sized balls rather than focusing on a target number of groups. Where clustering reflects density, the BM graph may better represent connectivity and density precisely because the balls themselves are not influenced by any data feature. As yet there is no algorithm to determine the optimal ball radius, but as demonstrated it is often more useful to understand data from multiple radii. An ongoing theoretical work stream seeks to provide more answers on the determination of optimal radii.

Notwithstanding the opportunities for development there remains ample scope for the application of BM in Finance. Whether simply using the algorithm to get an early impression of the data, or conducting more detailed examination of the BM graphs there is much which can be done. Value in visualisation is well understood, the present work highlights how more of the potential within data may be unlocked through BM.

\bibliography{tdaforef}
\bibliographystyle{apalike}

\end{document}